\documentclass[aps,prfluids,print,superscriptaddress]{revtex4-2}
\usepackage{graphicx}
\usepackage{tikz}
\usepackage{mathrsfs}
\usepackage{todonotes,comment,lineno}
\usepackage{amsmath,amsfonts}
\usepackage[font=small,skip=0pt,justification=raggedright]{subcaption}
\usepackage[font=small,skip=0pt,justification=raggedright]{caption}
\usepackage{todonotes}
\usepackage{siunitx}
\usepackage[normalem]{ulem}

\newcommand{\bx}{\mathbf{x}}
\newcommand{\bw}{\mathbf{w}}
\newcommand{\bz}{\mathbf{z}}
\newcommand{\by}{\mathbf{y}}
\newcommand{\bk}{\boldsymbol{\kappa}}

\newcommand{\momG}[1]{\int \zeta^{#1} G(\zeta) d \zeta}

\newcommand{\pbar}{\overline{\psi}}
\newcommand{\N}{N}
\newcommand{\C}{\mathscr{C}}

\pgfmathdeclarefunction{gauss}{2}{%
    \pgfmathparse{1/(#2*sqrt(2*pi))*exp(-\left( x-#1)^2)/(2*#2^2))}%
}

\newcommand{\solidrule}[1][1cm]{\rule[0.5ex]{#1}{0.75pt}}
\newcommand{\dashedrule}{\mbox{%
      \solidrule[2mm]\hspace{1mm}\solidrule[2mm]\hspace{1mm}\solidrule[2mm]}}

\newcommand{\solidruled}[1][0.5cm]{\rule[0.5ex]{#1}{0.75pt}}
\newcommand{\dashedruled}{\mbox{%
      \solidrule[1mm]\hspace{0.5mm}\solidrule[1mm]\hspace{0.5mm}\solidrule[1mm]}}

\definecolor{mplblue}{RGB}{31,119,180}
\definecolor{mplred}{RGB}{214,39,40}
\definecolor{mplgreen}{RGB}{44,160,44}
\definecolor{mplorange}{RGB}{255,127,14}
\definecolor{mplpurple}{RGB}{148,103,189}
\definecolor{mplbrown}{RGB}{140,86,91}
\definecolor{mplpink}{RGB}{227,119,194}
\definecolor{mplyellow}{RGB}{188,189,34}
\definecolor{mplcyan}{RGB}{23,190,207}
\definecolor{mplblack}{RGB}{0,0,0}
\definecolor{mplwhite}{RGB}{255,255,255}

\newcommand\rdmst{\bgroup\markoverwith{\textcolor{red}{\rule[0.5ex]{2pt}{0.4pt}}}\ULon}
\newcommand\gryst{\bgroup\markoverwith{\textcolor{red}{\rule[0.5ex]{2pt}{0.4pt}}}\ULon}

\def\cF{\mathcal{F}}

\def\rdmr#1{{#1}}
\def\gry#1{{#1}}
\def\rdm#1{{#1}}
\def\gryr#1{{#1}}

\begin{document}

\title{Effects of resolution inhomogeneity in large-eddy simulation}

\author{Gopal R. Yalla}
\affiliation{The Oden Institute for Computational Engineering and Sciences, The University of Texas at Austin}
\author{Todd A. Oliver}
\affiliation{The Oden Institute for Computational Engineering and Sciences, The University of Texas at Austin}
\author{Sigfried W. Haering}
\affiliation{The Oden Institute for Computational Engineering and Sciences, The University of Texas at Austin}
\affiliation{Sandia National Laboratories}
\author{Bj{\"o}rn Engquist}
\affiliation{The Oden Institute for Computational Engineering and Sciences, The University of Texas at Austin}
\affiliation{Department of Mathematics, The University of Texas at Austin}
\author{Robert D. Moser}
\affiliation{The Oden Institute for Computational Engineering and Sciences, The University of Texas at Austin}
\affiliation{Department of Mechanical Engineering, The University of Texas at Austin}

\date{\today}

\begin{abstract}
Large Eddy Simulation (LES) of turbulence in complex geometries is often
conducted using strongly inhomogeneous resolution. The issues
associated with resolution inhomogeneity are related to the noncommutativity of
the filtering and differentiation operators, which introduces a commutation term
into the governing equations. Neglect of this commutation term gives rise to
commutation error.  While the commutation error is well recognized, it is often
ignored in practice. Moreover, the commutation error \rdm{arising from the
implicit part of the filter} (i.e., projection onto the underlying
discretization) has not been well investigated.  Modeling the commutator between
numerical projection and differentiation is crucial for correcting errors
induced by resolution inhomogeneity in practical LES settings, which typically
rely solely on implicit filtering. \gryr{Here, we employ a multiscale
asymptotic analysis to investigate the characteristics of the commutator. 
This provides a statistical description of the commutator, which
can serve as a target for the statistical characteristics of a
commutator model.
Further, we investigate how commutation error manifests in simulation
and demonstrate its impact on the convection of a packet of homogeneous
isotropic turbulence through an inhomogeneous grid. A connection is made
between the commutation error and the propagation properties of the
underlying numerics. A modeling approach for the commutator is proposed that
is applicable to LES with filters that include projections to the discrete
solution space and that respects the numerical properties of the LES
evolution equation. It may also be useful in addressing other LES modeling
issues such as discretization error.}
\end{abstract}

\maketitle

\section{Introduction \label{sec:intro}}
The greatest impediment to the use of computer simulation for reliable
prediction of many high Reynolds number complex turbulent flows is representing
the effects of turbulence. Large Eddy Simulation (LES) has long been considered
a promising computationally tractable solution to modeling turbulent flows.
However, several challenges of LES modeling must still be addressed before this
promise is fulfilled. In particular, LES models are typically formulated under
the assumptions of isotropic unresolved turbulence in equilibrium with the large
scales, homogeneous isotropic filtering/resolution, and accurate representation
of all resolved scales by the underlying numerics. All of these assumptions are
typically violated in practice when simulating high Reynolds number complex
turbulent flows. In the work reported here, we address the issues that arise
from inhomogeneous filtering/resolution. \rdm{Note that by inhomogeneity
of an LES filter or resolution, we mean that the filter or resolution
characteristics vary in space; this should not be confused with
homogeneity or inhomogeneity of the turbulence.}

The challenges posed by resolution inhomogeneity arise because, in this case, the
filter that defines the resolved scales does not commute with
spatial differentiation. This effect is represented by a commutation term,
which should appear in the LES evolution equation.
When this effect is neglected it gives rise to commutation error,
which was first analyzed in detail by \citet{ghosal1995basic}. Since then, several investigators have
acknowledged the significant impact commutation error can have on an LES solution
\citep{van1995family,vasilyev1998general,marsden2002construction,haselbacher2003commutative,
iovieno2003variable,sagaut2006large,meneveau2000scale,girimaji2013closure,haering2015anisotropic,fureby1997mathematical,hamba2011analysis}.
Despite this, the commutation term is often neglected in practice
because of the modeling challenges involved.  Nevertheless, modeling the
commutation term is crucial for developing robust subgrid stress (SGS) models
for practical LES applications. 

Most of the previous work to address commutation error
has been in the context of explicit filtering. Explicit filters,
which may be employed in addition to the discrete projection that
defines the implicit filter, are used to minimize the effects of
numerical discretization errors by defining a filter width larger than
the discretization scale
\citep{germano1986differential,lund2003use,ghosal1996analysis,chow2003further,kravchenko1997effect,carati2001modelling,gullbrand2003effect,bose2011explicitly,bose2010grid}.
Approximately commuting with differentiation is then viewed as a desirable
property of explicit filters to minimize commutation error. In this context, \citet{van1995family} introduced
a one-parameter family of analytical filters that commute with differentiation
up to a given order in filter width. \citet{vasilyev1998general} 
developed a set of constraints for constructing discrete
filters that commute with differentiation up to a desired order. 
\citet{marsden2002construction} extended the work of \citep{vasilyev1998general}
to unstructured meshes. 

\gryr{However, in an LES, the projection onto a finite dimensional LES solution space
that is inherent in numerical discretization is ultimately responsible for
discarding information about the small-scale turbulence
\citep{Langford1999,hughes2000large}. 
This discrete projection (often referred to as the implicit filter) should therefore
be considered as part of the filter.
The commutator between filtering and differentiation that arises due
to spatially nonuniform numerical discretization
and the commutation error that arises from
neglecting it are the fundamental issues introduced by resolution
inhomogeneity in LES.
They are of particular importance in practical LES
as many applications rely solely on implicit filtering. 
Further, the commutation analysis of
\citet{ghosal1995basic} only applies to smooth formally invertible
filters, not filters that include a discrete projection.
Similarly, the commutative property of an explicit filter (such as those
mentioned above) would only reduce the additional commutation error introduced
by the explicit filter applied in addition to the discrete projection.
These explicit filters do not represent the commutator associated with the implicit filter and so, in
general, do not reduce the corresponding commutation error.}
Neither the commutation
error nor commutator models applicable to
implicit filtering
have been well investigated and are the focus of this work.

In this paper, we analyze the commutator for filters including
discrete projection in Sec.~\ref{sec:ICE} and explore the impacts of
the commutation error in Sec.~\ref{sec:impact}. Strategies to model
the commutation term are explored in Sec.~\ref{sec:modelform}, and
conclusions are offered in Sec.~\ref{sec:conclusion}.

\section{Analysis of the Inhomogeneous Commutator\label{sec:ICE}}
\rdm{To define the large scales of turbulence to be simulated in an
  LES, we define a filter operator $\cF$ which maps turbulent fields
  to large-scale (LES) fields. If $\cF$ is shift invariant (commutes
  with arbitrary shift operators), then the filter is said to be
  \emph{homogeneous} and it commutes with derivative operators. To
  define the large scales, $\cF$ should be smoothing, but to enable
  solution on a computer, we will also insist that the range of $\cF$
  be finite dimensional, thereby including the discretization for numerical
  solution. Thus, as discussed above, $\cF$ includes what is commonly
  called the implicit filter (projection onto the discretization) in addition to any
  explicit filter that may be employed. Unless the discretization is a
  Fourier truncation, such a filter is not homogeneous. Instead, it
  may be ``discretely homogeneous,'' that is, invariant to spatial
  shifts by integral multiples of a discretization scale $\Delta$. For
  example, projections to finite volume, finite element and spline
  solution spaces on uniform grids are discretely homogeneous, while
  the same projections on nonuniform grids are discretely
  inhomogeneous (see \citep{moser2020statistical}). The consequences of
  such resolution inhomogeneity are the subject of the current
  paper. Resolution inhomogeneity is a property of the filter, and
  should not be confused with inhomogeneity of turbulence, which
  refers to turbulence whose \emph{statistics} are not invariant to
  spatial shifts.}

\rdm{Assuming that mean quantities in an LES are well resolved, the
filtered fluctuating Navier-Stokes momentum equations read
\begin{equation}
\frac{\partial \overline{u'_i}}{\partial t} +U_j\overline{\frac{\partial
u'_i}{\partial x_j}}+\overline{\frac{\partial u'_iu'_j}{\partial x_j}}
-\frac{\partial \langle u'_iu'_j\rangle}{\partial x_j}=
-\overline{\frac{\partial p'}{\partial
x_i}}+\nu\overline{\frac{\partial^2 u'_i}{\partial x_j\partial x_j}},
\label{eq:FNS}
\end{equation}
where over-line represent filtering, that is
$\overline{\cdot}=\cF(\cdot)$, $\langle\cdot\rangle$ represents the
mean or expected value, $U_i$ is the mean velocity, and $u'_i$ and
$p'$ are the fluctuating velocity and pressure respectively. To obtain
the fluctuating LES equations, one would like to interchange the order
of filtering and spatial differentiation, but with inhomogeneous
filters, this would introduce commutation error for each of the
filtered derivative terms in (\ref{eq:FNS}). These errors would be
small for the viscous term and the errors in the pressure term would
be subsumed in the treatment of continuity. The remaining terms
representing convection by the mean and the fluctuations are thus of
primary interest here. In many turbulent flows the mean velocity
relative to the grid is much larger than the fluctuating velocity, so
arguably the commutation error for the mean convection term is most
important. Further, there are numerous other issues
associated with modeling $\overline{u'_iu'_j}$ in the fluctuating
convection term, including modeling the subgrid stress
and filtering the nonlinear products used in the modeling
\cite{carati2001modelling}.}
\gryr{For these reasons, we are particularly focused on the effects of resolution
	inhomogeneity on the mean convection term $U_j \overline{\partial
u'_i/\partial x_j}$ here, although the results may be insightful for the nonlinear
term as well.}

\rdm{For the purposes of analysis, it is useful to further simplify the mean convection and} consider
the effects of the resolution inhomogeneity on the filtered one-dimensional advection equation:
\begin{equation}
	\overline{\frac{\partial u}{\partial t} + U \frac{\partial u}{\partial
        x}} = 0 
	\label{eq:Filt1DAdvEq}
        ,
\end{equation}	
where $U$ is the constant convection velocity.
Furthermore, let $\delta/\delta x$ denote a
discrete derivative operator defined on the discrete solution space.
Then (\ref{eq:Filt1DAdvEq}) can be written as
\begin{equation}
    \frac{\partial \overline{u}}{\partial t} + U\frac{\delta \bar{u}}{\delta
    x} = -U\C(u)
    \label{eq:fullcomm1DAdvEq}
    ,
\end{equation} 
where $\C(u) = \overline{\partial u/\partial x}
- \delta \overline{u}/\delta x$
is the commutation term (similar to a commutator).
The commutation term $\C(u)$ can be decomposed into an inhomogeneous and homogeneous
part as $\C(u) = \C^I(u) + \C^H(u)$
where, 
\begin{equation}
\C^I(u)=\overline{\frac{\partial u}{\partial x}}
-\overline{\frac{\partial {u}}{\partial \xi}}\frac{d\xi}{d x}\qquad
\C^H(u)= \overline{\frac{\partial {u}}{\partial \xi}}\frac{d\xi}{d x}
-\frac{\delta \overline{u}}{\delta x}
\label{eq:commerror_split}
,
\end{equation}
and $\xi$ is the new spatial coordinate in which the grid or resolution is uniform
as in \citep{ghosal1995basic}. The homogeneous part represents the effects
of the numerical discretization error in $\delta/\delta x$ and is
non-zero even if the resolution is homogeneous. The inhomogeneous part
characterizes the effects of the inhomogeneous resolution \rdm{and is
 zero when the resolution is homogeneous.} 
This formulation can be extended to three dimensions as well as
to the full Navier-Stokes equations
\gry{\citep{moser2020statistical}}. \rdm{However, for the nonlinear terms in
  the Navier-Stokes equations there is an additional complication in
  the definition of the discrete derivative operator
  \citep{carati2001modelling}, which impacts the
  definition of the commutator \citep{moser2020statistical}.}

\rdmr{There is also a subtlety to this decomposition when the filter
  is discretely homogeneous in $\xi$ since discretely
  homogeneous filters can be decomposed into a homogeneous filter
  followed by sampling on a uniform grid. If the homogeneous filter is
  defined to include a Fourier cutoff with cutoff wavenumber less than
  the Nyquist wavenumber for the grid, the sampling does not discard
  information, and this is usually what is intended when defining the
  filter. When applying this to the decomposition
  (\ref{eq:commerror_split}), we can choose to include the effect of sampling as
  part of $\C^I$ or $\C^H$. Here, we will generally choose the
  latter, so that $\C^I$ is expressed in terms of a non-invertible
  homogeneous filter in $\xi$.}

In numerical analysis one generally aspires to make the resolution
sufficiently fine so that $\C^H$ is negligible, but an LES is by
definition under resolved and so $\C^H$ generally needs to be
considered. Models for $\C^H$ are not available, so it is often
neglected leading to discretization error \cite{ghosal1996analysis,chow2003further,kravchenko1997effect}.
Although discretization error is widely acknowledged by LES practitioners, 
the inhomogeneous term $\C^I$ has received less attention.
The effects of neglecting $\C^I$,
as well as modeling strategies for $\C^I$, are not well understood
and are the primary focus of the current
paper.

%\begingroup
%\color{blue}

\subsection{Asymptotic Analysis of the Inhomogeneous Commutator}
In the seminal work of \citet{ghosal1995basic}, the commutator is estimated
through Taylor series analysis allowing the commutator to be
analyzed. However there were limitations of that work. First, the
analysis uses an approximate inversion of the filter operator, and as
such is formally only applicable to invertible filters, and thus not
to filters including discrete projections as considered here. And
second, in simplifying the expansions, an ad hoc ordering is employed
which resulted in the neglect of terms that a more careful analysis
would identify as important. Here we pursue a similar program using
asymptotic analysis with the goals of placing the results of \citet{ghosal1995basic}
for invertible filters on firmer ground
(Sec.~\ref{sec:ghosal_review}), and of developing statistical
characterizations of the commutator applicable to non-invertible
filters (Sec.~\ref{sec:statistics}). The results for invertible
filters are of interest here because they can provide guidance on
appropriate forms and dependencies for a model of the commutator.
This may be valuable because filters that include discrete projections
can be considered to be limits of sequences of invertible filters.

\subsubsection{Series Representation of the Inhomogeneous Commutator}
  \label{sec:ghosal_review}

While \citet{ghosal1995basic} employ an ad hoc ordering, their results can be
consistently interpreted asymptotically. As described in
Appendix~\ref{sec:multiscale_deconvolution}, the analysis in \citep{ghosal1995basic} employs a mapping of
the physical space $x$ to a mapped space $\xi$ in which the resolution
is uniform to define the filtering operator.  A Taylor series expansion yields a
series representations for $\C^I$ that is valid asymptotically for
$\Delta_\xi\rightarrow 0$, where $\Delta_\xi$ is the uniform
resolution in $\xi$ space. This expansion is in
terms of the derivatives of the unfiltered field $u$. To express the
commutator in terms of the derivatives of $\overline u$, the filter is
inverted through another asymptotic expansion in $\Delta_\xi$. But, to
properly order the expansion, the way in which the derivatives of $\overline u$ scale
with $\Delta_\xi$ must be determined. In \citet{ghosal1995basic}, it is assumed
that $u=e^{i\kappa x}$ and their analysis is consistent with the
assumption that $\kappa\sim \Delta_\xi^{-p}$ for $0<p<1$ (see
Appendix~\ref{sec:app_ghosal}). However,
this is not necessarily consistent with the way the derivatives of
$\overline u$ scale when $u$ is the turbulent velocity.

Assuming the resolution in physical space $\Delta(x)$ is in the
inertial range of a high Reynolds number turbulence, the Kolmogorov
hypotheses imply that
\begin{equation}
\frac{\partial^n \overline u}{\partial x^n}\sim \Delta_\xi^{1/3-n}
\end{equation}
(see Appendix~\ref{sec:multiscale_deconvolution}). With this ordering, the lowest order
expansion for the commutator is given by
\begin{equation}
	\begin{split}
    \C^I(u) &= - M_2 \Delta \frac{d \Delta}{d x}
	\frac{\partial^2 \overline{u} }{\partial {x}^2 } + \left( \frac{M_2^2}{2} -
    \frac{M_4}{6} \right) \Delta^3 \frac{d \Delta}{d
	x}\frac{\partial^4 \overline{u} }{\partial {x}^4}  +
    \dots + C_N \Delta^{N-1} \frac{d \Delta}{d
	x}\frac{\partial^N \overline{u}}{\partial {x}^N} + \dots +
	\mathcal{O}(\Delta_\xi^{4/3}) \\
	&= \frac{d\Delta}{dx}\sum_{n=1}^{\infty}
		C_{2n}\Delta^{2n-1} \frac{\partial^{2n} \overline{u}}{\partial
{x}^{2n}} + \mathcal{O}(\Delta_\xi^{4/3})
	\end{split}
    \label{eq:origCommError}
	,
\end{equation}
where $M_k$ is the $k^{\rm th}$ order moment of the filter kernel, $N$ is
even, and in general, the coefficient $C_j$ on the $j^{th}$ order term
depends on the moments of the filter up to order $j$. In \citet{ghosal1995basic},
only the first term in this series is retained because the other terms
are higher order in $\kappa\Delta\sim \Delta_\xi^{1-p}$, but clearly
this would not be consistent with filtering turbulence in a Kolmogorov
inertial range as (\ref{eq:origCommError}) is.

An alternative approach to developing a series representation of
$\C^I$ is formulated for a different, though related asymptotic
limit. Consider the situation in which the derivative $d\Delta/dx$ is
order $\epsilon$, where $\epsilon$ is asymptotically small. 
In this case, a multiscale asymptotic
analysis of $\overline{u}$ in terms of a fast variable $\tilde x$ and
slow variable $w=\epsilon x$ yields the simple result
$\C^I=-\epsilon\partial \overline u/\partial w+\mathcal{O}(\epsilon^2)$ (see
Appendix~\ref{sec:multiscale}), which can be expressed directly as a
convolution operator applied to the unfiltered field, where the kernel
is in terms of the filter kernel and its derivative (\ref{eq:msCommErr}). As
with the analysis discussed above, a Taylor series representation of
the filter inverse can be applied to produce a series representation
of the commutator in terms of the filtered field and its
derivatives. However, the asymptotic interpretation may be
different. In particular, the $\epsilon\rightarrow0$ limit can be
approached by allowing the length scale $L_\Delta$ over which the
resolution changes
($\frac1{L_\Delta}\sim\frac{1}{\Delta}\frac{d\Delta}{dx}$) to grow
while $\Delta$ remains constant. In this case derivatives of
$\overline u$ as well as $\Delta$ are order one in
$\epsilon$. Alternatively, $L_\Delta$ can remain constant while
$\Delta$ goes to zero, which is equivalent to the previous analysis. In this
case, $\Delta\sim\epsilon$ and for inertial range turbulence, the
$n^{th}$ derivative of $\overline u$ scales as $\epsilon^{1/3-n}$. In
either case, one obtains
\begin{equation}
	\begin{split}
    \C^I(u) = \frac{d\Delta}{dx}\sum_{n=1}^{\infty}
		C_{2n}\Delta^{2n-1} \frac{\partial^{2n} \overline{u}}{\partial
\tilde{x}^{2n}} + \mathcal{O}(\epsilon^q)
	\end{split}
    \label{eq:newCommError}
	,
\end{equation}
where $q=2$ when the asymptotic limit is taken with constant $\Delta$ 
while $q=4/3$ when it is taken at constant $L_\Delta$ (see
Appendix~\ref{sec:multiscale_deconvolution}). This is the
same series as in (\ref{eq:origCommError}).

\rdm{Despite the fact that the above analyses are predicated on the
use of an invertible filter and we are concerned with filters that
include a discrete projection, the characteristics of the commutator
expression provide insights relevant to modeling of the commutation
term.  First, note that to leading order this approximation is
proportional to $d\Delta/d x$, and is a series in the even $x$
derivatives of $\bar{u}$. The lowest order term appears as a viscous
term, which is dissipative when $d\Delta/dx >0$ (i.e. convecting from
fine to coarse resolution), and the higher order terms are
hyperviscous. Similarly, these terms would be anti-dissipative when
convecting from coarse to fine resolution, and thus will create
resolved energy in this case. Clearly this commutator expression is characterizing
the transfer of energy between resolved and unresolved scales as a
consequence of the resolution inhomogeneity. In addition, since each
of the terms in (\ref{eq:origCommError}) and (\ref{eq:newCommError}) are of the same
asymptotic order, they are all equally important, and indeed,
depending on the characteristics of the filter, the higher order
derivative terms could dominate. This suggests that a model of the
commutator formulated as a differential operator should include as
high-order derivatives as possible. It is also interesting to observe
that the asymptotically higher order terms include dispersive terms in
addition to dissipative ones, and that higher order derivatives of
$\Delta$ appear (see Appendix~\ref{sec:app_ghosal}).}

Finally, this analysis may provide clarity on some of the existing
literature surrounding commutation error. In particular, the
deconvolution analysis in \citep{ghosal1995basic} has often been used
to motivate the development of smooth explicit filters whose first
$N-1$ moments are zero so that the commutation error is of explicit
order $\Delta^N$
(e.g, \citep{vasilyev1998general,marsden2002construction,van1995family,haselbacher2003commutative}).
However, this is only meaningful if the derivatives of $\overline u$
scale sufficiently weakly with $\Delta$, as discussed above, so that
the first terms in (\ref{eq:origCommError}) dominate
asymptotically. Unfortunately, for high Reynolds number turbulence,
each term in (\ref{eq:origCommError}) is of the same asymptotic order.  Therefore,
this analysis suggests that constructing filters so that the
coefficients $C_j$ for $j<N-1$ in (\ref{eq:origCommError}) vanish
would likely not render the commutator negligible. The commutator will
thus need to be modeled.

%\endgroup
\subsubsection{\gry{Statistical Analysis of the Inhomogeneous
Commutator}}
\label{sec:statistics}

\rdm{Because an LES filter always includes a projection to the
finite-dimensional numerical solution space, either explicitly or
implicitly, the information in a filtered turbulent field is not
sufficient to determine the evolution of that filtered turbulence
\citep{langford1999optimal}. As a consequence, one can only expect
LES models, including models of the commutator, to match statistical
characteristics of the quantity being
modeled \citep{moser2020statistical}.}
\gry{The challenge lies in identifying the important
statistical characteristics and developing models capable of representing them.}
\rdm{Here we analyze \emph{a priori} statistical properties
of the commutator in terms of statistical characteristics of the
unfiltered turbulence, to inform potential commutator models.}

\rdm{The finite-dimensional projection inherent to LES
filters determines the information available in an LES upon which to base a
model, and so a deconvolution analysis is ill-suited to determining statistical
properties.  Instead, we apply the multiscale asymptotic
analysis discussed in Sec.~\ref{sec:ghosal_review} and detailed in
Appendix~\ref{sec:multiscale_deconvolution} to characterize the statistics of the
commutator. In Appendix~\ref{sec:multiscale_spectral}, such an
analysis is applied to an inhomogeneous
three-dimensional isotropic filter characterized by a slowly varying filter width $\Delta$.
After performing a Fourier transform in
the fast variable for which the filter is homogeneous, 
the commutator between filtering and differentiation
applied to
turbulent velocity fluctuations $u_j$ can be written explicitly as
\begin{equation}
		\widehat{\C_i^I}(u_j) =
		-\frac{\partial \Delta}{\partial x_i} \widehat
		G'(\Delta|\bk|)|\bk|\widehat{u}_j(\bk),
	\label{eq:comm_hat1}
\end{equation}
where $\bk$ is the wavenumber vector, $\widehat G$ is the Fourier
transform of the isotropic filter kernel $G$, and $\widehat G'$ is the
derivative of $\widehat G$ with respect to its argument. 
Note that the ``local Fourier transform''
analysis in \citep{vasilyev2004local} holds in this multiscale
asymptotic sense.}

\rdm{The commutator is a linear operator,
	and (\ref{eq:comm_hat1}) shows that it is proportional to the gradient
of $\Delta$ and its spectrum is proportional to $\widehat G'$. The
commutator thus acts on the wavenumbers over which the filter spectrum
rolls off from order one to zero. These are generally the smallest
resolved scales of the LES. For a Fourier cutoff filter in which  $\widehat G$
is discontinuous at the cutoff wavenumber $\kappa_c$, $\widehat G'$ is
a Dirac delta function at $\kappa_c$, so in this case the commutator acts only at
the slowly varying cutoff.}

\rdm{While (\ref{eq:comm_hat1}) is an explicit expression for the
commutator, it requires knowledge of the unfiltered quantity, which is
generally not available. If the turbulence is being convected by a mean
velocity $U_i$, then the commutator $\C_i^I(u_j)$ arising from the mean
convection term enters the $u_j$
evolution equation as $U_i\C_i^I(u_j)$, and (\ref{eq:comm_hat1}) can be
used to determine the contribution of the commutator to the evolution
of the filtered spectrum tensor, and in particular the
three-dimensional filtered energy spectrum $\overline E(\bw,\kappa)$, to obtain
\begin{equation}
  \tilde\C^I(\overline{E}) = -U_k\frac{\partial
    \Delta}{\partial x_k}\widehat G(\Delta \kappa)\widehat G'(\Delta\kappa)\kappa E(\kappa)\\
\label{eq:comm_E}
\end{equation}
where $\tilde\C$ indicates the contribution of the commutator to the
evolution equation for its argument. This contribution still requires
knowledge of the unfiltered turbulence, in this case the unfiltered
spectrum, but at least in high Reynolds number isotropic turbulence,
Kolmogorov inertial range theory provides a good model for $E$.  This
is useful because \emph{a priori} consistency of a commutator model
with (\ref{eq:comm_E}) is a necessary condition for LES prediction of
the energy spectrum \citep{meneveau1994statistics,moser2020statistical}. Similarly, integrating
(\ref{eq:comm_E}) over $\kappa$ yields the contribution of the
commutator to the evolution of the resolved turbulent kinetic energy
$k^>$, and a necessary condition for LES prediction of $k^>$. For the
special case of a Fourier cutoff filter, the result simplifies
to
\begin{equation}
  \tilde\C^I(k^>)= U_k \frac{\partial \kappa_c}{\partial x_k}
  E(\kappa_c),
	\label{eq:comm_k1}
\end{equation}
which is consistent with the result obtained by \citet{moser2020statistical} by
other means. For coarsening resolution ($U_k\partial \kappa_c/\partial
x_k<0$), the commutator transfers energy to unresolved scales
with the dissipation occurring only at the cutoff
wavenumber. Similarly, for refining resolution ($U_k\partial \kappa_c/\partial
x_k>0$), the commutator transfers energy from the subgrid to the resolved
turbulence at the cutoff wavenumber. Further, when this spectral analysis is applied to the full
nonlinear terms in the filtered Navier-Stokes equations, an additional 
commutator contributes to the evolution of the spectrum tensor, which
can be determined in terms of $\widehat S_{ijk}(\bk)$, the Fourier
transform of the two-point third-order correlation tensor.}

\subsection{Numerical Analysis of Commutation Error\label{sec:numerics}}
  \gryr{When convecting through a coarsening grid, the resolved
    fluctuations in a fine region will be moving into a coarse region
    in which not all the resolved scales can be
    represented. Similarly, solution scales that cannot be resolved in
    a coarse region will become resolvable as the solution convects
    into a finer resolution region.  The previous subsections show how
    the inhomogeneous commutator $\C^I$ is responsible for
    transferring energy between the subgrid and resolved turbulence
    in both these cases. However, notice that the injection of energy
    into the resolved scales is required for the refining resolution
    case to maintain consistency with the definition of the filter,
    but that neglecting this effect will not lead to numerical inconsistencies
    since the coarse region solution is perfectly well represented in
    the fine region. This is not true for flow through a coarsening
    grid. For this reason, our investigation of commutation error
    (neglect of $\C$) in this paper is particularly focused on flow
    through coarsening grids because of the numerical
consistency issues inherent to this case.}

\rdm{ The impact of the commutator and specifically its neglect is affected by
	the characteristics of the discrete derivative operator, which is accounted
	for in the homogeneous commutator $\C^H$. Here, by recalling results from
	numerical analysis
	\cite{trefethen1982group,vichnevetsky1981energy,vichnevetsky1983group,vichnevetsky1981propagation,long2011numerical,vichnevetsky1987wave2,frank2004spurious,ascher2004multisymplectic},
	we consider the impact of neglecting both commutators, as is typical in LES,
	in the case of a filter consisting of just the projection to the
finite-dimensional discrete solution space (i.e. only an implicit filter).} 
  Neglecting the commutator in (\ref{eq:fullcomm1DAdvEq}) gives:
\begin{equation}
    {\frac{\partial \overline{u}}{\partial t} + U \frac{\delta
        \overline{u}}{\delta
	x}} =  0 .
	\label{eq:1DAdvEq}
\end{equation}
We begin by recalling, as an example, the solution of
(\ref{eq:1DAdvEq}) using a second-order centered finite difference
scheme on a uniform mesh with mesh size $\Delta$. The numerical first derivative
is then given by $\delta u_j/\delta x=\tfrac{1}{2\Delta}(u_{j+1}-u_{j-1})$.

\gry{It is well recognized that, for initial conditions of the form $e^{i \kappa x}$, solutions of
	(\ref{eq:1DAdvEq}) take the form $e^{i (\kappa x - \omega t)}$ and propagate at a phase velocity that depends on their
wavenumber \citep{trefethen1982group}.}
The relation $\omega =
\omega(\kappa)$ is called the \textit{dispersion relation}. Individual waves
propagate at a phase speed given by $c(\kappa) = \omega(\kappa)/\kappa$;
however, the evolution of a wave packet, which can be decomposed into
Fourier modes with wavenumbers ranging over a relatively narrow band,
is governed by the \textit{group velocity}: 
\begin{equation}
	\mathcal{G}(\kappa) = \frac{d \omega}{d \kappa}(\kappa)
        . 
	\label{eq:groupvel}
\end{equation}
The group velocity is the velocity at which information and energy
propagate and as such is of great importance in LES.

Substituting the form $u_j = e^{i(\kappa x_j - \omega t)}$ into (\ref{eq:1DAdvEq})
with the second order centered difference scheme
yields the dispersion relation and group velocity:
\begin{equation}
    \omega(\kappa)=U\kappa'(\kappa) = U\frac{\sin(\kappa\Delta)}{\Delta }
    \qquad\mbox{and}\qquad
    \mathcal{G}(\kappa) = U\cos(\kappa \Delta )
    \label{eq:CDdispersion}, 
\end{equation}
where $\kappa'$ is the spectrum of the numerical derivative operator,
which is often referred to as the \textit{effective} (or
\textit{modified}) wavenumber.  Notice that at the Nyquist wavenumber
for the grid, $\kappa_c=\pi/\Delta$, both $\kappa'$ and $\omega$ are
zero. As a consequence there is a wavenumber $\kappa_a\in(0,\kappa_c)$
at which $\omega$ is maximized with value $\omega_{\text{max}}(\Delta)=\omega(\kappa_a)$
($\kappa_a=\kappa_c/2$ for second order
central difference) so that the group velocity is zero. Therefore, for
$\kappa\in(\kappa_a,\kappa_c)$ the group velocity is negative so that
wave packets with wavenumbers in this range will propagate upstream
against the convection velocity. Also note that for any frequency
$\omega<\omega_{\text{max}}(\Delta)$, there are two wavenumbers that
will evolve with that frequency, one with positive and one with
negative group velocity. The wavenumber with positive group velocity
($\kappa<\kappa_a$) is a consistent approximation to a solution of the
advection equation while the other ($\kappa>\kappa_a$) is spurious.
As pointed out by \citet{vichnevetsky1981energy}, a general solution
to (\ref{eq:1DAdvEq}) can therefore be decomposed as $u=p+q$, where $p$
has a forward propagation and is a consistent approximation, and $q$
propagates backwards and is spurious.

\begin{figure}[t!]
\centering 
\begin{subfigure}{0.75\textwidth}
\centering
\hspace{-2em}\includegraphics[width=1\textwidth,height=0.25\textheight]{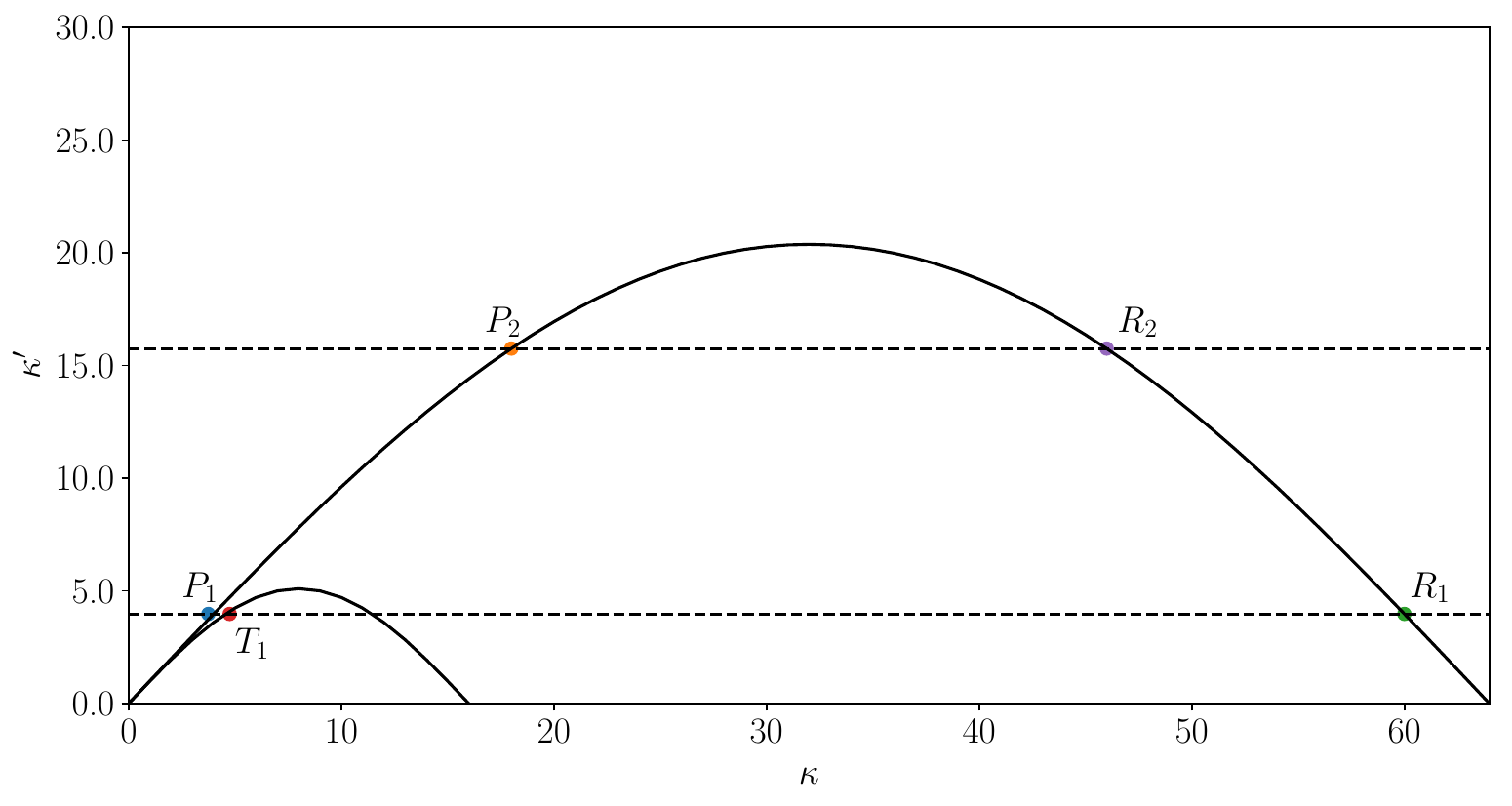}
\caption{}
\label{fig:sharpchangeA}
\end{subfigure}
\vskip\baselineskip
\begin{subfigure}{0.45\textwidth}
\centering
\includegraphics[width=1\textwidth]{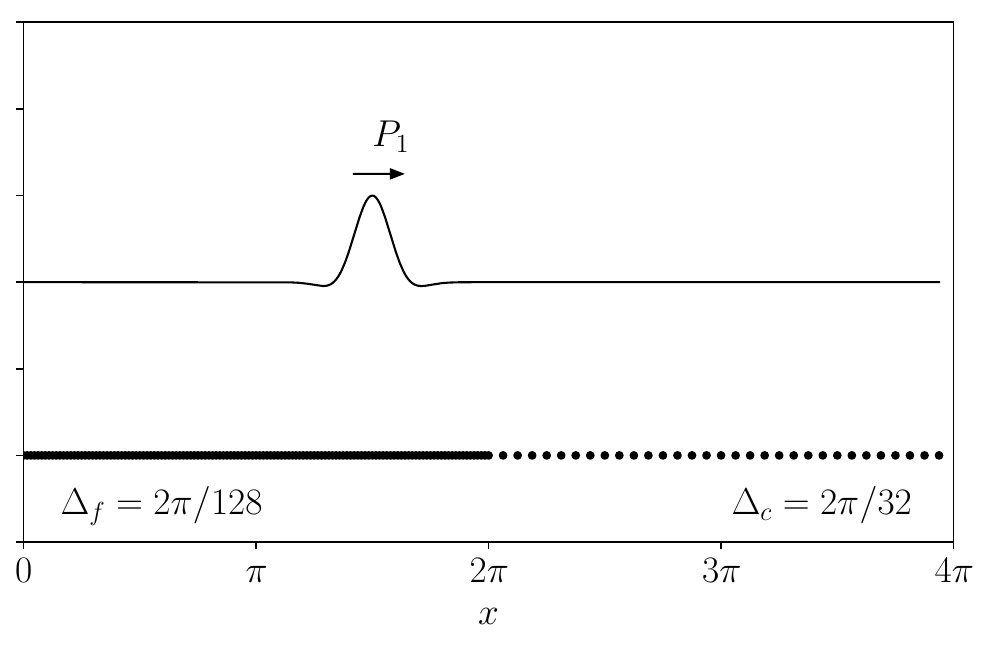}
\caption{}
\label{fig:sharpchangeB}
\end{subfigure}
\hspace{1em}
\begin{subfigure}{0.45\textwidth}
\centering
\includegraphics[width=1\textwidth]{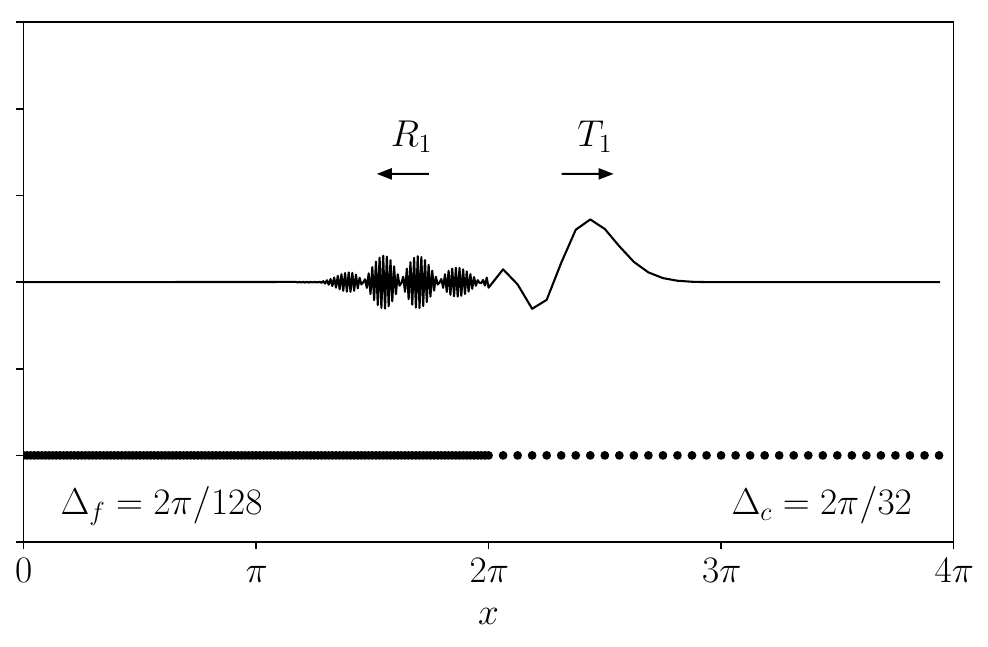}
\caption{}
\label{fig:sharpchangeC}
\end{subfigure}
\vskip\baselineskip
\begin{subfigure}{0.45\textwidth}
\centering
\includegraphics[width=1\textwidth]{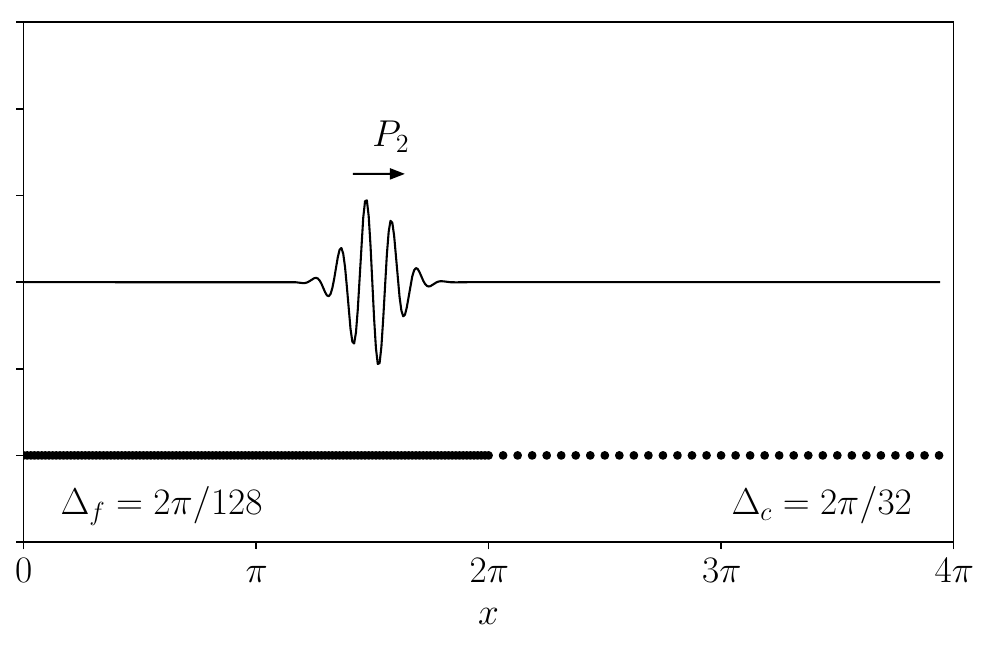}
\caption{}
\label{fig:sharpchangeD}
\end{subfigure}
\hspace{1em}
\begin{subfigure}{0.45\textwidth}
\centering
\includegraphics[width=1\textwidth]{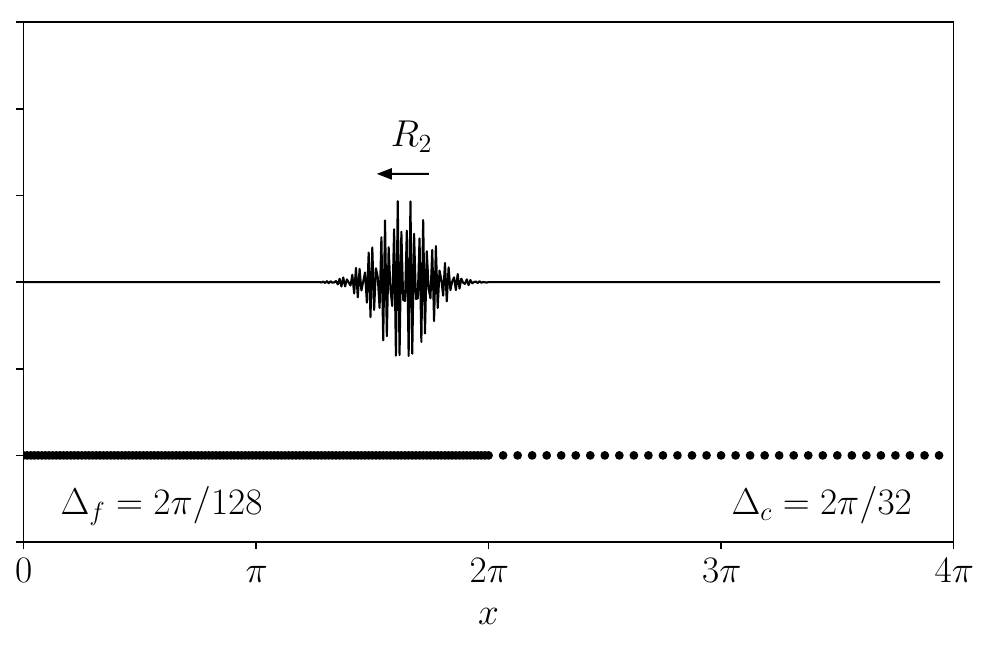}
\caption{}
\label{fig:sharpchangeE}
\end{subfigure}
\caption{\small 
    (a) The second order centered difference dispersion relation for both the
    fine ($\Delta_f = 2\pi/128$) and coarse ($\Delta_c = 2\pi/32$) regions of the
    grid,  (b) An incident wave $P_1$ (\textcolor{mplblue}{$\bullet$}) that can
    be resolved in both the fine and coarse regions.
    (c) The subsequent reflected wave $R_1$ (\textcolor{mplred}{$\bullet$}) and
    transmitted wave $T_1$ (\textcolor{mplgreen}{$\bullet$}) after the $P_1$
    wave has propagated through the resolution change.
    (d) An incident wave $P_2$ (\textcolor{mplorange}{$\bullet$}) that can
    only be resolved in the fine region.
    (e) The subsequent reflected wave $R_2$ (\textcolor{mplpurple}{$\bullet$})
    after the $P_2$ wave has propagated through the resolution change.
}  \label{fig:sharpchange} 
\end{figure}

Consider next a grid \gry{for a domain of length $L=4\pi$} with a sharp change in resolution from $\Delta_f=2\pi/128$ to
$\Delta_c=2\pi/32$ as shown in Fig.~\ref{fig:sharpchange}, 
and two different initial
conditions given by 
\begin{equation} 
  u_i(x,t=0)=\cos(\eta_ix)e^{-5(x-3\pi/2)^2},
\label{eq:simpleICs} 
\end{equation}
with $\eta_1=4$ and $\eta_2=18$, which we refer to as wave packets
$P_1$ and $P_2$, respectively. \rdm{The energy spectrum of these
  wave packets is the sum of three Gaussian functions of wavenumber, with
  standard deviation of $\sqrt{5}$. They are centered around
  $\kappa=\pm\eta_i$ and 0. As a consequence, more than 99\% of the
  energy resides in wavenumbers with $|\kappa|<\eta+5.8$. In the
  fine region, $\kappa^f_a=32$, so both wave packets have virtually all
  of their energy in wavenumbers $|\kappa|<\kappa^f_a$}.
Both wave packets are
thus well resolved in the fine region and propagate as expected with
approximately the convection velocity (Figs.~\ref{fig:sharpchangeB} and
\ref{fig:sharpchangeD}). The packet $P_1$ is centered around the wavenumber
$\kappa=4$ which can be supported on
the coarse as well as the fine grid (Fig.~\ref{fig:sharpchangeA}), \rdm{with more
than 90\% of the energy residing in wavenumbers with $|\kappa|<\kappa^c_a=8$.}  The packet therefore
mostly propagates into the coarse region (Fig.~\ref{fig:sharpchangeC}) in a wave
packet centered around a slightly larger wavenumber with a slightly
lower group velocity ($T_1$ in
Fig.~\ref{fig:sharpchangeA}). But because of the resolution change, some of $P_1$ 
is also reflected back into the fine region in a wave packet centered
around a much higher wavenumber ($R_1$ in Fig.~\ref{fig:sharpchangeA}) which has
negative group velocity. 
\gry{Since (\ref{eq:1DAdvEq}) with a central difference derivative scheme is an
energy preserving approximation of the advection equation, the energy from the
incident wave is split between the reflected wave and the transmitted wave
\citep{vichnevetsky1981energy,vichnevetsky1983group,vichnevetsky1981propagation,long2011numerical}.}
The $P_2$ packet is centered around the wavenumber
$\kappa = 18$ on the fine grid, which cannot be supported on
the coarse grid ($P_2$ in Fig.~\ref{fig:sharpchangeA}), and indeed
virtually none of the energy resides in wavenumbers with $|\kappa|<\kappa^c_a$. It therefore
cannot propagate into the coarse region and instead is entirely reflected back
into the fine region (Fig.~\ref{fig:sharpchangeE}) in a packet centered around a
much larger wavenumber ($R_2$ in Fig.~\ref{fig:sharpchangeA}). \gry{In both cases,
it is effective wavenumbers that are preserved through the resolution change
(Fig.~\ref{fig:sharpchangeA}).}
The reflected waves $R_1$ and $R_2$ are entirely spurious. 

\gry{Because the system is linear, the above results can be extended to
grids with gradually changing resolution.  In this case, a local wavenumber
$\kappa_j$ and a local group velocity $\mathcal{G}_j$ \gryr{can be defined by
substituting a given frequency $\omega$ (or $\kappa'$) and the local grid spacing $\Delta(x_j)$
into (\ref{eq:CDdispersion}).}} \gryr{As above, there will be two possible values of $\kappa_j$,
$\kappa_j^{(1)}$ and $\kappa_j^{(2)}$, satisfying $\kappa_j^{(1)} <
\kappa_j^{(2)}$ and $(\kappa^{(2)}_j \Delta(x_j)) = \pi - (\kappa_j^{(1)}
\Delta(x_j))$, with group velocities $G_j$ and $-G_j$, respectively. }
There are three main results of such an analysis \citep{vichnevetsky1987wave2,long2011numerical} that will
be relevant for our purposes. First, no reflections occur if the local group
velocity is uniform and nonzero, as expected. \gry{Second, a total reflection
occurs for all wavenumbers that become unresolvable on the grid \gryr{(i.e.,
exceed the Nyquist wavenumber)}, and the reflection
occurs at the point where the local group velocity vanishes
($\mathcal{G}_j = 0$). Thirdly, no reflections occur for wavenumbers that can be
resolved throughout the
domain if $\Delta(x_j)$ varies over length scales that
are long compared to the wavelength $\lambda=2\pi/\kappa_j$
($\frac{\lambda}{\Delta}\frac{d\Delta}{dx}\ll 1$). Thus, wave packets analogous to $P_1$ will
be completely transmitted through a sufficiently smooth resolution
change.}

\gry{The behavior described here is representative of all energy-conserving 
numerical schemes with two wavenumbers per effective wavenumber.} These are among the most
common numerical schemes used in turbulence applications (e.g., centered
difference, B-splines, finite volume), however, other
numerics with different propagation properties are possible. 
For instance, consider the box scheme whose semi-discretization of
(\ref{eq:1DAdvEq}) is given by
\begin{equation} 
    \frac{\partial }{\partial t} \left( \frac{u_j + u_{j+1}}{2}
    \right) + U \frac{u_{j+1}-u_{j}}{x_{j+1}-x_j} = 0 
\label{eq:boxscheme} ,
\end{equation} 
which is also energy preserving.
Instead of reflecting unresolvable scales of motion at higher
wavenumbers into the fine region, the box scheme transmits unresolvable scales
at lower wavenumbers through the coarse region (similar to an aliasing effect)
\citep{frank2004spurious,ascher2004multisymplectic}. 
The result is still spurious numerical oscillations. 
\rdm{These results from numerical analysis have profound consequences for LES.}
 \rdm{When LES turbulence convects into a more coarsely resolved
  region, the spectral characteristics of the numerical derivative
  operator $\delta/\delta x$ dictate that neglecting the inhomogeneous
  commutator can produce non-physical fine-scale noise propagating
  upstream, spoiling the solution far from the resolution
  change. This is explored in the next section.}

\section{Impacts of Resolution Inhomogeneity on LES \label{sec:impact}}
\rdm{The commutator analysis of Sec.~\ref{sec:ICE} indicates that the
combined effects of neglecting the inhomogeneous commutator and the
dispersion characteristics of the numerical derivative operator could
have a profound impact on an LES of turbulence flowing through a
domain with varying spatial resolution. To characterize this impact,
we consider a simple case of such a flow, making two simplifications
to clearly expose the effects. As discussed in Sec.~\ref{sec:ICE}, we
consider commutation error for mean convection since this is commonly
the dominant effect. This is consistent with the Taylor frozen field hypothesis.
Moreover, a localized packet of turbulent fluctuations is used to expose the
non-local effects of commutation error. Again we consider filters that
consist of only an implicit projection to the discrete solution
space.}

\subsection{A Numerical Experiment \label{sec:setup}}

\begin{figure*}[t!]
\centering
\begin{subfigure}[b]{0.45\textwidth}
\centering
\includegraphics[width=1\textwidth]{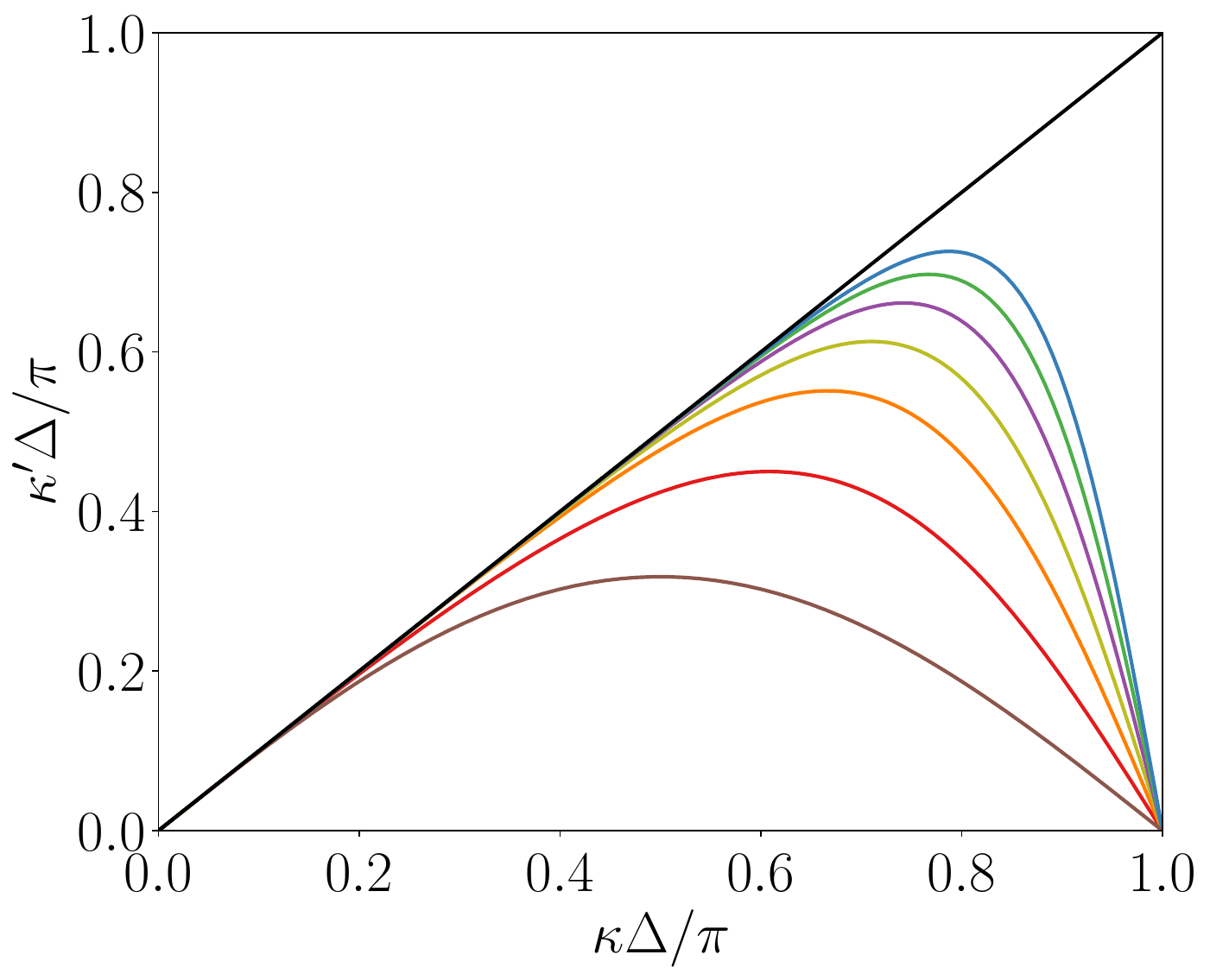}
\caption{}
\label{fig:BsplineEff1}
\end{subfigure}%
\hspace{1em}
\begin{subfigure}[b]{0.45\textwidth}
\centering
\includegraphics[width=1\textwidth]{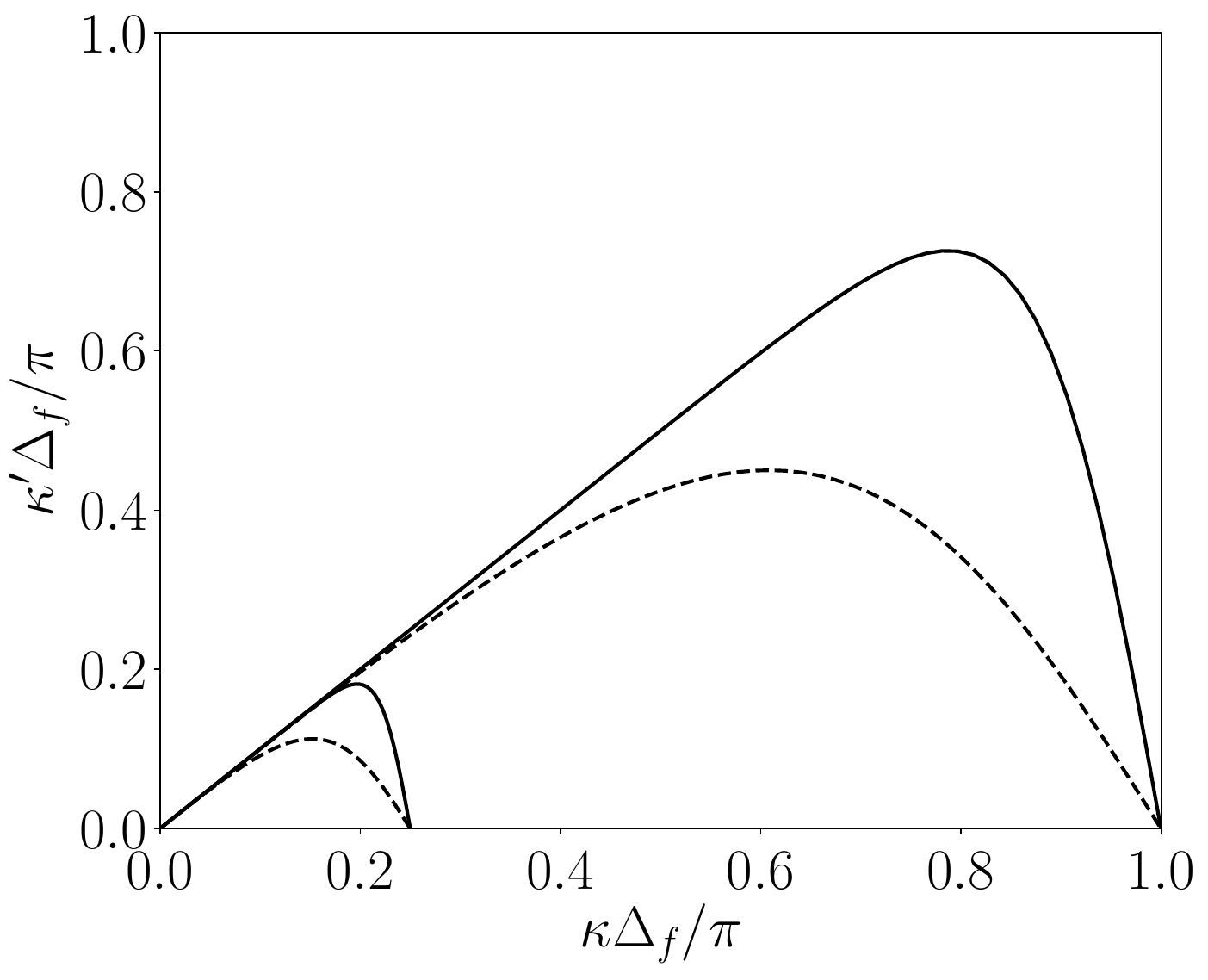}
\caption{} 
\label{fig:Bspline12832}
\end{subfigure}
\caption{(a) The spectrum of the first derivative operators
    for a B-spline collocation method of several orders, a second-order centered
    difference method, and a spectral method;
    \textcolor{mplred}{\solidrule{}} ($B_1^2$), 
    \textcolor{mplorange}{\solidrule{}} ($B_1^3$),
    \textcolor{mplyellow}{\solidrule{}} ($B_1^4$), 
    \textcolor{mplpurple}{\solidrule{}} ($B_1^5$),
    \textcolor{mplgreen}{\solidrule{}} ($B_1^6$),
    \textcolor{mplblue}{\solidrule{}} ($B_1^7$),
    \textcolor{mplbrown}{\solidrule{}} ($B_1^{CD}$),
    \protect\solidrule{} (Spectral).
	(b) The consistently normalized spectrum of the $B_1^2$ and $B_1^7$ operators for 
    the fine region of
    the domain with spacing $\Delta_f$
    and the coarse region
	of the domain with spacing $\Delta_c = 4\Delta_f$;
    \protect\solidrule{} ($B_1^7$), 
	\protect\dashedrule{} ($B_1^2$).
}
\label{fig:BsplineEff}

\end{figure*}

\rdm{Under the frozen field hypothesis and neglecting commutators, the LES
equations for turbulence flowing at constant velocity $U_x$ in the $x$
direction simplify to:}
\begin{equation}
    \frac{\partial \overline{\mathbf{u}'}}{\partial t} + U_x \frac{\delta
        \overline{\mathbf{u}'}}{\delta x} = 0 
   \label{eq:pureconvection}
   .
\end{equation}
\rdm{The resolution in the $x$ direction is made to vary with $x$, while
the resolution in the other directions is constant, and periodic
boundary conditions are imposed in all three directions.} 
The filter is defined as a projection onto a periodic B-spline
representation in the $x$-direction and Fourier spectral
representations in the $y$- and $z$-directions. The $\delta/\delta x$
operator in (\ref{eq:pureconvection}) is defined as B-spline
collocation. B-spline collocation is a convenient numerical treatment for 
inhomogeneous resolution offering a range of orders of accuracy, and have been advocated for use in
turbulence simulations
\citep{kravchenko1997effect,kravchenko1999b,shariff1998two,kwok2001critical,bazilevs2007variational,lee2015direct}.
Let $B_n^k$ denote the $n$th derivative B-spline operator of order
$k$, and $B_n^{CD}$ denote the $n$th derivative second-order centered
difference operator. 
The spectra of the $B_1^k$ operators
(Fig.~\ref{fig:BsplineEff}) have similar propagation properties as the
$B_1^{CD}$ operator
discussed in Sec.~\ref{sec:numerics}, in that there are generally two wavenumbers $\kappa$ that have
the same effective wavenumber $\kappa'$, one with positive group
velocity and the other with negative group velocity. Further, with
increasing $\kappa$, the negative group velocities get larger in magnitude
(larger negative slopes on the right side of Fig.~\ref{fig:BsplineEff1}).

For the results presented here, a third-order low storage Runge-Kutta method is used for time advancement
\citep{spalart1991spectral}.
Note that the spurious reflection/transmission phenomena described in Sec.~\ref{sec:numerics} 
depend only on spatial discretization \citep{vichnevetsky1987wave}. 

A two dimensional slice of the numerical grid is shown in Fig.~\ref{fig:tikzgrid}a.  
The domain in the propagation direction is
divided into a uniform fine region of size $2\pi$, a uniform coarse region of
size $6\pi$, and two \rdm{transition} regions \rdm{of approximate size
$2\pi$ in which the
resolution is inhomogeneous}.  The fine resolution spacing between
B-spline knot points is $\Delta_{f} = 2\pi/128$, and
the coarse \rdm{knot} spacing is $\Delta_{c} = 2\pi/32$. \rdm{In the
transition regions, the knot spacing is designed to vary as a Sigmoid
function between $\Delta_f$ and $\Delta_c$ over a distance in $x$ of
order $1/\alpha$. To this end,} the mapping function
$g(\xi): [0,1] \to [x_{ {\rm start}},x_{\rm end}]$ is defined implicitly through the differential
equation:
\begin{equation} 
	\Delta(x)\equiv \frac{d g(\xi)}{d \xi} \Delta_\xi =
	\frac{\Delta_f}{1+e^{\alpha g(\xi)}} +
	\frac{\Delta_c}{1+e^{-\alpha g(\xi)}}
	\label{eq:gridmapping} ,
\end{equation}
\rdm{where, $\Delta_\xi=1/N_\xi$ is the uniform resolution in $\xi\in [0,1]$, with
$N_\xi$ the number of knot intervals in the transition region. The knot
points $x_j$ are then defined as $x_j=g(j\Delta_\xi)$ for
$j=0,1,\ldots N_\xi$.} The parameter $\alpha$ controls the sharpness of the
grid change, \rdm{with the transition thickness defined by
$(\Delta_c-\Delta_f)/(d\Delta/dx)=4/\alpha$.  To generate the knot
points used here, (\ref{eq:gridmapping}) was solved numerically for
$g(\xi)$ using
a standard Runge–Kutta–Fehlberg method and
$g(0)=x_{\rm start}=-\pi$, $\alpha=4$ and $N_\xi=75$. With these
parameters, $g(1)\approx 3.1996$, defining a transition region grid on
an interval slightly larger than $2\pi$.}

The domain in
the two spectral directions is $[0,2\pi]$ and with an effective uniform grid
spacing of $\Delta_f$. Thus, LES turbulence will be convected through an
\textit{anisotropic}, inhomogeneous grid --- a common
scenario in practice for structured grids.  \rdm{Moreover, in this configuration the three dimensional commutation error simplifies to the one
dimensional case, which will expose the implications
of the numerical analysis in Section~\ref{sec:numerics} for
commutation error in LES.}

The initial condition is taken to be a `packet' of well-resolved,
homogeneous, isotropic turbulence. This packet is analogous to the
wave packets studied in the one-dimensional examples in
Sec.~\ref{sec:numerics}.  \gry{To create this packet, a spectral
LES of infinite Reynolds number homogeneous, isotropic turbulence was
performed in a $2\pi\times2\pi\times2\pi$ domain with 64 Fourier modes in
each direction. A Smagorinsky model was used to represent the subgrid
stress in this simulation and a negative viscosity forcing
that isotropically injects energy over a wavenumber shell of radius $0
< |\boldsymbol{\kappa}| \le 2$ was introduced to allow the turbulence
to become statistically stationary. The energy injection rate and
therefore, the equilibrium dissipation rate was set to $1$.} A
representative instantaneous velocity field from the LES was then
introduced into the fine region of the the B-spline/spectral
simulation and modulated with a Gaussian so that the fluctuations go
smoothly to zero. \gryr{Note that this procedure does not produce a divergence free
velocity, however, this is not an issue for the linear problem solved here; in
fact, a divergence free projection would distort the desirable properties of
the packet.} The resolution used in the spectral
simulation ensures that the modulated packet is well-resolved by the
B-splines in the fine resolution region. Specifically, an isotropic
grid spacing of $2 \pi/64$ in the fully spectral simulation
corresponds to $\kappa_{max} \Delta_f
\approx 1.5$ in the B-spline simulation, where $\kappa_{max}=32$ is the largest
nonzero wavenumber in the turbulence packet. 
As seen in Fig.~\ref{fig:BsplineEff1}, $(\kappa\Delta) = 1.5$ is in
the positive group velocity regime.

\subsection{Results \label{sec:results1}}

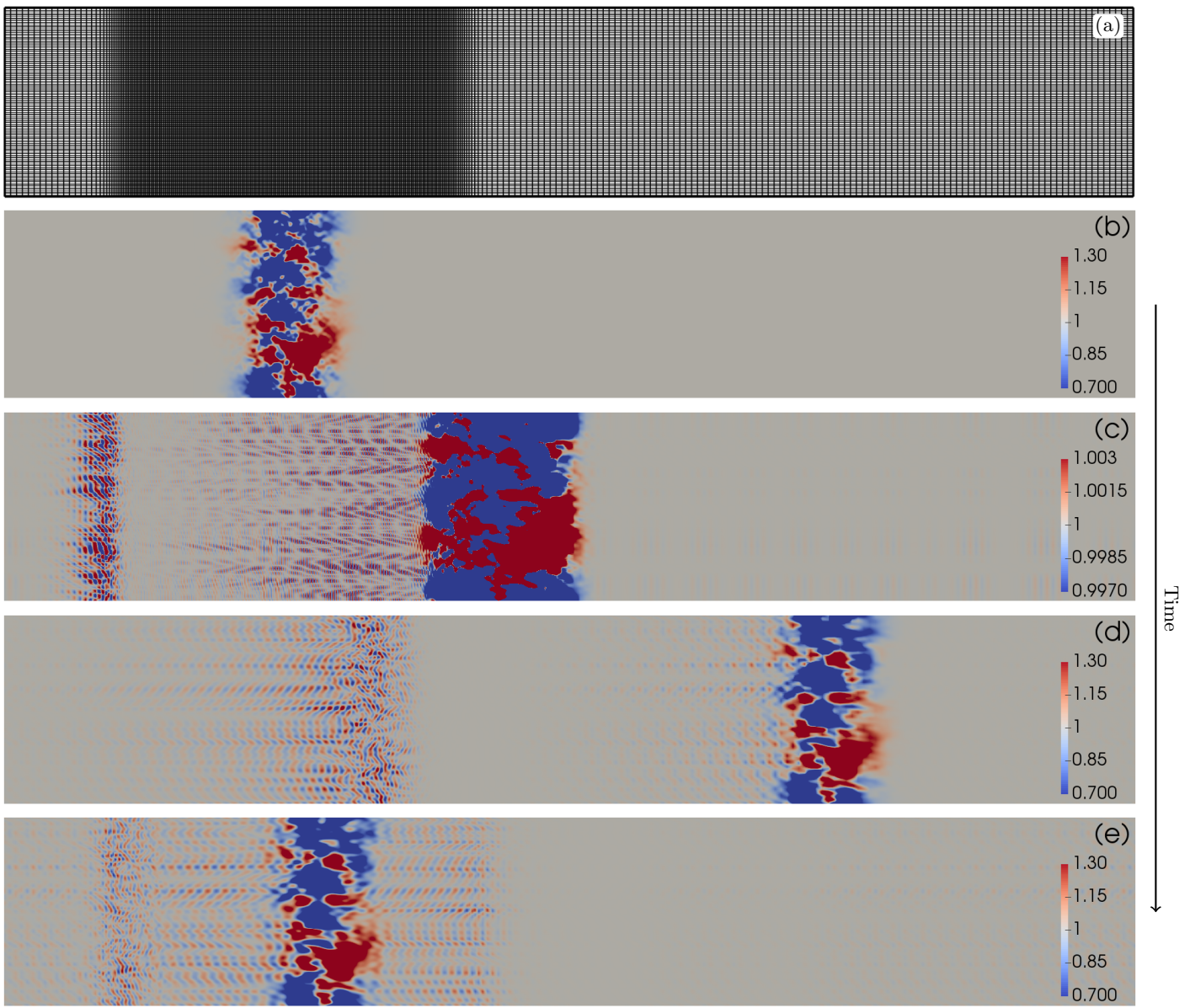
\begin{figure}[t!]
    \centering
    \begin{tikzpicture}
		\node at (0,0)
		{\includegraphics[width=1\textwidth]{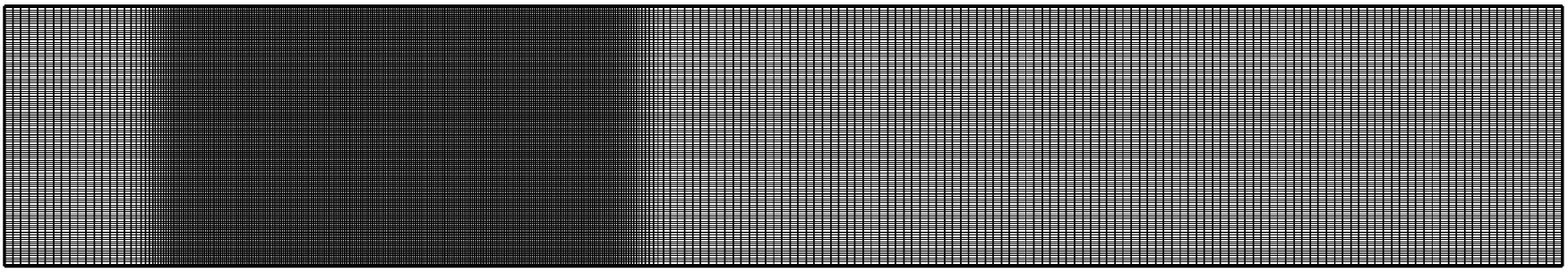}};
		\node at (0,-3.2)
		{\includegraphics[width=1\textwidth]{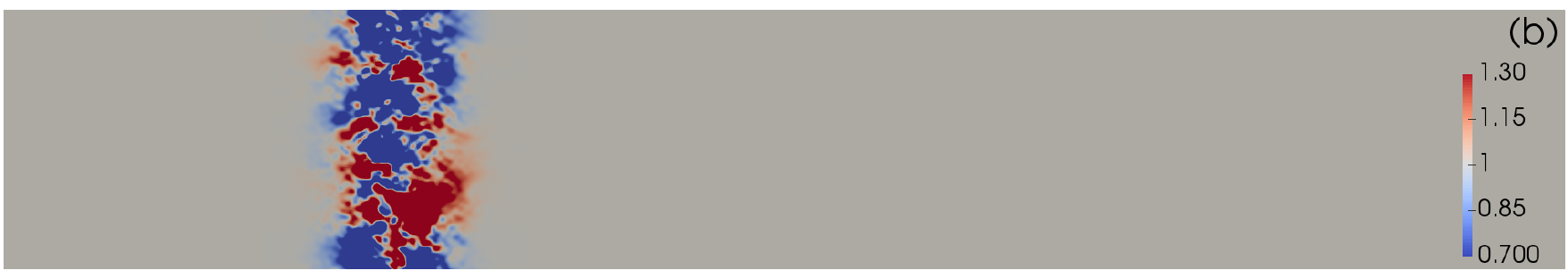}};
		\node at (0,-6.4)
		{\includegraphics[width=1\textwidth]{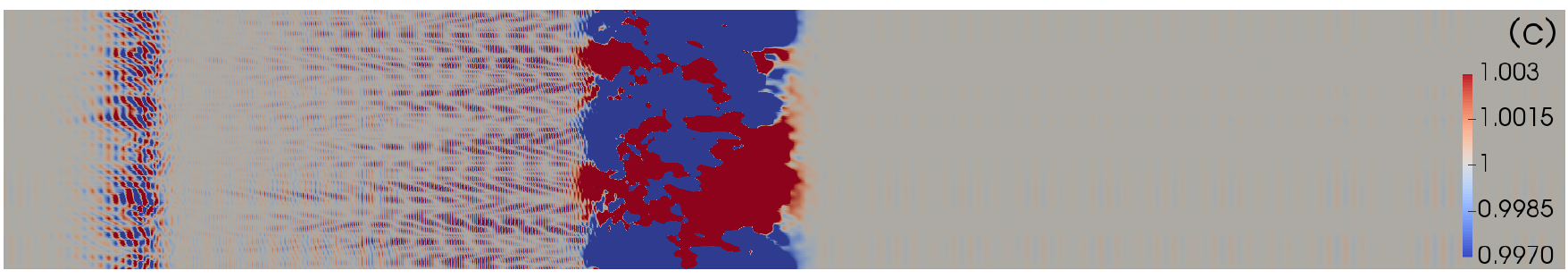}};
		\node at (0,-9.6)
		{\includegraphics[width=1\textwidth]{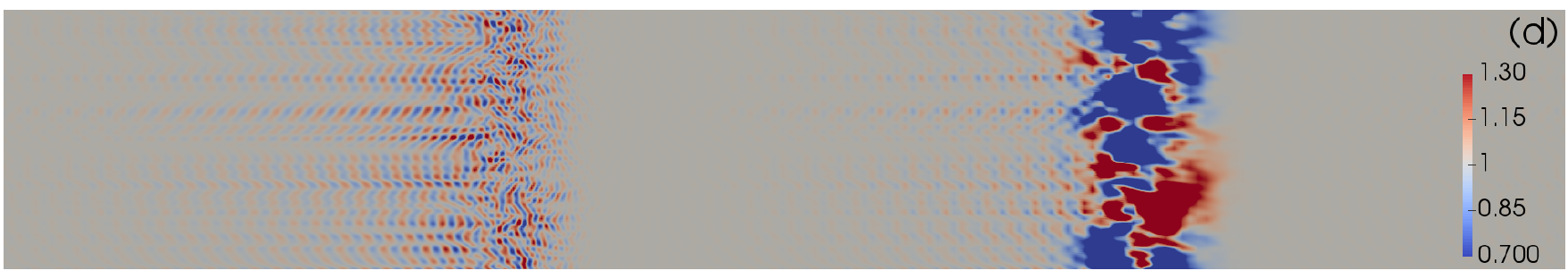}};
		\node at (0,-12.8)
		{\includegraphics[width=1\textwidth]{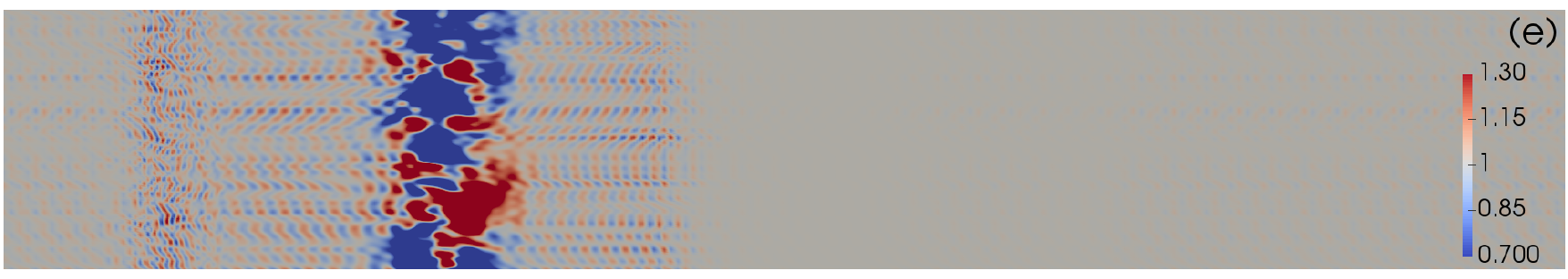}};
		\draw[->,thick] (9.25,-3.2) -- (9.25,-12.8);
		\node[thick,rotate=270] at (9.5,-8) {Time};
        \node[fill=white,rounded corners=2pt,inner sep=1pt] at (8.5,1.2) {(a)};
    \end{tikzpicture}
    \caption{Convection of a packet of homogeneous, isotropic turbulence
    through an anisotropic, inhomogeneous grid with seventh-order B-splines and
a convection velocity of 1. The packet is moving through the resolution change
to the right. (a) A slice of the numerical grid. (b), (c), (d),
and (e) show the streamwise velocity field at times $0.00$, $7.03$, $11.72$,
and $39.06$, respectively. In part (c) the color scale is different to
		emphasize small amplitude fluctuations to highlight the spurious high
wavenumber reflections moving to the left through the fine region of the grid.
\gryr{The results for the second-order B-spline case are qualitatively
    similar to those shown here albeit with more dispersion in the
higher wavenumbers and a wider range of reflected
scales.}}  \label{fig:tikzgrid}
\end{figure}

\begin{figure*}[t!]
\centering
\begin{subfigure}[b]{0.45\textwidth}
\centering
\includegraphics[width=1\textwidth]{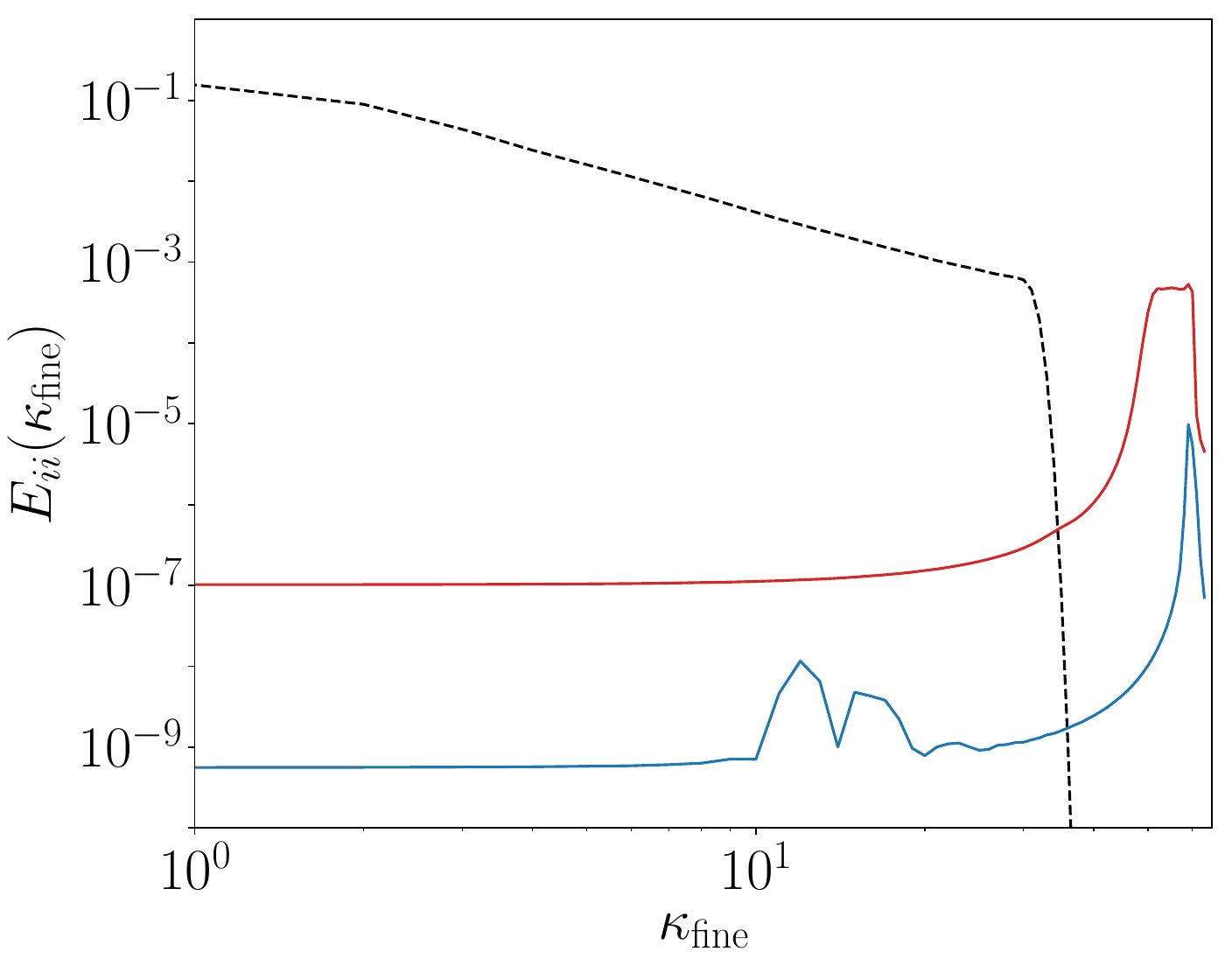}
\caption{$t = 7.03 (\textcolor{mplblue}{\solidruled{}}),
11.72$ (\textcolor{mplred}{\solidruled{}})}
\label{fig:nodissPile}
\end{subfigure}%
\hspace{1em}
\begin{subfigure}[b]{0.45\textwidth}
\centering
\includegraphics[width=1\textwidth]{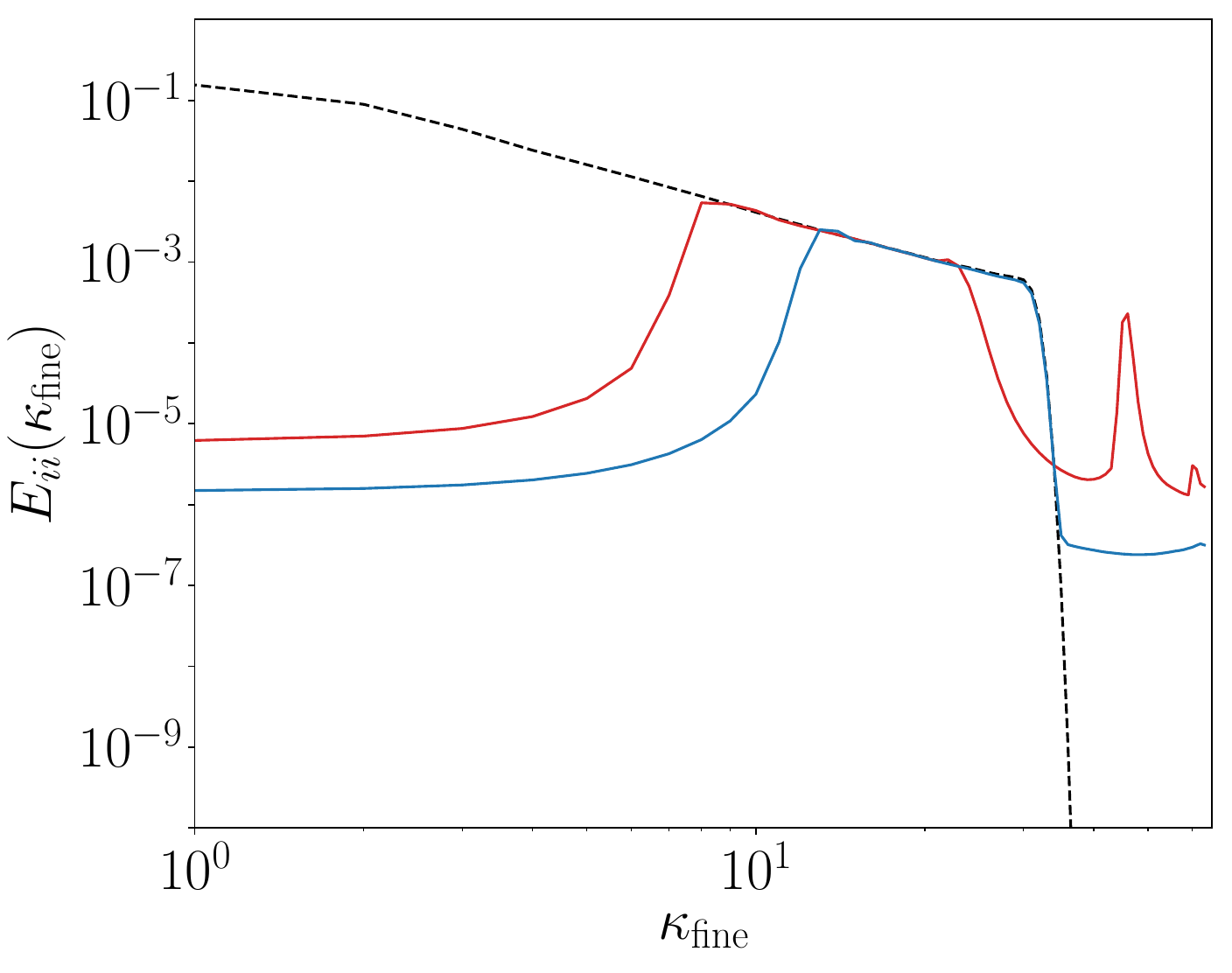}
\caption{$t = 15.63 (\textcolor{mplblue}{\solidruled{}}),
23.44$ (\textcolor{mplred}{\solidruled{}})}
\label{fig:nodissref}
\end{subfigure}
\vskip\baselineskip
\begin{subfigure}[b]{0.45\textwidth}
\centering
\includegraphics[width=1\textwidth]{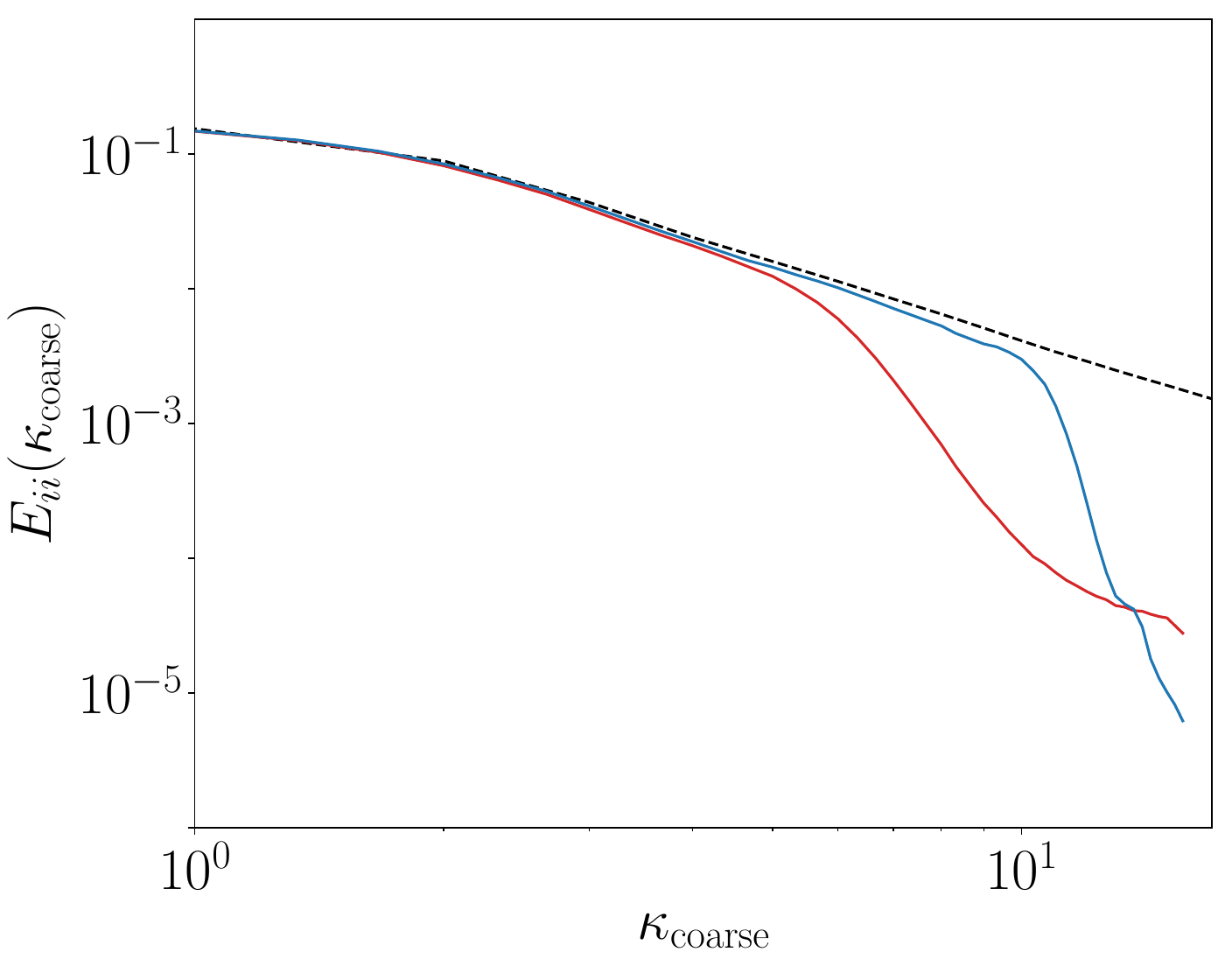}
\caption{$t = 18.75 (\textcolor{mplblue}{\solidruled{}}),
11.72$ (\textcolor{mplred}{\solidruled{}})}
\label{fig:nodisscoarse}
\end{subfigure}%
\hspace{1em}
\begin{subfigure}[b]{0.45\textwidth}
\centering
\includegraphics[width=1\textwidth]{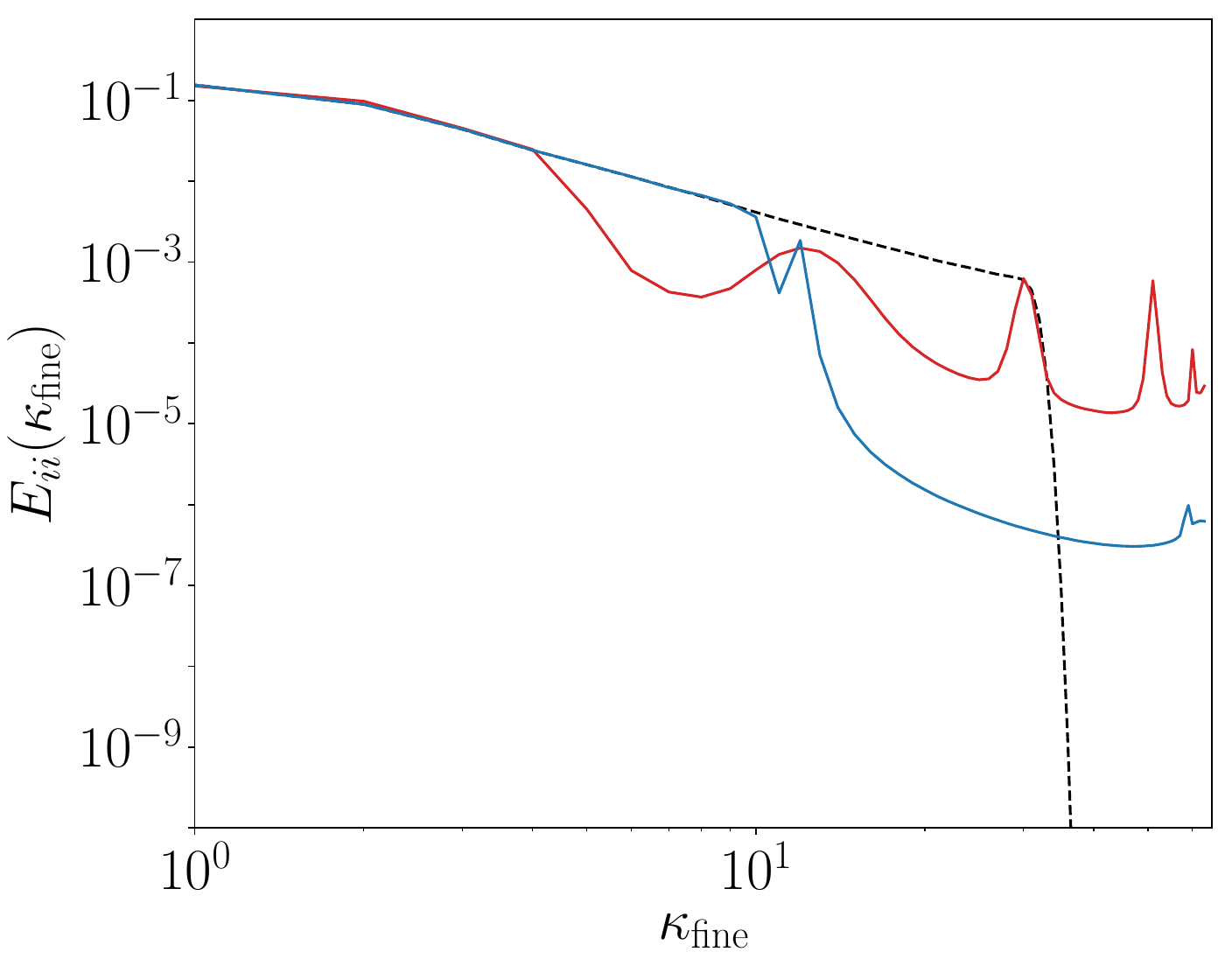}
\caption{$t = 39.06 (\textcolor{mplblue}{\solidruled{}}),
39.84$ (\textcolor{mplred}{\solidruled{}})}
\label{fig:nodissend}
\end{subfigure}
\caption{One dimensional energy spectra, $E_{ii}$, in the direction of
	inhomogeneity at different times $t$; $\kappa_{\rm{fine}}$
	and $\kappa_{\rm{coarse}}$ represent the wavenumbers in the fine and coarse regions
of the domain, respectively.     
	(a) High wavenumber
    reflections propagating backward through the fine region corresponding to
    the energy in the fine region in Fig.~\ref{fig:tikzgrid}c.
    (b) The subsequent reflections propagating forward through the fine region
    corresponding to the energy in the fine region in Fig.~\ref{fig:tikzgrid}d.
    (c) The spectra of the resolved turbulence packet 
    in the coarse region corresponding to the energy in the coarse region in
    Fig.~\ref{fig:tikzgrid}d.
    (d) The spectra of the turbulence packet after one flow through corresponding
    to the energy in the fine region in Fig.~\ref{fig:tikzgrid}e.
	\textcolor{mplblue}{\solidrule{}} (7th order B-splines), 
    \textcolor{mplred}{\solidrule{}} (2nd order B-splines), 
    \protect\dashedrule{} (Initial spectra of the turbulence packet shown in
        Fig.~\ref{fig:tikzgrid}b).
}
\label{fig:nodissSpectra}
\end{figure*}

The effects of resolution inhomogeneity on the spatial structure (see
Fig.~\ref{fig:tikzgrid}) and on the one dimensional energy spectra (see
Fig.~\ref{fig:nodissSpectra}) of the turbulence packet are examined at several
stages of the simulation for a single flow-through.  Seventh-order B-splines and
second-order B-splines are used to illustrate the behavior of higher- and
lower-order methods.  Based on the numerical analysis in
Sec.~\ref{sec:numerics}, the consistently normalized spectra of the $B_1^2$ and $B_1^7$
operators in the fine and coarse regions of the domain 
are sufficient to predict the behavior of the commutation error (see Fig~\ref{fig:Bspline12832}).  
To see this,
let the wavenumbers $|\kappa|\in [0,16]$ be referred to as the \textit{coarse
wavenumbers}, wavenumbers $|\kappa| \in (16,32]$ be the \textit{fine
wavenumbers}, and wavenumber $|\kappa| \in (32,64]$ be the \textit{spurious
wavenumbers}, and recall that the fine region of the domain is capable of
representing the fine, coarse, and spurious wavenumbers, while the coarse region
is only capable of representing the coarse wavenumbers.  The initial packet of
turbulence only contains fine and coarse wavenumbers, so any energy transferred
to higher wavenumbers by the resolution inhomogeneity is indeed spurious. 

As the turbulence packet convects into the coarse region of the domain, all of
the energy in the fine wavenumbers is transferred to scales with negative group
velocity in the spurious wavenumber regime (see Fig~\ref{fig:tikzgrid}c).
As in Sec.~\ref{sec:numerics}, this energy transfer occurs between wavenumbers that
share an effective wavenumber. The corresponding energy spectra at this stage of the
simulation show a pile up of energy in the largest wavenumbers in the fine
region of the domain (see Fig~\ref{fig:nodissPile}). Notice that, for each
numerical scheme, the energy is concentrated in a narrow band of wavenumbers
that corresponds to the region with negative slope in the effective
wavenumbers shown in Fig.~\ref{fig:Bspline12832}. \gry{The reflections in the second
order B-spline case occur over a wider range of wavenumbers and are collectively
more intense than for seventh-order case, as more energy is being reflected
(see Fig~\ref{fig:nodissPile}).}
Furthermore, the propagation
speed of the reflections is much greater for seventh-order B-splines than
second-order B-splines, as indicated by the slopes of the effective wavenumbers.
Interestingly, we observed that, for a B-spline collocation method, the ratio of
the group velocity to the convection velocity of the highest wavenumber
reflections for each B-spline order matches the order of the B-spline (e.g., the
Nyquist wavenumber propagates at negative $N$ times the convection velocity for
$N$th order B-splines).  This appears to be a special property of B-spline
collocation that
deserves proof and is consistent with the work of
\citet{vichnevetsky1991nonlocal}, who demonstrated that an infinite speed of
reflection occurs for spectral numerics.

\gry{Once the reflected fluctuations reach the resolution change on the left side of
the fine region, they are reflected back into the fine region with positive
group velocity with their initial wavenumbers. This re-reflection can be
tracked from Fig.~\ref{fig:tikzgrid}c in which the reflected wavepacket
consisting of spurious wavenumbers is visible on the left-hand side as
it propagates upstream (to the left), to Fig.~\ref{fig:tikzgrid}d in
which the re-reflected wavepacket consisting now of fine wavenumbers
is visible propagating down-stream.} 
These secondary reflections
occur in the fine wavenumber regime but are as erroneous as the spurious
reflections that created them.  For both B-spline orders, the energy spectra in
the fine resolution region for the initial turbulence packet and the reflected
scales of motion match for all fine wavenumbers (see Fig.~\ref{fig:nodissref}).
This indicates a total reflection occurs for all scales that are only
representable on the fine grid, which agrees with the analysis of the $P_2$-type
waves discussed in Sec.~\ref{sec:numerics}. 
Without the commutator $\C^I$, this cycle of reflection between
fine and spurious wavenumbers repeats. The energy initially contained in the
fine wavenumber regime gets trapped in the fine resolution
region.

The only fluctuation scales that make it through to the coarse region of the
domain are those that can be represented on the coarse grid, i.e., the coarse
wavenumbers (see Fig.~\ref{fig:tikzgrid}d).
The energy spectra at the initial time, and after the packet has convected into the coarse
region, match \textit{almost} identically for all coarse wavenumbers (see
Fig.~\ref{fig:nodisscoarse}).
A relatively small fraction of the energy in the coarse wavenumbers also gets trapped in
the fine region, as shown in Fig.~\ref{fig:nodissref}. This behavior is also
predicted by the numerical analysis of the $P_1$-type waves discussed in
Sec.~\ref{sec:numerics}, and would vanish in the limit of a smoothly varying
grid.

\gry{The numerical experiment described here focuses on the idealized
case of frozen turbulence consistent with Taylor's hypothesis to
emphasize the impact of commutation error. 
The scales trapped in the fine
region of the domain are physically incorrect and numerically
problematic. An increase in high wavenumber energy can lead to
numerical instabilities, and the trapped low wavenumber energy can
corrupt otherwise meaningful statistics. Moreover, it is reasonable
to expect that in an LES nonlinear effects would
magnify these problems as erroneous fluctuations would interact with and
contaminate incoming turbulence.  Consider, for instance,
the turbulence packet after one flow through (see
Fig.~\ref{fig:tikzgrid}e).  As the coarsely resolved packet re-enters
the fine region (without any active forcing), the spectrum gets
corrupted by the trapped energy (see Fig.~\ref{fig:nodissend}).
Furthermore, a shift in energy from lower wavenumbers to higher
wavenumbers would be particularly damaging in real turbulence as the
former are more responsible for momentum transport while the latter
are more responsible for dissipation. The nonlocal wavenumber interactions introduced by
resolution inhomogeneity may corrupt the energy cascade, which, in
homogeneous isotropic turbulence, is known to be dominated by
interactions local in wavespace \citep{waleffe1992nature}.  Lastly,
notice that unlike the effects of discretization error, the effects of
resolution inhomogeneity do not improve with higher-order numerics.
Further study of these effects in an actual LES is warranted, but is
out of scope for this paper.  However, it is clear that a model for
the inhomogeneous part of the commutator is needed to mitigate the
effects of the commutation error. }

\section{Commutator Modeling \label{sec:modelform}}
\gryr{In this section we propose an approach to modeling the inhomogeneous
	commutator based on the characteristics of the commutator and the commutation
	error explored in Sec.~\ref{sec:ICE}. As previously discussed, a
	model for the commutator is responsible for transferring energy between
	resolved and unresolved scales as a consequence of the resolution
	inhomogeneity.  In the coarsening grid case, a commutation model must
	transfer the energy in newly unresolvable scales to the subgrid scales.  In
	the refining grid case, a model for $\C^I$ would have to transfer energy
	from the subgrid to the resolved turbulence, presumably through some type of
	forcing.  Notice that the requirements of a commutation model in the coarsening
	and refining cases are fundamentally different. It has been suggested that a
	``good'' commutation model should handle both of these scenarios (e.g.,
	\citep{girimaji2013closure}), however, because of these different
	requirements, this may not be appropriate. A commutation model for the
	coarsening and refining grid cases may need to be developed independently.
We pursue this modeling approach here for the case of flow through coarsening grids to
address the issues discussed in Sec.~\ref{sec:impact}.}

	\gryr{A common mechanism for providing the transfer of energy from resolved
		to subgrid scales is a viscosity-based model, as suggested by the second order term in
	(\ref{eq:origCommError}), which is equipped with the viscosity
	$\nu_{comm} = U \Delta \partial \Delta/\partial x$. However, as indicated by
	(\ref{eq:comm_hat1}), a commutation model should ideally only affect
	wavenumbers near the cutoff wavenumber. This property preserves wavenumbers that are
	well resolved throughout the resolution change while removing those that are
	not. As such, a hyperviscosity is a more appropriate model for the
	commutator, as also indicated by the leading order terms in
	(\ref{eq:origCommError}). Specifically, the leading order terms in
	(\ref{eq:origCommError}) suggest the following form for a general one
	dimensional hyperviscosity commutation model:}
\begin{equation}
    \begin{split}
        U\C^I(u) \simeq
	    (-1)^{N/2}C U\Delta \frac{d \Delta}{d x} \left( \Delta^{\N-2}
		\frac{\partial^\N \overline{u}}{ \partial x^\N}
    \right)
    \label{eq:largekcomm}
    \end{split}
    , 
\end{equation}
for some constant $C$ and even order $\N$ ($\N$ is assumed to be a positive even
integer for the remainder of this paper). 

For any finite value of $\N$ in (\ref{eq:largekcomm}), there
is a trade-off between removing high wavenumber scales in fine regions of the
grid that are approaching unresolvability, and preserving the well-resolved
scales in coarse regions of the grid. Larger values of $\N$ lead to sharper
filters, which perform better in the context of this trade-off than smaller
values of $\N$. Accordingly, it is desirable to make $\N$ as large as possible.
Again, this is consistent with (\ref{eq:comm_hat1}). 
However, the number of available derivatives of the filtered field limits how
large $\N$ can be, i.e., the underlying numerics constrain $\N$ based on the
number of accessible derivative operators.  For example, CFD codes typically
only have access to second derivative operators so that $\N$ would be limited to
$2$.  Furthermore, larger values of $N$ require not only higher order numerics
but also additional boundary conditions, 
which are often mentioned as a problem with hyperviscosity models
\citep{cook2005hyperviscosity,vasilyev1998general}.  
With this discussion in mind, we let (\ref{eq:largekcomm}) serve as the
foundation for developing a one dimensional commutation model. 
The following subsections propose strategies for improving the model.

\subsection{The $B_2-B_1B_1$ Model \label{sec:B2B1B1}}

Let $F_\N(u)\approx \partial^\N u/\partial x^\N$ be some numerical
operator that approximates the $\N$th derivative.
The commutation model (\ref{eq:largekcomm}) can then be written as: 
\begin{equation} 
    \C^I(u) = (-1)^{N/2} C \Delta \frac{d \Delta}{d x} \left( \Delta^{\N-2}
	F_\N (\overline{u}) \right) 
    \label{eq:largekcomm1} 
    .  
\end{equation}

\begin{figure*}[t!]
\centering
\includegraphics[width=0.50\textwidth]{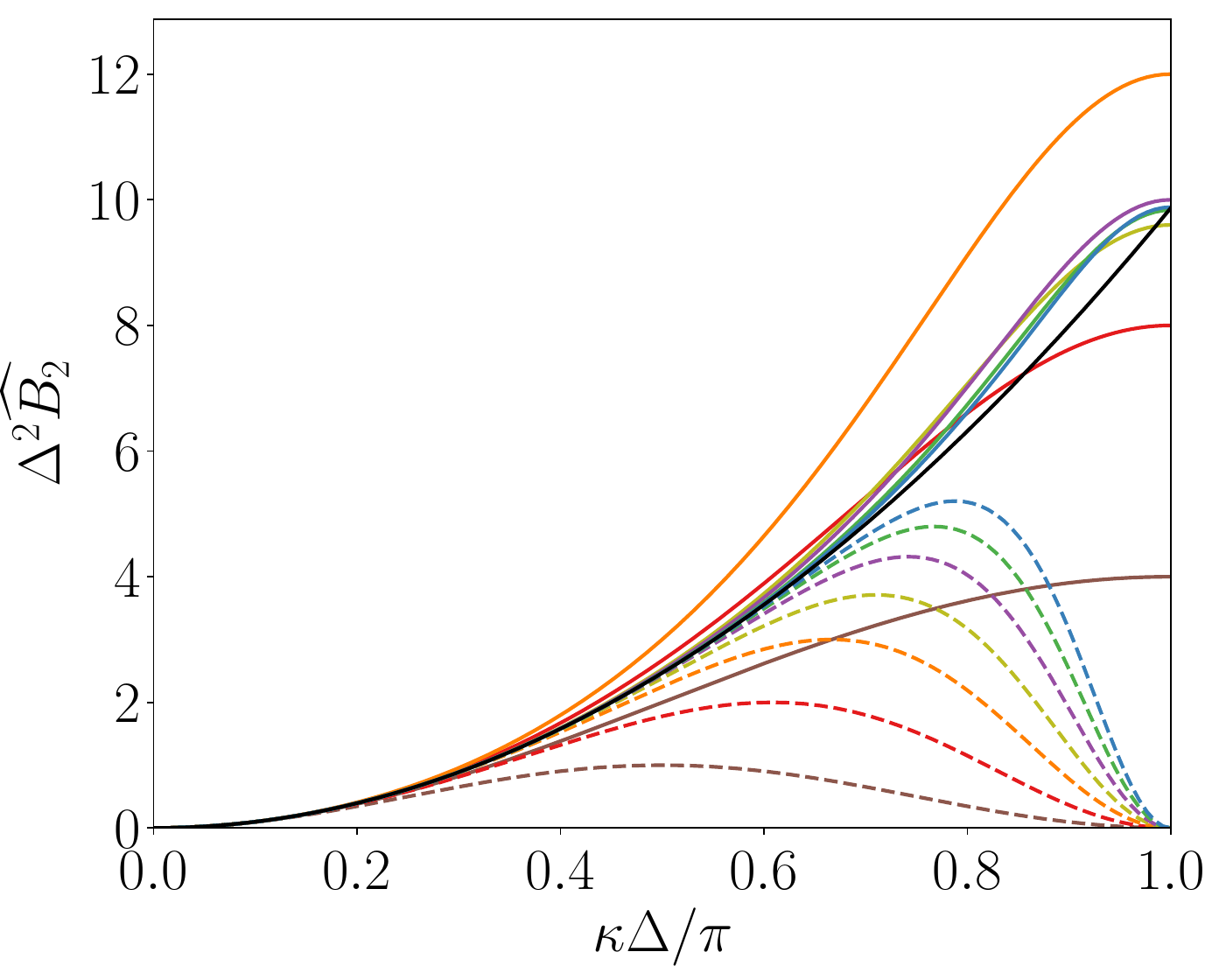}
\caption{The spectrum of the second derivative operators ($\widehat{B_2}$)
    for a B-spline collocation method of several orders, a second-order centered
    difference method, and a spectral method. Repeated application of the first
    derivative operators are shown by the corresponding dashed lines;
    \textcolor{mplred}{\solidruled{}},\textcolor{mplred}{\dashedruled{}} 
        ($B_2^2,B_1^2B_1^2$)
    \textcolor{mplorange}{\solidruled{}},\textcolor{mplorange}{\dashedruled{}} 
        ($B_2^3,B_1^3B_1^3$)
    \textcolor{mplyellow}{\solidruled{}},\textcolor{mplyellow}{\dashedruled{}} 
        ($B_2^4,B_1^4B_1^4$)
    \textcolor{mplpurple}{\solidruled{}},\textcolor{mplpurple}{\dashedruled{}} 
        ($B_2^5,B_1^5B_1^5$)
    \textcolor{mplgreen}{\solidruled{}},\textcolor{mplgreen}{\dashedruled{}} 
        ($B_2^6,B_1^6B_1^6$)
    \textcolor{mplblue}{\solidruled{}},\textcolor{mplblue}{\dashedruled{}} 
        ($B_2^7,B_1^7B_1^7$)
    \textcolor{mplbrown}{\solidruled{}},\textcolor{mplbrown}{\dashedruled{}} 
    ($B_2^{CD},B_1^{CD}B_1^{CD}$)
    \protect\solidrule{} (Spectral).
}
\label{fig:BsplineEff2}
\end{figure*}

As mentioned above, it is desirable to take $N$ large, but the underlying
numerics often limit $N$. However, lower-order numerical operators can be
designed to mimic higher-order filters without increasing the order of the
differential equation. For instance, consider the operator given by the difference between the numerical second
derivative operator, $B_2$, and repeated application of the numerical first
derivative operator, $B_1B_1$, (for a general numerical scheme). Figure~\ref{fig:BsplineEff2} 
shows each of these operators for second-order centered difference
numerics and several orders of B-splines. A simple Taylor expansion 
for second-order centered difference numerics gives: 
\begin{equation}
    \begin{split}
        (B_2^{CD} - B_1^{CD} B_1^{CD})u 
    &\equiv \frac{-u_{j+2} + 4 u_{j+1} - 8 u_j + 4 u_{j-1} + u_{j-2}}{4
    \Delta^2} = - \frac{\Delta^2}{4} \frac{d^4 u}{dx^4} +  \mathcal{O}(\Delta^4) \sim -\Delta^2 F_4(u)
    \end{split}
    \label{eq:B2B1B1CD}
	.
\end{equation}
Similarly, it can be shown that $(B_2^7 - B_1^7B_1^7)u \sim \Delta ^8 \frac{d^{10}
u}{dx^{10}} \approx \Delta^8 F_{10}(u)$ and $(B_2^2 - B_1^2B_1^2)u \sim -\Delta^2
\frac{d^{4} u}{dx^{4}}  \approx -\Delta^2 F_4(u)$.
In all of these cases, \gryr{$(-1)^{N/2+1}(B_2-B_1B_1) \sim \Delta^{N-2}F_N$} for some
value of $N$ \gryr{(and positive constant of proportionality)}, which
corresponds exactly to the form of (\ref{eq:largekcomm1}); i.e., 
\begin{equation} 
    \gryr{\C^I(u) = -C \Delta \frac{d \Delta}{d x} \left( B_2-B_1B_1
	\right) \bar{u}}
    \label{eq:largekcomm2} 
     .
\end{equation} 
The $B_2-B_1B_1$ operator has the effect of higher
order differential operators without changing the order of the differential
equation. This avoids the need to explicitly define higher order derivative
approximations and for additional boundary conditions.  Furthermore, the
$B_2-B_1B_1$ operator is
particularly useful because the first and second differential operators are
already required by the governing equations, and are thus readily available in
practical applications.

Aside from approximating higher order derivatives, the $B_2-B_1B_1$ operator has
several desirable properties that make it useful for commutation modeling. Compare
the second derivative operator, $B_2$, with repeated application of the first
derivative operator, $B_1B_1$, in Fig.~\ref{fig:BsplineEff2}; for numerically
well-resolved wavenumbers, the $B_2$ and $B_1B_1$ operators are almost
identical, and they cancel out. However, for insufficiently resolved
wavenumbers, their difference is nonzero and can be used to filter out higher wavenumbers.  In
essence, the $B_2 - B_1 B_1$ operator acts as an indicator for the scales whose
dynamics are not sufficiently representable by the underlying numerics. This property is
particularly useful for reducing commutation error as the $B_2-B_1B_1$ model
is specifically formulated to damp wavenumbers with negative group velocity, which is where the commutation
error manifests for many typical numerical schemes. Moreover, the operator naturally adapts to the underlying numerics.

\subsection{Model Coefficient \label{sec:modelconst}}

\gryr{In an LES, the statistical analysis in Sec.~\ref{sec:statistics} can be used to set the model
coefficient $C$ to produce the correct rate of energy transfer to the subgrid
scales (e.g., evaluating (\ref{eq:comm_E}) or~(\ref{eq:comm_k1}) for a
Kolmogorov spectrum).}
However, for the simple case of linear convection considered here, 
it is useful to examine how the behavior of the model changes as the coefficient
varies. In Appendix~\ref{sec:A2}, a coefficient for this purpose is derived,
which is repeated here,
\begin{equation}
    C = \frac{(-1)^{\frac{N-2}{2}}\log(\varepsilon)}{ 2\left( \frac{ \Delta_c-
    \Delta_f}{\Delta_c} \right) \left( \Delta^\N \hat{F}_\N(\kappa_{a}) \right)} 
	,
    \label{eq:modelconst}
\end{equation}
where $\varepsilon$ is a tolerance
level indicating the target fraction of incident energy that will be reflected,
and ${\hat{F}}_\N(\kappa)$ is the spectrum
of $F_\N$ evaluated at the wavenumber $\kappa$.
Setting the parameter $\varepsilon$ involves a tradeoff between dissipating erroneous reflections in
fine regions of the grid and preserving well-resolved wavenumbers in coarse
regions of the grid. 

\gryr{Furthermore, this choice of coefficient may indicate how the numerical properties of the
	commutation error discussed in Sec.~\ref{sec:numerics} can
be exploited to improve the model. To elaborate, notice that we use the
value of $\hat{F}_\N$ at the apex wavenumber $\kappa_a$. This choice is made to
take advantage of how the commutation error manifests numerically. Specifically,
the constant is designed to quickly damp high wavenumbers \textit{after} they have
been reflected. Targeting reflections yields a 
larger separation between the scales that must be filtered out, and those
that need to be preserved. 
This approach is especially advantageous for low values of $\N$ for which the filters produced
from (\ref{eq:largekcomm}) are not particularly sharp. In
essence, it is more advantageous to use a model to correct for the absence of
$\C^I$ in this problem, than to model $\C^I$ directly. This strategy works
particularly well
with the $B_2-B_1B_1$ filters described in the previous section, which target
the poorly resolved wavenumbers. In LES, more work is
needed to see if a similar exploit can be performed. For example, nonlinear
interactions may require scales to be removed before reflection, but this would lead to
more dissipation of the resolved turbulence. }

\begin{figure*}[t!]
    \centering
    \includegraphics[width=0.5\textwidth]{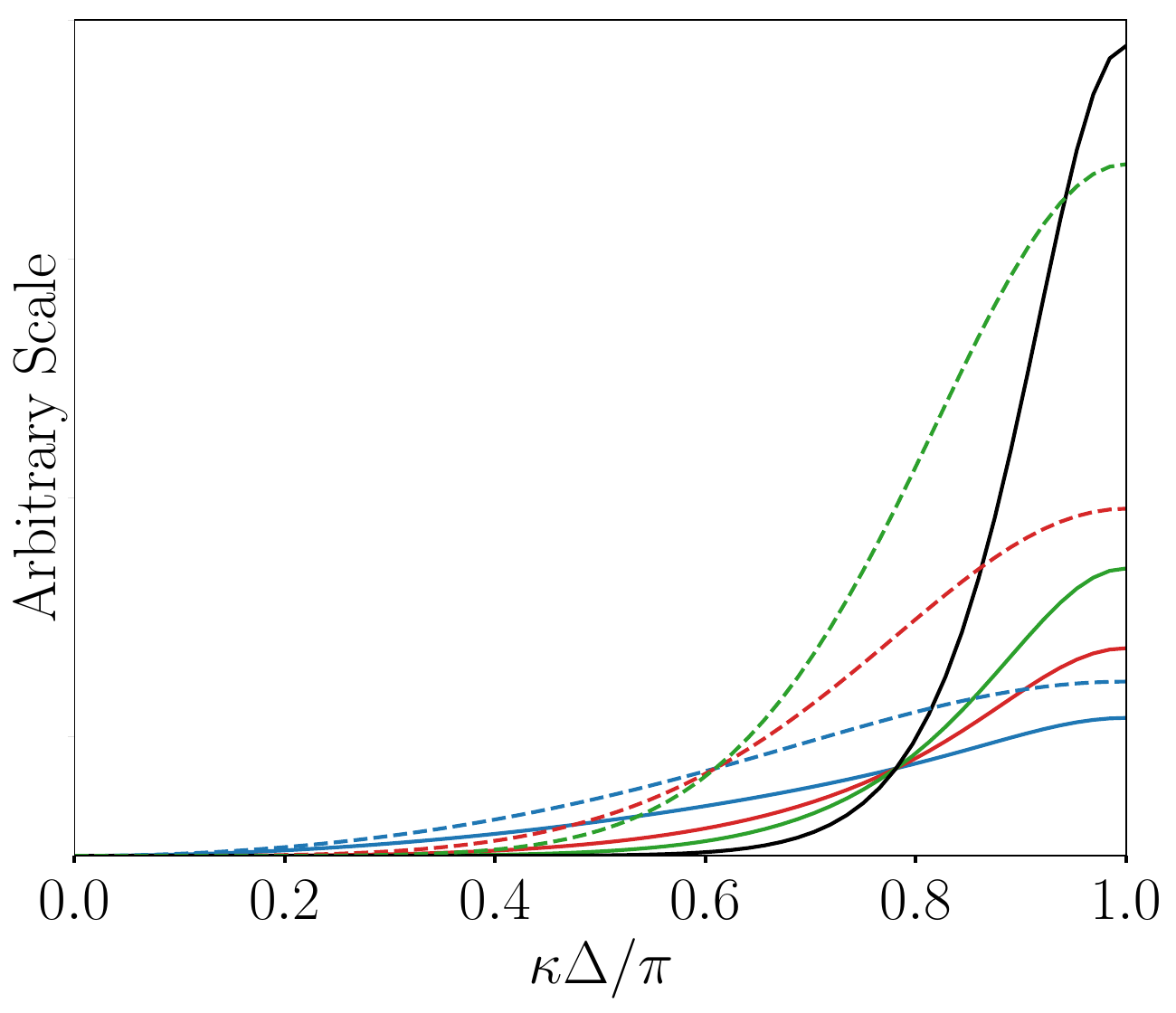}
    \caption{
		\gry{Spectra of the operator $C F_N$ for various forms of $F_N$ and the constant $C$ determined by
			(\ref{eq:modelconst}). The scale is arbitrary and depends on 
	$\varepsilon$ and the resolution $\Delta$. }
    \textcolor{mplblue}{\solidrule{}} ($F_2 = B_2^7$),
    \textcolor{mplred}{\solidrule{}} ($F_4 = B_4^7$),
    \textcolor{mplgreen}{\solidrule{}} ($F_6 = B_6^7$),
    \textcolor{mplblack}{\solidrule{}} ($\Delta^8 F_{10}\sim B_2^7-B_1^7B_1^7$),
    \textcolor{mplblue}{\dashedrule{}} ($F_2=B_2^2$),
    \textcolor{mplred}{\dashedrule{}} ($\Delta^2 F_4 \sim B_2^2-B_1^2B_1^2$),
    \textcolor{mplgreen}{\dashedrule{}} ($F_6 = (B_2^2)^3$).
}
    \label{fig:filters}
\end{figure*}

\gryr{The spectrum of the operator $C F_\N$ for second- and seventh-order
	B-splines with this choice of coefficient and several different choices of
	$\N$ is shown is shown in Fig.~\ref{fig:filters}. For an
arbitrary tolerance value of $\varepsilon$, the model coefficient creates an
intersection point at $\kappa_{a}$ between different values of $\N$. 
This intersection point shifts depending on the order of the underlying
numerics. Figure~\ref{fig:filters} shows how as $N$ increases, the poorly
resolved scales are dissipated more rapidly and the well resolvable scales are better preserved. }

\subsection{Model Results}
\label{sec:results}

\begin{figure}[h!]
    \centering

    \begin{tikzpicture}
        \node at (0,0)
        {\includegraphics[width=1\textwidth]{aniso_grid2.pdf}};
		\node at (0,-3.2)
        {\includegraphics[width=1\textwidth]{7B_nodiss_0_aniso1.pdf}};
		\node at (0,-6.4)
        {\includegraphics[width=1\textwidth]{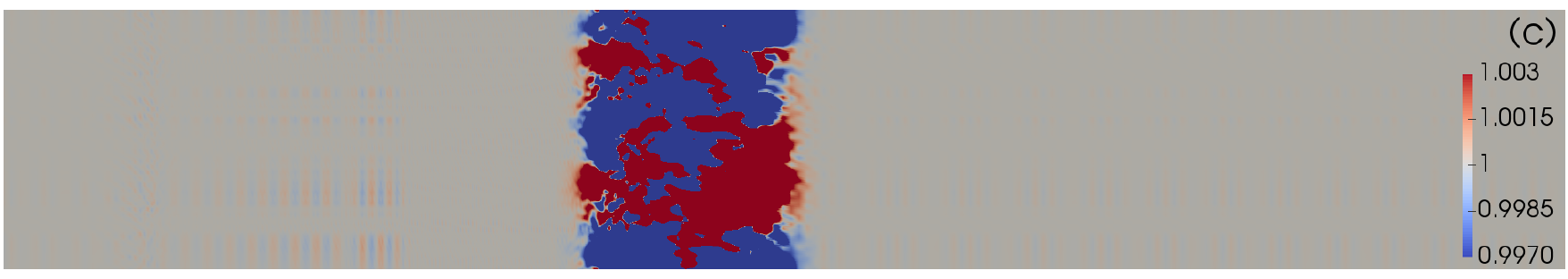}};
		\node at (0,-9.6)
        {\includegraphics[width=1\textwidth]{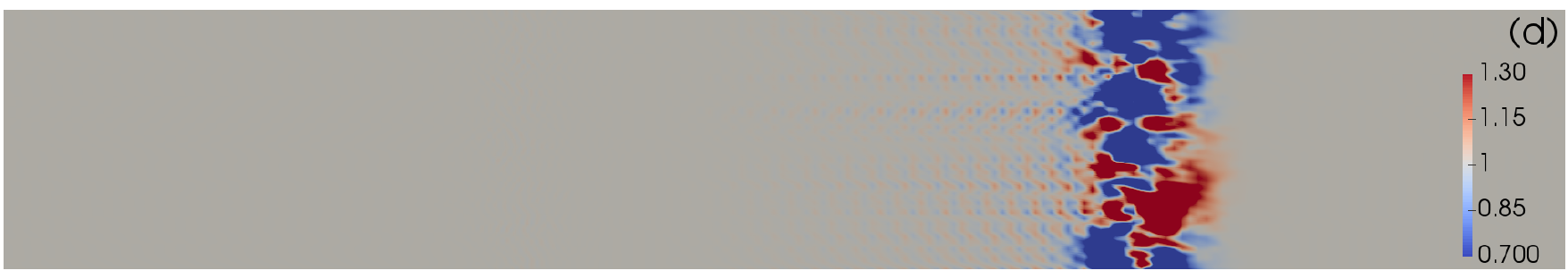}};
		\node at (0,-12.8)
        {\includegraphics[width=1\textwidth]{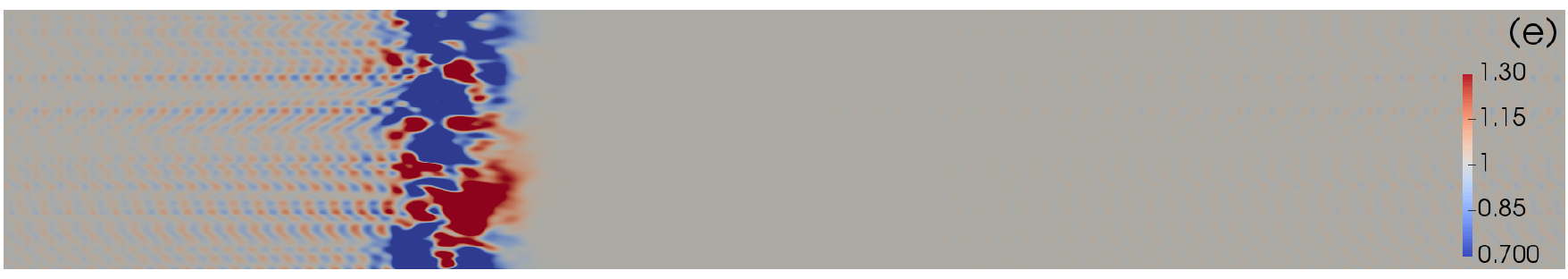}};
        
		\draw[->,thick] (9.25,-3.2) -- (9.25,-12.8);
		\node[thick,rotate=270] at (9.5,-8) {Time};
        \node[fill=white,rounded corners=2pt,inner sep=1pt] at (8.5,1.2) {(a)};
    \end{tikzpicture}
    \caption{Convection of a packet of homogeneous, isotropic turbulence through
        an anisotropic, inhomogeneous grid with seventh-order B-splines, a
        convection velocity of 1, and a commutation model with $\Delta^8F_{10}
		\sim
        (B_2^7-B_1^7B_1^7)$ and $\varepsilon=0.001$. 
		The packet is moving through the resolution change
		to the right. (a) A slice of the numerical grid. (b),
        (c), (d), and (e) show the streamwise velocity field at times $0.00$,
    $7.03$, $11.72$, and $39.06$, respectively.
	Figure (c) is scaled to highlight the absence of the spurious high
	wavenumber reflections, as in Figure~\ref{fig:tikzgrid}c.
	}
\label{fig:realResults}
\end{figure}

\gryr{The ability of the model to correct for the issues related to resolution
	inhomogeneity is tested in the same setting described in Sec.~\ref{sec:impact}. 
The commutation model is introduced into equation (\ref{eq:pureconvection}) as
\begin{equation}
    \frac{\partial \overline{\mathbf{u'}}}{\partial t} + 
    U_x \frac{\delta \overline{\mathbf{u'}}}{\delta x}
    = (-1)^{\frac{N}{2}+1}C \Delta_x \left(U_x \frac{\partial \Delta_x}{\partial x} \right) 
    \left( \Delta_x^{N-2} F_\N (\overline{\mathbf{u'}}) \right)
    \label{eq:model}
    ,
\end{equation}
where the constant $C$ is given by (\ref{eq:modelconst}), and the operator
$F_\N$ is an approximation of the $\N$th derivative in the $x$-direction ($F_\N \approx
\partial^\N /\partial_x^\N)$. 
Recall that in this setting the local grid spacing is
$\boldsymbol{\Delta}(x) = (\Delta_x(x),\Delta_y,\Delta_z)$ 
and the dependence on $x$ in (\ref{eq:model}) arises because 
the resolution inhomogeneity is only in the $x$-direction.
}

\begin{figure*}[t!]
\centering
\begin{subfigure}[t]{0.45\textwidth}
\centering
\includegraphics[width=1\textwidth]{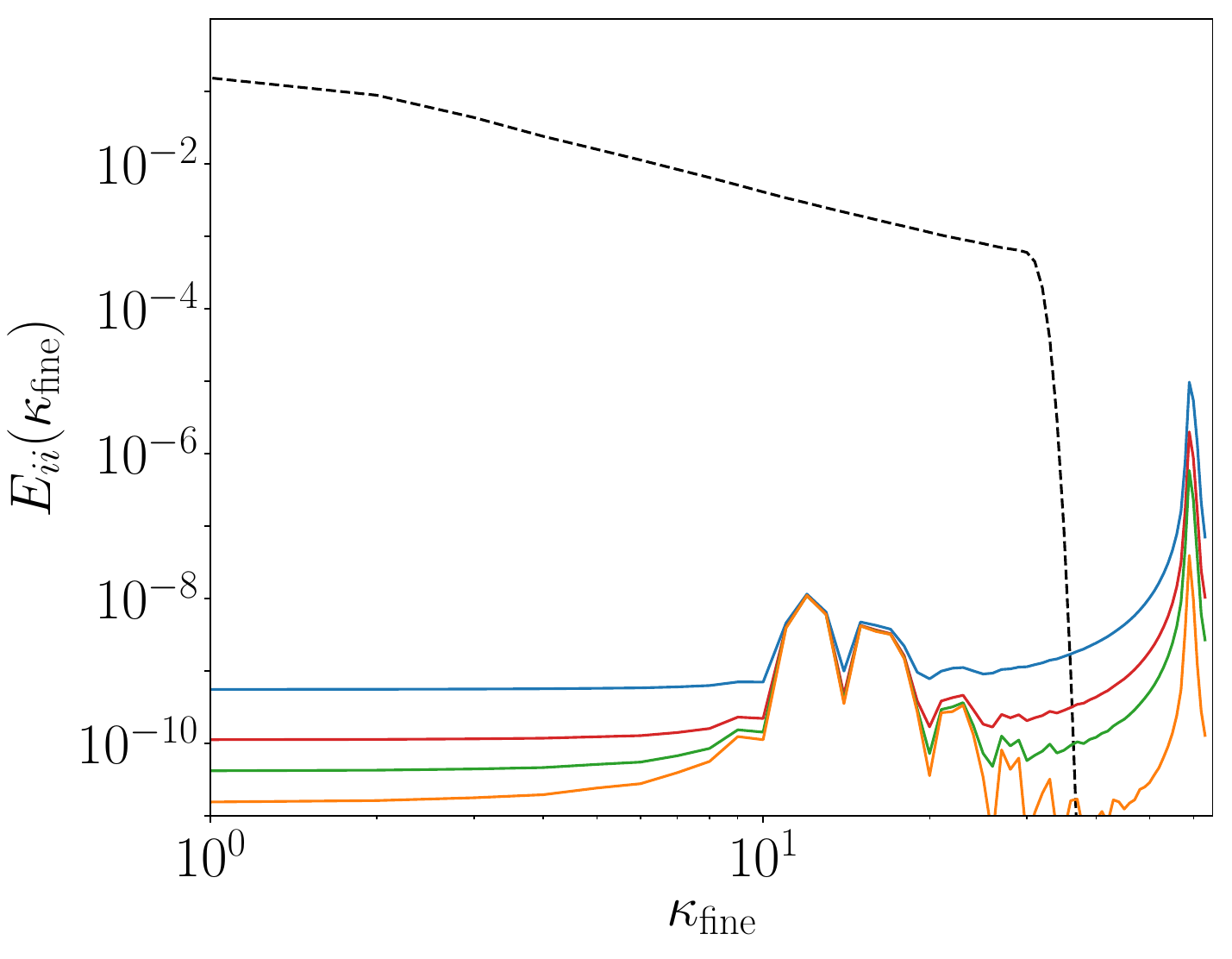}
\caption{$t = 7.03$}
\label{fig:7Bpile}
\end{subfigure}%
\hspace{1em}
\begin{subfigure}[t]{0.45\textwidth}
\centering
\includegraphics[width=1\textwidth]{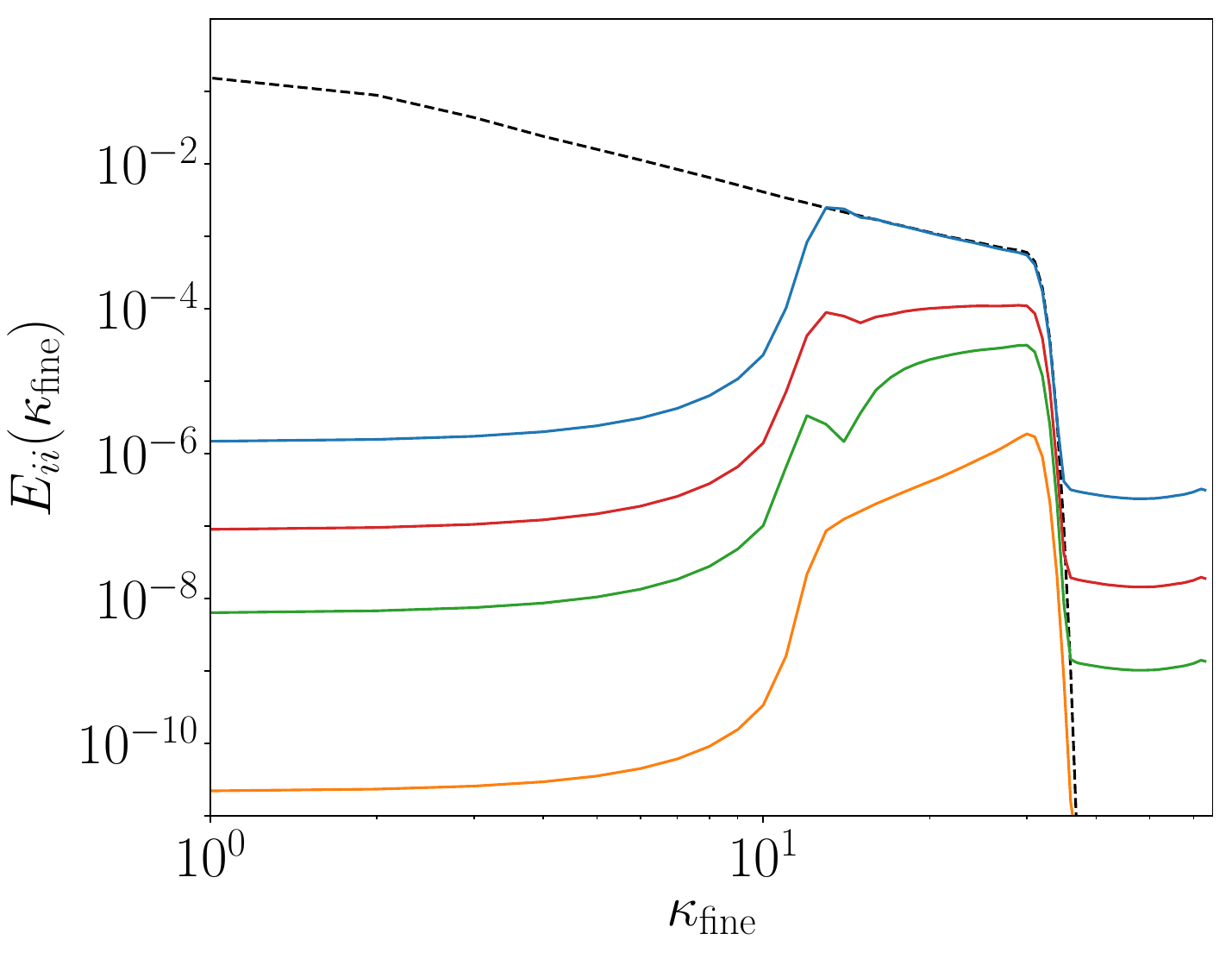}
\caption{$t = 15.63$}
\label{fig:7Bref}
\end{subfigure}
\vskip\baselineskip
\begin{subfigure}[t]{0.45\textwidth}
\centering
\includegraphics[width=1\textwidth]{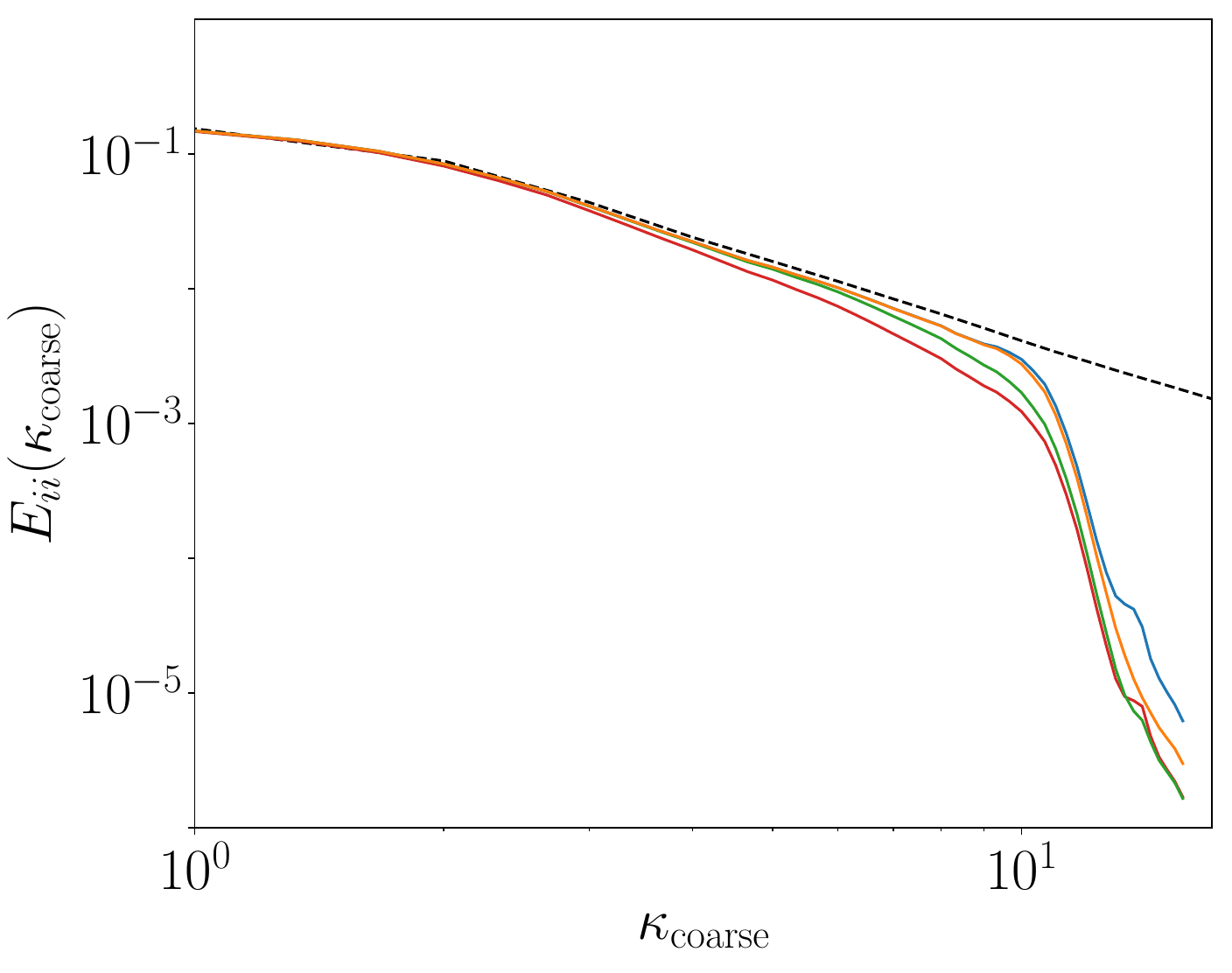}
\caption{$t = 11.72$}
\label{fig:7Bcoarse}
\end{subfigure}%
\hspace{1em}
\begin{subfigure}[t]{0.45\textwidth}
\centering
\includegraphics[width=1\textwidth]{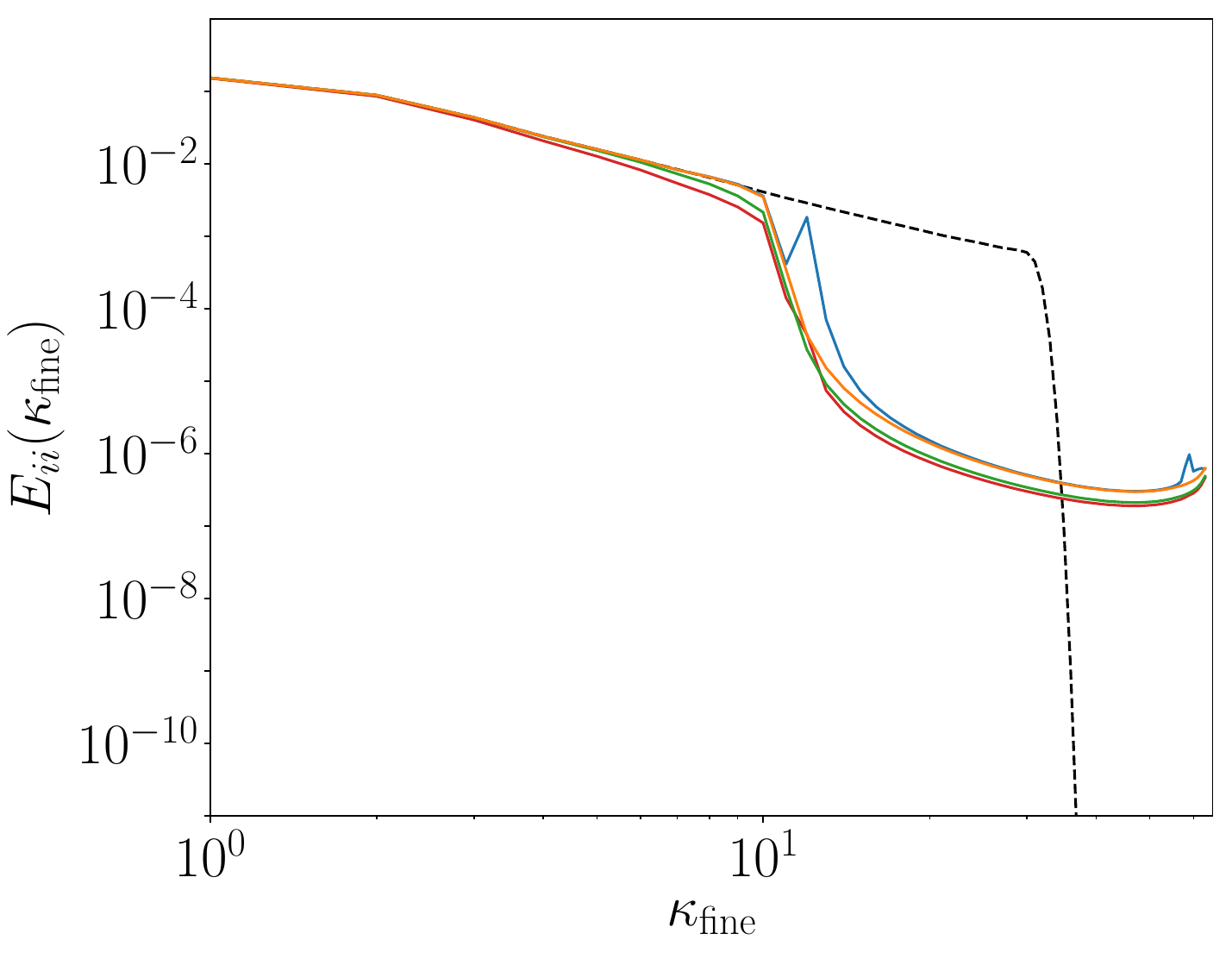}
\caption{$t = 39.06$}
\label{fig:7Bend}
\end{subfigure}
\caption{
    Energy spectra, $E_{ii}$, in the direction of convection at different times
    $t$ for 7th order B-splines;
	$\kappa_{\rm{fine}}$ and $\kappa_{\rm{coarse}}$ represent the wavenumbers in the fine and coarse
    regions of the domain, respectively.
    (a) High wavenumber
    reflections propagating backward through the fine region.     
    (b) The subsequent reflections propagating forward through the fine region.
    (c) The spectra of the resolved turbulence packet 
    in the coarse region.     
    (d) The spectra of the turbulence packet after one
    flow through.    
    \textcolor{mplblue}{\solidrule{}} (No model), 
    \textcolor{mplred}{\solidrule{}} ($F_2 = B_2^7$, $\varepsilon  = 0.1$), 
    \textcolor{mplgreen}{\solidrule{}} ($F_4 = B_4^7$, $\varepsilon = 0.01$), 
    \textcolor{mplorange}{\solidrule{}} ($\Delta^8 F_{10} \sim (B_2^7-B_1^7B_1^7)$, $\varepsilon = 0.001$),
    \protect\dashedrule{} (Initial Packet) 
}
\label{fig:7Bresults}
\end{figure*}

\begin{figure*}[t!]
\centering
\begin{subfigure}[b]{0.45\textwidth}
\centering
\includegraphics[width=1\textwidth]{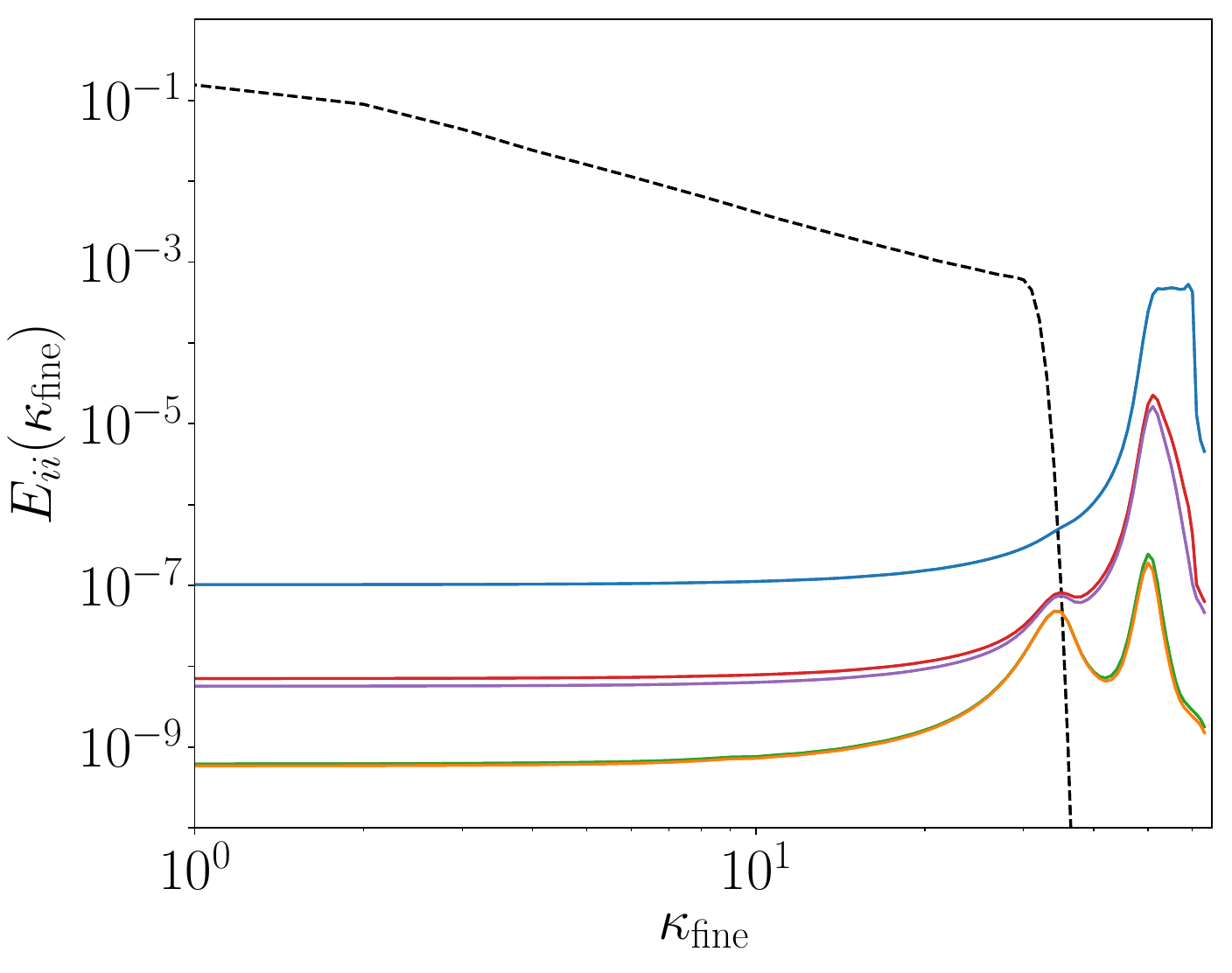}
\caption{$t=11.72$}
\label{fig:2Bpile}
\end{subfigure}%
\hspace{1em}
\begin{subfigure}[b]{0.45\textwidth}
\centering
\includegraphics[width=1\textwidth]{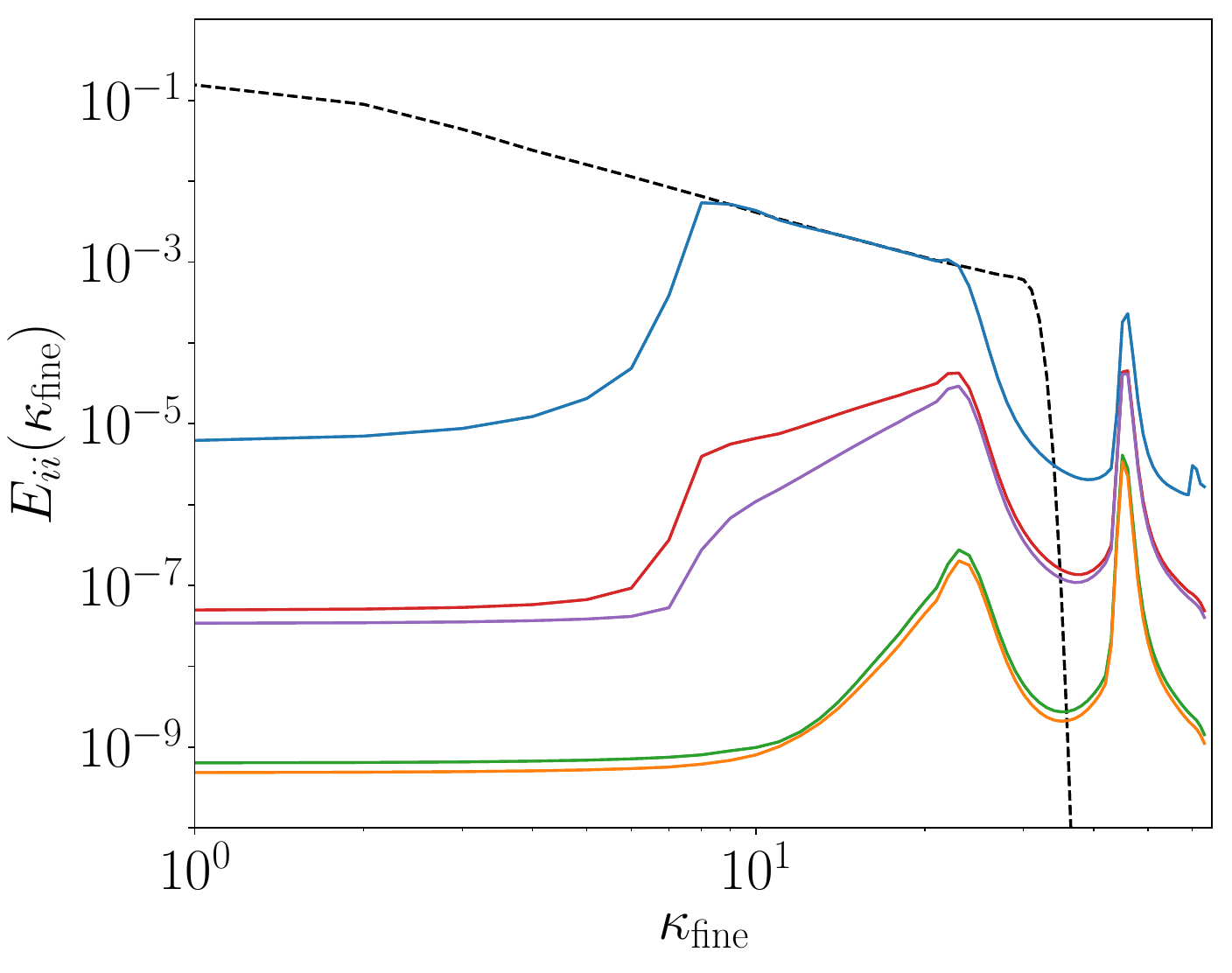}
\caption{$t=23.44$}
\label{fig:2Bref}
\end{subfigure}
\vskip\baselineskip
\begin{subfigure}[b]{0.45\textwidth}
\centering
\includegraphics[width=1\textwidth]{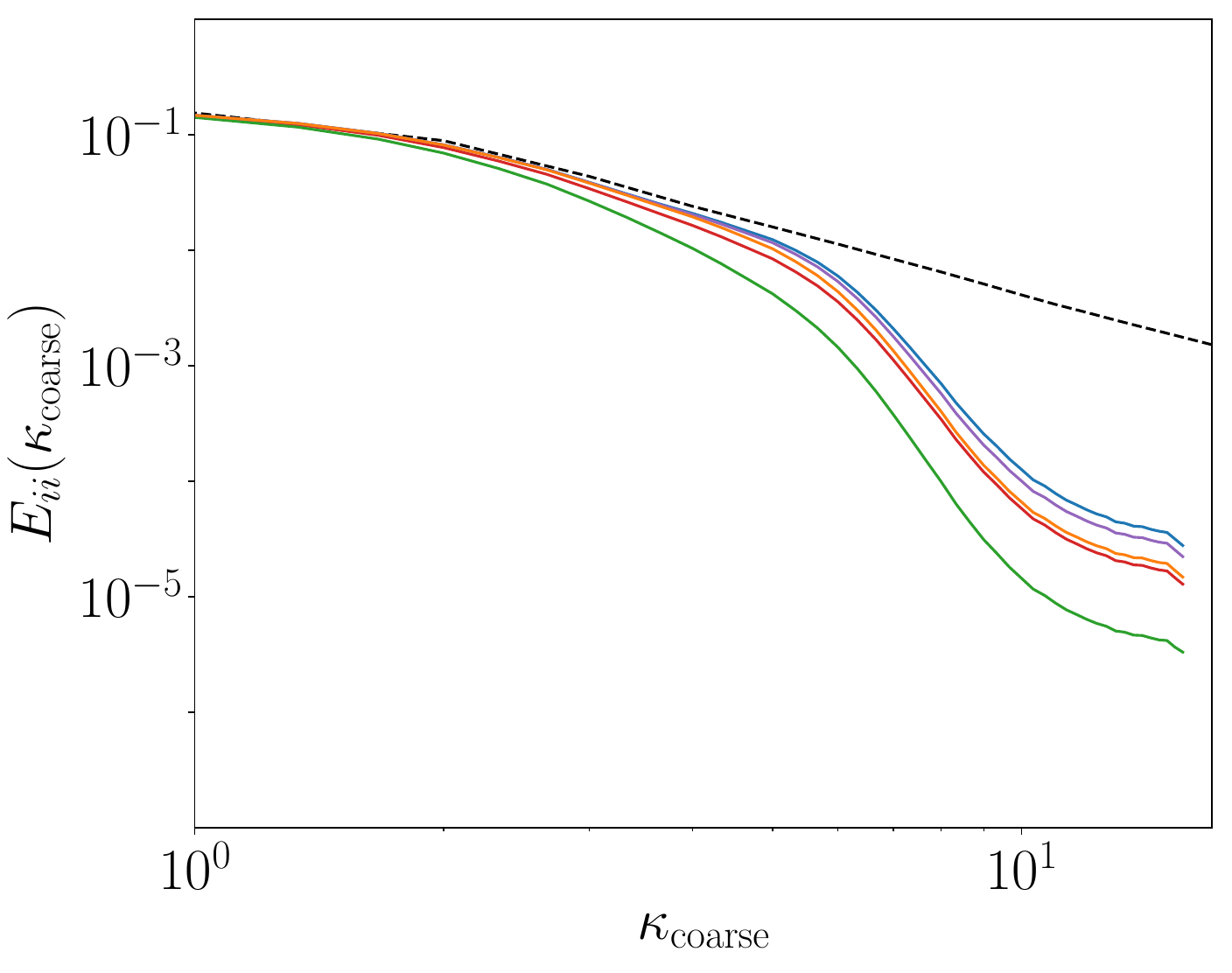}
\caption{$t=11.72$}
\label{fig:2Bcoarse}
\end{subfigure}%
\hspace{1em}
\begin{subfigure}[b]{0.45\textwidth}
\centering
\includegraphics[width=1\textwidth]{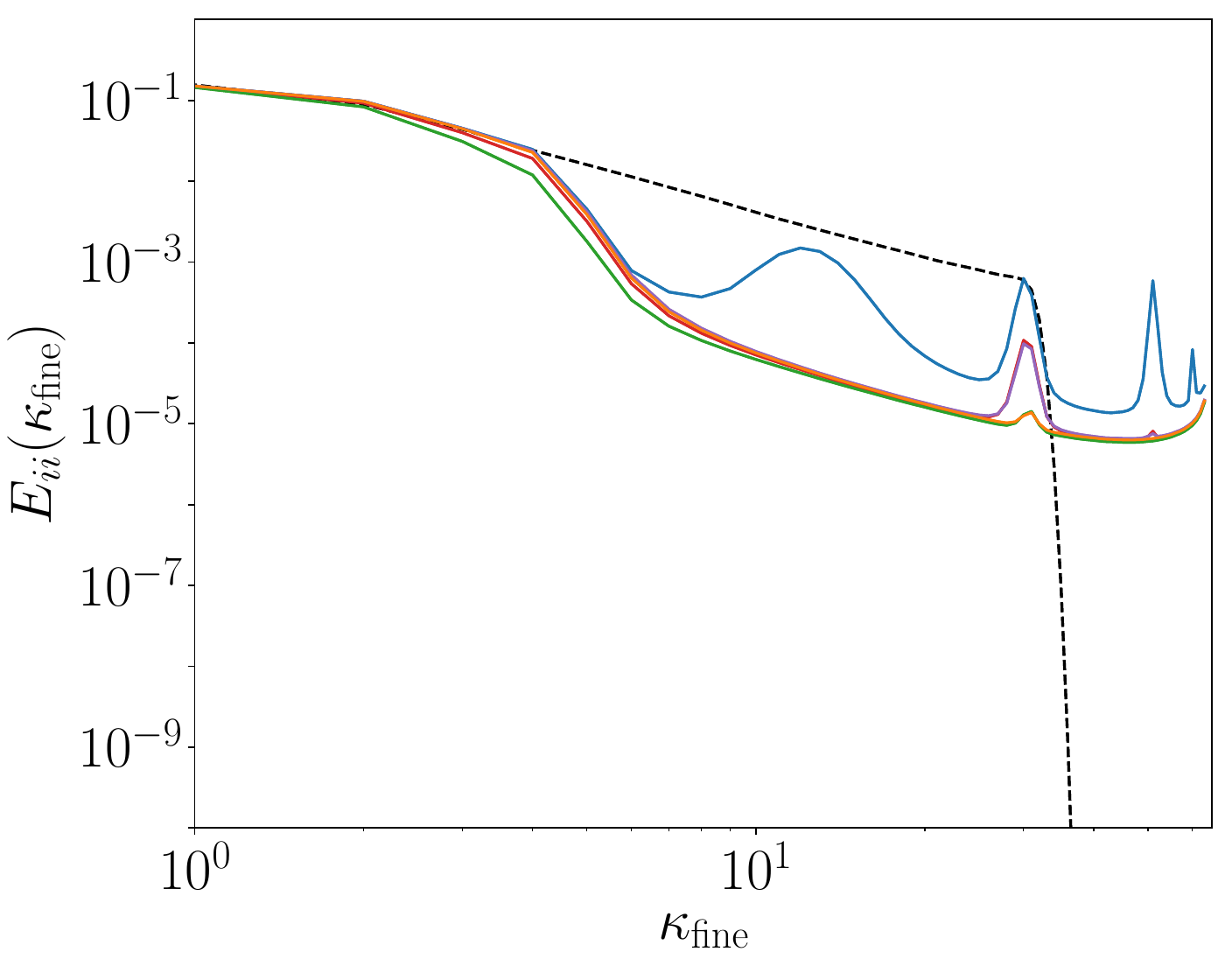}
\caption{$t=39.84$}
\label{fig:2Bend}
\end{subfigure}
\caption{
    Energy spectra, $E_{ii}$, in the direction of convection at different times
    $t$ for 2nd order B-splines;
	$\kappa_{\rm{fine}}$ and $\kappa_{\rm{coarse}}$ represent the wavenumbers in the fine and coarse
    regions of the domain, respectively.
	(a) High wavenumber
    reflections propagating backward through the fine region.     
    (b) The subsequent reflections propagating forward through the fine region.
    (c) The spectra of the resolved turbulence packet 
    in the coarse region.     
    (d) The spectra of the turbulence packet after one
    flow through.    
    \textcolor{mplblue}{\solidrule{}} (No model), 
    \textcolor{mplpurple}{\solidrule{}} ($\Delta^2 F_4 \sim -(B_2^2-B_1^2B_1^2)$, $\varepsilon = 0.1$), 
    \textcolor{mplorange}{\solidrule{}} ($\Delta^2 F_4 \sim -(B_2^2-B_1^2B_1^2)$, $\varepsilon =
    0.001$),
    \textcolor{mplgreen}{\solidrule{}} ($F_2 = B_2^2$, $\varepsilon = 0.001$), 
    \textcolor{mplred}{\solidrule{}} ($F_2 = B_2^2$, $\varepsilon  = 0.1$), 
    \protect\dashedrule{} (Initial Packet)}
\label{fig:2Bresults}
\end{figure*}

\def\arraystretch{1.25}
\begin{table}[b!]
\caption{\gry{Model Constants.}}
\begin{center}
\begin{tabular}{@{\hspace{1em}}c@{\hspace{1em}}@{\hspace{1em}}c@{\hspace{1em}}@{\hspace{1em}}c@{\hspace{1em}}@{\hspace{1em}}c@{\hspace{1em}}}
    \hline
    \hline
    $\N$ & $F_\N$ & $\varepsilon$ & $C$ \\
    \hline
    $2$  & $B_2^7$ & $0.1$ & $0.24$ \\
    $4$  & $B_4^7$ & $0.01$ & $0.07$\\
	$10$  & $\Delta^{-8} (B_2^7-B_1^7B_1^7)$ & $0.001$ & $4.31$ \\
    $2$  & $B_2^2$ & $0.1$ &   $0.38$ \\
    $2$  & $B_2^2$ & $0.001$ & $1.15$ \\
	$4$  & \gryr{$-\Delta^{-2}(B_2^2-B_1^2B_1^2)$} & $0.1$ &     $0.76$ \\
	$4$  & \gryr{$-\Delta^{-2}(B_2^2-B_1^2B_1^2)$} & $0.001$ &   $2.28$ \\
    \hline
    \hline
\end{tabular}
\end{center}
\label{tab:const}
\end{table}

For the seventh-order B-spline results, three different choices of $\N$ and
$\varepsilon$ are tested: $\N=2$ corresponding to the second derivative operator
$F_2 = B_2^7$ with $\varepsilon = 0.1$, $\N = 4$ corresponding to the fourth
derivative operator $F_4 = B_4^7$ with $\varepsilon = 0.01$, and $\N = 10$
corresponding to the $\Delta^8 F_{10} \sim B_2^7-B_1^7B_1^7$ operator with
$\varepsilon = 0.001$.  For the second-order B-spline results, two choices of
$\N$ are tested for both $\varepsilon = 0.1$ and $\varepsilon = 0.001$: $\N = 2$
corresponding to the $F_2 = B_2^2$ operator, and $\N = 4$ corresponding to the
\gryr{$\Delta^2 F_4 \sim -(B_2^2 - B_1^2 B_1^2)$} operator.  These values, along with the
model coefficients, are listed in Table \ref{tab:const}. The one-dimensional
energy spectra in the fine and coarse regions of the domain are shown in
Fig.~\ref{fig:7Bresults} and Fig.~\ref{fig:2Bresults} for seventh- and
second-order B-splines, respectively.  The results of the model in physical
space for the seventh order B-spline case with $\Delta^8 F_{10} \sim
(B_2^7-B_1^7B_1^7)$ and $\varepsilon = 0.001$ are shown in
Fig.~\ref{fig:realResults}.

Compare these results with the pure convection case (i.e., the no model case)
examined in Sec.~\ref{sec:impact}.  The model significantly corrects the
spatial structure and the energy distribution of the turbulence packet as it
flows through the inhomogeneous grid. In all cases, the model reduces the
spurious high wavenumber reflections by (at least) a factor around
$\varepsilon$, as desired (see Figs~\ref{fig:7Bref} and~\ref{fig:2Bref}). Recall that the largest
initial wavenumber with positive group velocity has the smallest reflected
wavenumber and is dissipated the least by the model, so the value of
$\varepsilon$ should be validated at these wavenumbers in the spectra
results. 
Moreover, the model preserves the resolvable turbulence in the coarse region as
much as possible. The seventh-order results show that higher order filters (i.e.,
larger values of $\N$) preserve the resolvable turbulence 
while dissipating the reflections more strongly. In particular, the $B_2^7-B_1^7
B_1^7$ model matches the ideal spectra in the coarse region almost exactly and
is still able to reduce reflections by at least three orders of magnitude (see
Fig~\ref{fig:7Bcoarse}). Similarly, in the second-order B-spline results, the
$\N=4$ cases match the
original spectra in the coarse region more closely than the $\N=2$ cases for the same value
of $\varepsilon$ (see Fig~\ref{fig:2Bcoarse}).  Finally, the model mitigates the effect of erroneous
reflections on incoming turbulence, as demonstrated by examining the
turbulence packet after one flow through (see Figs.~\ref{fig:7Bend}
and~\ref{fig:2Bend}).  Even a modest reduction
in the reflections --- such as that from the low $\N$ and $\varepsilon$ cases
--- yields much better spectra than the pure convection case.  The
spectra after one flow through match quite well with the initial packet's
spectrum for all coarse wavenumbers.

\section{Conclusion \label{sec:conclusion}}

\gryr{
Practical LES of high Reynolds number turbulent flows often requires
inhomogeneous resolution. 
The inhomogeneous part of the commutator $\C^I$ is responsible for transferring energy between
resolved and unresolved scales as a consequence of the resolution inhomogeneity,
and so it must be
modeled. However, $\C^I$ is often ignored in practice leading to commutation
error.  In this paper, we investigate the commutator and corresponding commutation error
as related to filters that include a discrete projection.
}

\gryr{
The impact of the commutation error that occurs as turbulence convects through coarsening grids 
is governed by the propagation
properties of the underlying numerics (see Sec.~\ref{sec:numerics}). For many
conservative numerical schemes such as those considered here, the energy in
newly unresolvable scales is unphysically transferred to higher wavenumbers in
the fine region of the grid, instead of to the subgrid scales in the coarse
region of the grid. The result is a non-physical reflection of unresolvable
scales back into the fine region of the grid at higher wavenumbers with negative
group velocities.
The nonlocal wavenumber interactions introduced by
resolution inhomogeneity may be especially problematic in LES of turbulence (see
Sec.~\ref{sec:impact}).} Since the implicit filter combined with the numerical
derivative operators define the scales in the resolved field whose dynamics are
accurately represented, LES modeling in general cannot be pursued independently
of the properties of the numerics (e.g., 
\citep{fureby1997mathematical,hughes2000large,bazilevs2007variational,grinstein2007implicit,ghosal1996analysis}). As \citet{meneveau2000scale} mentioned in their review
paper, ``our understanding of the interplay between numerical and
modeling issues is presently quite limited.'' The work here aims to address this
interplay in the context of commutation error.

\rdm{The statistical analysis of the commutation term $\C^I$ developed in
	Sec.~\ref{sec:statistics} yields a quantitative measure of the
magnitude of $\C^I$ and therefore how important it is to model, as a
function of the resolution gradient and the convection velocity.
Furthermore, a commutator
model can be formulated to match important statistical
features of the commutator \emph{a priori}, such as its spectrum.
For example, the dependence of the commutator spectrum on the
derivative of the Fourier transformed filter kernel shows that a
commutator model should act at the high wavenumbers over which the
filter rolls off.  Similarly, the parameters in a model of the commutator can be
calibrated to match known statistical characteristics \emph{a priori}
(e.g., evaluating (\ref{eq:comm_E}) for a Kolmogorov spectrum).
It is
important to consider the statistical characteristics of the
commutation term because \emph{a priori} consistency of certain statistical
characteristics of an LES model is a necessary condition for accurate
\emph{a posteriori} statistics of an LES
solution \citep{meneveau1994statistics,moser2020statistical}.}

\rdm{The series approximation of $\C^I$ from Sec.~\ref{sec:ghosal_review}, is
also useful for informing commutation models, despite the fact that this
analysis is formally only applicable to invertible filters. In
particular, (\ref{eq:newCommError}) shows that asymptotically, the commutator is
expressible in terms of even derivatives of the filtered field, is
proportional to the resolution gradient and proportional to the
convection velocity. This places significant constraints on any
operator intended to model the commutator. Furthermore, when applied
to inertial range turbulence, the fact that all the terms in the
series (\ref{eq:newCommError}) are of the same asymptotic order implies that
high order derivatives of the filtered field are as important as low
order derivatives, suggesting that practical models expressed in terms
of derivatives of the filtered field should include derivatives of as
high an order as feasible. Indeed, this observation motivated the
formulation of the model proposed in Sec.~\ref{sec:modelform}. The asymptotic
ordering of the terms in (\ref{eq:newCommError}) also suggests that using
``commuting filters'' whose low-order moments vanish, which has often
been proposed based on the analysis of \citep{ghosal1995basic}, is not
sufficient to make the commutator negligible. }
\rdm{This is not to say that explicit filters are not useful for other
purposes, such as eliminating energy in scales with negative group
velocity due to numerical dispersion, which will also mitigate the effects of
commutation error.}

\rdm{For the case of flow through a coarsening grid, a practical formulation of
	a high-order dissipative model is proposed in Sec.~\ref{sec:modelform}. The
	model is based on the analysis in Sec.~\ref{sec:ICE} and is formulated to be
	proportional to $B_2-B_1B_1$ ($B_2$ and $B_1$ are numerical second and first
derivative operators), which avoids several practical complications with
hyperviscosity models.} In setting the
model constant, there is a trade-off between eliminating the spurious
reflections in the fine region and preserving the dynamically
meaningful scales in the coarse region (see Sec.~\ref{sec:modelconst}).
Furthermore, the $B_2-B_1B_1$ operator could also be useful for addressing 
discretization error, which can dominate over the LES models and must
therefore be considered in LES \citep{ghosal1996analysis,chow2003further,kravchenko1997effect}.  
The $B_2-B_1B_1$ operator dissipates scales whose
dynamics are poorly represented in an LES and adapts naturally to
the underlying numerics without needing to define ad hoc filter widths or
explicit filters. This aligns with previous work suggesting the use of hyperviscosities for
mitigating the effects of discretization error \citep{cook2005hyperviscosity}.

Finally, the commutator modeling pursued here has focused on the case
when the turbulence flows from fine resolution to coarse
resolution.
However, the other situation (flowing from coarse to fine resolution)
is also of interest. Modeling $\C^I$ in this case is challenging
because resolved fluctuations must be created. Models based on
negative dissipation \citep{ghosal1995basic}  and forcing \citep{haering2020active}, 
have been proposed, but more work is required. \gryr{As in the coarsening
	resolution case, the analysis in Sec.~\ref{sec:ghosal_review}
	and~\ref{sec:statistics} may be useful in developing an appropriate model.}

\appendix

\section{Multiscale Analysis of the Commutator}
\label{sec:multiscale}
In this Appendix, we consider a multiscale asymptotic analysis of the
inhomogeneous part of the commutator. This analysis is used to obtain the
leading order terms in a series representation of the commutator
in the case of invertible filters (Appendix~\ref{sec:multiscale_deconvolution}) and the statistical characteristics of the
commutator for a general filter (Appendix~\ref{sec:multiscale_spectral}). 
\subsection{Series Representation of the Commutator}
\label{sec:multiscale_deconvolution}
As in \cite{ghosal1995basic}, any smoothly nonuniform grid $x$ with spacing
$\Delta(x)$ can be mapped
to a uniform grid of spacing $\Delta_\xi$ through some invertible monotonic differentiable mapping function
$\xi = f(x)$. Let $G(\xi)$ be a symmetric filter kernel normalized on $\xi$
that decays 
sufficiently fast so that all moments of $G$ exist.  To define the filtering
operation applied to an arbitrary function $\psi(x)$, we first  make a change
of variables to $\xi$ ($\psi(\xi) \equiv \psi(f^{-1}(\xi))$) and then filter
$\psi(\xi)$ with the homogeneous filter defined by $G(\xi)$:
\begin{equation}
	\overline{\psi}(\xi)  =  \frac{1}{\Delta_\xi}
        \int
        G \left(\frac{\xi - \eta}{\Delta_\xi} \right) \psi(\eta) d\eta 
	\label{eq:filtmapxi}
	.
\end{equation}
The result is then transformed back to $x$ to obtain:
\begin{equation}
	\overline{\psi}(x) \equiv \overline\psi(f(x)) =
        \frac{1}{\Delta_\xi} \int
        G \left(\frac{f(x)-f(y)}{\Delta_\xi} \right) \psi(y) f'(y) dy 
	\label{eq:filtmap}
	.
\end{equation}
Therefore $\displaystyle{\frac{d \overline{\psi}}{dx} \equiv \frac{d\overline{\psi} }{d
\xi}\frac{d\xi}{ dx}}$, so that the inhomogeneous part of the commutator is
\begin{equation}
	\C^I(\psi) = \overline{\frac{ d\psi}{dx}} -
	{\frac{d{\overline{\psi}} }{d \xi}}\frac{d\xi}{ dx}
	\label{eq:inhomo_comm}
\end{equation}
as in \citep{moser2020statistical}.

Now, suppose that the resolution (filter width) is slowly varying in $x$, that is
$\frac{d \Delta}{d x}$ is order $\epsilon \ll 1$. Notice that this limit can be
approached in two ways. In particular, consider the length scale $L_\Delta$
defined as the inverse logarithmic derivative of the resolution
($\frac1{L_\Delta}\sim\frac{1}{\Delta}\frac{d\Delta}{dx}$). Then the
$\epsilon$ limit can be approached by (1) allowing $L_\Delta$ to grow
while $\Delta$ remains constant, or (2) letting $L_\Delta$ remain constant while
$\Delta$ goes to zero. In either case, (\ref{eq:filtmap}) is asymptotically
equivalent to 
\begin{equation}
	\overline{\psi}(x) = \frac{f'(x)}{\Delta_\xi} \int
	G \left(\frac{f'(x)(x-y)}{\Delta_\xi} \right) \psi(y) dy  + \mathcal{O}({\epsilon})
	\label{eq:filtmap_limit}
	.
\end{equation}

Further, in the case of an inhomogeneous filter with slowly varying resolution,
a filtered quantity will vary over a long and short length scale, the scale of
filter variation and the scale of resolved turbulent fluctuations, respectively.
As such, we use (\ref{eq:filtmap_limit}) to facilitate a multiscale asymptotic
analysis of the commutator in terms of a slow variable $w = \epsilon x$ and fast
variable $\tilde{x}$. In this case, $\Delta$ depends on
$w$, but not $\tilde x$. Since $f'(x) = \Delta_\xi/\Delta(x)$, we have 

\begin{equation}
	\overline{\psi}(w,\tilde{x}) = \frac{1}{\Delta(w)} \int G\left(
	\frac{\tilde{x}-y}{\Delta(w)}\right) \psi(y) dy
	.
	\label{eq:1Dmulti_filter}
\end{equation}
In what follows, the dependence of $\Delta$ on $w$ is implied though
no longer explicitly indicated. Using multiscale asymptotics, the derivative
of $\overline{\psi}$ with respect to $x$ is written
\begin{equation}
	\frac{d \overline{\psi}}{d x} =
			\frac{\partial \overline{\psi}}{\partial \tilde{x}} + \epsilon
		\frac{\partial \overline{\psi}}{\partial w} +\mathcal{O}(\epsilon^2)
		.
\end{equation}
Since the filter is homogeneous in $\tilde{x}$, $\displaystyle{\frac{\partial
\overline{\psi}}{\partial \tilde{x}} = \overline{\frac{ d\psi}{dx}}}$.
Therefore, to leading order the commutator is given by
\begin{equation}
	\C^I(\psi) = \overline{\frac{ d \psi}{d x}} -\frac{d\overline{\psi}}{\partial x} =  - \epsilon \frac{\partial \overline{\psi}}{\partial w},
\end{equation}
which can be computed as
\begin{equation}
	  \C^I(\psi) =
		  \frac{1}{\Delta^2}\frac{d\Delta}{d
		  x}\int\left(\frac{\tilde{x}-y}{\Delta}G'\left(\frac{\tilde{x}-y}{\Delta}\right)
            +
		G\left(\frac{\tilde{x}-y}{\Delta}\right)\right)\psi(y)\,dy 
.
\label{eq:msCommErr}
\end{equation}
where $G'$ is the derivative of $G$ with respect to its argument.
Introducing the variable $\zeta = (y-\tilde{x})/\Delta$ and expanding $\psi(y)$
in a Taylor series about $\tilde{x}$ gives
\begin{equation}
	\begin{split}
	  \C^I(\psi)
          &=
		  \frac{1}{\Delta}\frac{d\Delta}{d
		  x}\int\left(\zeta G'\left(\zeta\right)
		  + G\left(\zeta\right)\right) \left( \sum_{n=0}^\infty
		  \frac{(\zeta \Delta)^n}{n!}\frac{d^n \psi}{dx^n}(\tilde{x}) \right)
		  d\zeta  \\
		  &= \sum_{n=1}^{\infty} \left( \frac{-\Delta^{2n-1}}{(2n-1)!}  \frac{d\Delta}{dx} 
		  \frac{d^{2n} \psi}{dx^{2n}}(\tilde{x})\int
	  \zeta^{2n} G(\zeta) d \zeta \right)
	\end{split}
	,
	\label{eq:1Dmulti_comm}
\end{equation}
where we have used the fact that odd order moments of $G$ are zero. 
To express the commutator in terms of the filtered field $\overline{\psi}$, we first invert 
(\ref{eq:1Dmulti_filter}) using the same procedure to obtain
\begin{equation}
	\begin{split}
	\psi(\tilde{x}) = \overline{\psi}(w,\tilde{x}) - \sum_{n=1}^{\infty}
	\frac{\Delta^{2n}}{(2n)!}
		  \frac{d^{2n} \psi}{dx^{2n}}(\tilde{x}) \int
		  \zeta^{2n} G(\zeta) d \zeta
		  .
	\end{split}
	\label{eq:1Dmulti_invert}
\end{equation}
Then we can recursively substitute (\ref{eq:1Dmulti_invert}) into
(\ref{eq:1Dmulti_comm}) to obtain an expression for the commutator
in terms of $\overline{\psi}$. However, to properly order this
expansion, the way in which derivatives of $\overline{\psi}$ and
$\Delta$ scale with $\epsilon$ must be considered. When
$\epsilon\rightarrow0$ at constant $\Delta$, both $\overline\psi$ and
$\Delta$ are order one in $\epsilon$. However, when
$\epsilon\rightarrow0$ at constant $L_\Delta$, $\Delta\sim\epsilon$
and, in general, the derivatives of $\overline\psi$ scale with powers
of $\epsilon$. In high Reynolds number turbulence that has been
filtered at scale $\Delta$ in the inertial range, the Kolmogorov scale
similarity hypotheses for the statistics of velocity differences
imply that the statistics of the derivatives
of the filtered velocity $\overline u$ depend only on $\Delta$ and the
rate of kinetic energy dissipation $\varepsilon_k$. Dimensional analysis
then requires that the standard deviation of
$\partial^n \overline{u}/\partial x^n$ scales
as $\varepsilon_k^{2/3}\Delta^{1/3-n}$.
Thus, taking $\psi$ to be $u$, the $n^{th}$ derivative of
$\overline u$ in the series expansion will scale as $\epsilon^{1/3-n}$.
Regardless of how the limit of small $\epsilon$ is approached, one obtains
\begin{equation}
	\begin{split}
    \C^I(u) &= - M_2 \Delta \frac{d \Delta}{d x}
	\frac{\partial^2 \overline{u} }{\partial \tilde{x}^2 } + \left( \frac{M_2^2}{2} -
    \frac{M_4}{6} \right) \Delta^3 \frac{d \Delta}{d
	x}\frac{\partial^4 \overline{u} }{\partial \tilde{x}^4}  +
    \dots + C_N \Delta^{N-1} \frac{d \Delta}{d
	x}\frac{\partial^N \overline{u}}{\partial \tilde{x}^N} + \dots
+ \mathcal{O}(\epsilon^q) \\
	&= \sum_{n=1}^{\infty} \left(
		C_{2n}\Delta^{2n-1} \frac{d\Delta}{dx}\frac{\partial^{2n} \overline{u}}{\partial
	\tilde{x}^{2n}} \right) + \mathcal{O}(\epsilon^q)
	\end{split}
	\label{eq:1Dmulti_ordering}
	,
\end{equation}
where $q=2$ when the asymptotic limit is taken with constant $\Delta$ (the
leading order series being order $\epsilon$) and 
$q=4/3$ when it is taken at constant $L_\Delta$ (the leading order series
being order $\epsilon^{1/3}$). 
Here we let $M_k$ denote the $k$th order moment of the filter kernel, $N$ is
even, and in general, the coefficient $C_j$ on the $j^{th}$ order term depends on the moments of the
filter up to order $j$. 

\subsection{Spectral Characteristics of the Commutator}
\label{sec:multiscale_spectral}

We turn our attention now to the spectral characteristics of the commutator.
However, to make a connection to the statistical properties of the commutator in
LES of turbulence, we consider instead a three-dimensional isotropic inhomogeneous
filter, defined similarly to (\ref{eq:filtmap_limit}) as
\begin{equation}
	\overline{\psi}(\bx) = \frac{1}{\Delta(\bx)^3}\int  G\left( \frac{|\bx -
		\by|}{\Delta(\bx)}
	\right) \psi(\by)  d\by
	,
\end{equation}
where $G$ is now a scalar function on $[0,\infty)$ satisfying $4\pi\int_0^\infty G(r)r^2\,dr=1$.
The same multiscale expansion holds as above for the case where 
$\frac{\partial \Delta}{\partial x_i} \sim \mathcal{O}(\epsilon)$. 
The filtering operation can be expressed as
\begin{equation}
	\overline{\psi}(\bw,\tilde{\bx}) = \frac{1}{\Delta(\bw)^3} \int G\left(
		\frac{|\tilde{\bx}-\by|}{\Delta(\bw)}
	\right) \psi(\by) d \by
	,
        \label{eq:multiscaleubar}
\end{equation}
where $\mathbf{w} = \epsilon \bx$ is the slow variable
and $\tilde{\bx}$ is the fast
variable, and the commutator, 
$\displaystyle{\C_i^I(\psi) = \overline{\frac{\partial \psi}{\partial x_i}} -\frac{\partial
\overline{\psi}}{\partial x_i} =  - \epsilon \frac{\partial
\overline{\psi}}{\partial w_i}}$, can be computed as 
\begin{equation}
	\begin{split}
	  \C_i^I(\psi)
          &=
          \frac{\epsilon}{\Delta^4}\frac{\partial\Delta}{\partial
              w_i}\int\left(\frac{|\tilde\bx-\by|}{\Delta}G'\left(\frac{|\tilde\bx-\by|}{\Delta}\right)
            +
            3G\left(\frac{|\tilde\bx-\by|}{\Delta}\right)\right)\psi(\by)\,d\by \\
	&\equiv \int \mathcal{C}_i(\bw,\tilde{\bx}-\by) \psi(\by) d\by 
	\end{split}
	,
        \label{eq:comm_int}
\end{equation}

Furthermore, because the filter is homogeneous in the fast variable, it is useful to
consider the Fourier transform of $\overline{\psi}$ in the fast variable: 
\begin{equation}
	\widehat{\overline{\psi}}(\bw,\bk) = \frac{1}{(2\pi)^3} \int
	\overline{\psi}(\bw,\tilde{\bx}) e^{-i \bk \cdot
	\tilde{\bx}}\,d\tilde{\bx}
	.
\end{equation}
Applying the convolution theorem to (\ref{eq:multiscaleubar}) yields
$\displaystyle{\widehat{\overline{\psi}}(\bw,\bk) = \widehat{\psi}(\bk)
  \widehat{G}(\Delta|\bk|)}$, where $\widehat \psi$ is the Fourier transform
of $\psi$ and $\widehat G(|\bk|) = \frac{1}{(2\pi)^3}\int
G(|\bz|)e^{-i\bk\cdot\bz}\,d\bz$ is the Fourier transform of the
filter kernel, which depends only on $|\bk|$ because $G(|\bz|)$ is
isotropic. Note that because the unfiltered quantity $\psi$ does not
depend on $\Delta$, it also does not depend on $\bw$.
The Fourier transform of the commutator is thus
given by
\begin{equation}
		\widehat{\C_i^I}(\psi) = -\epsilon\frac{\partial \widehat{\overline{\psi}}}{\partial w_i}(\bw,\bk) =
		-\widehat{\psi}(\bk)
		\frac{\partial \widehat G(\Delta|\bk|)}{\partial x_i} =
                -\widehat \psi(\bk)\widehat
                G'(\Delta|\bk|)|\bk|\frac{\partial \Delta}{\partial x_i}
			\equiv - \widehat{\mathcal{C}}_i(\bw,\bk)\widehat{\psi}(\bk)
	.
	\label{eq:comm_hat}
\end{equation}
where $\widehat G'$ is the derivative of $\widehat G$ with respect to
its argument.

While (\ref{eq:comm_int}) and (\ref{eq:comm_hat}) provide explicit
expressions for the commutator, they require knowledge of the
unfiltered field $\psi$ or its Fourier transform. If $G$ were invertible,
we could relate $\psi$ and $\overline{\psi}$ as in (\ref{eq:1Dmulti_invert}), but 
this is not the case for non-invertible filters, such as those that
include a finite dimensional  projection.
As such, this information is
generally not available in an LES, however, we may have theory or models
for the statistics of $\psi$, which could allow us to determine the
statistics of the commutator.

Consider, for example, homogeneous, isotropic turbulence flowing
through an inhomogeneous grid at a velocity $U_i$ that is much greater
than the fluctuations $u_i$ so that Taylor's frozen field hypothesis
holds, as in Sec.~\ref{sec:impact}. In this case, Kolmogorov theory
provides a model for the spectrum tensor $\phi_{ij}(\bk)=\int \langle
\widehat u_i(\bk')\widehat u^*_j(\bk)\rangle\, d\bk'$, $*$ denotes
complex conjugate, $\widehat u_i$ is the Fourier transform of the
velocity, and $\phi_{ij}$ is also the Fourier
transform of the two-point correlation tensor. The spectrum tensor of
the filtered velocity is given by $\overline{\phi}_{ij}(\bk)=\widehat
G^2(\Delta|\bk|)\phi_{ij}(\bk)$.
The commutator arising
from the convection term in the filtered evolution equation is
$U_k\C_k(u_j)$, and its contribution to the evolution of
$\overline{\phi}_{ij}$ is given by
\begin{equation}
  \tilde\C^I(\overline{\phi}_{ij})=
  U_k\int\langle \widehat{\overline{u}}_i(\bk')\widehat{\C_k^I}(u_j)^* \rangle
  \, d\bk'+
  U_k\int\langle \widehat{\overline{u}}^*_j(\bk)\widehat{\C_k^I}(u_i) \rangle \, d\bk'=
  -2U_k\frac{\partial \Delta}{\partial x_k} \widehat
  G(\Delta|\bk|)\widehat G'(\Delta|\bk|)|\bk|\phi_{ij}(\bk)
  \label{eq:comm_phi}
\end{equation}
where the $\tilde\C$ nomenclature indicates the contribution of the
commutator to the evolution equation for its argument.
The contribution of the commutator to the evolution of the
filtered three-dimensional energy spectrum
$\overline{E}(|\bk|)=2\pi|\bk|^2\overline{\phi}_{ii}(\bk)$ 
and resolved turbulent
kinetic energy $k^>=\int_0^\infty \overline{E}(\kappa)\, d\kappa$ can
easily be obtained from (\ref{eq:comm_phi}) as:
\begin{align}
	\tilde\C^I(\overline{E})& = -2U_k\frac{\partial
    \Delta}{\partial x_k}\widehat G(\Delta \kappa)\widehat G'(\Delta\kappa)\kappa E(\kappa)\\
	\tilde\C^I(k^>) & = -2U_k\frac{\partial\Delta}{\partial x_k} 
\int_0^\infty \widehat G(\Delta\kappa)\widehat G'(\Delta\kappa)\kappa
E(\kappa)\,d\kappa\label{eq:comm_k}
\end{align}
Note that unlike the analysis in Appendix~\ref{sec:multiscale_deconvolution}, the analysis outlined here does
not rely on deconvolution, and so is applicable to noninvertible
filters that include implicit truncation. For example, if $G$ is a
Fourier cutoff and $G'$ is interpreted in the sense of distributions, then
(\ref{eq:comm_k}) simplifies to
\begin{equation}
  \tilde\C^I(k^>)= U_k \frac{\partial \kappa_c}{\partial x_k}
  E(\kappa_c),
\end{equation}
where $\kappa_c$ is the cutoff wavenumber. Note that this multiscale analysis can also be
generalized to the case of spatially varying anisotropic
resolution. 

\section{Generalizing the Analysis of \citet{ghosal1995basic}}
\label{sec:app_ghosal}

\citet{ghosal1995basic} did not employ a multiscale asymptotic
analysis such as that in Appendix~\ref{sec:multiscale}, however, their
analysis can be interpreted asymptotically. In this appendix, we
explore the relationship between the series analysis of
Appendix~\ref{sec:multiscale_deconvolution} and that of Ghosal and
Moin, and extend the latter to characterize the asymptotically higher
order terms. 

Recall, the filtering of an arbitrary function $\psi(x)$ was defined in
(\ref{eq:filtmap}) as
\begin{equation}
	\overline{\psi}(x) = \frac{1}{\Delta_\xi} \int
	G \left(
	\frac{f(x)-f(y)}{\Delta_\xi} \right) \psi(y) f'(y) dy 
	\label{eq:filtmap1}
	.
\end{equation}
As in \citet{ghosal1995basic}, we work directly with
(\ref{eq:filtmap1}) and obtain
\begin{equation}
	\begin{split}
            \C^I(\psi) &= \int
			G(\zeta) \psi'(y) \left[
			1-\frac{f'(x)}{f'(y)}
		\right] d \zeta 
	\end{split}
	\label{eq:commerror}
\end{equation}
for the inhomogeneous part of the commutator, 
where we have introduced the variable $\zeta = (f(y)-f(x))/\Delta_\xi$. 

To expand (\ref{eq:commerror}) in a series of explicit powers of
$\Delta_\xi$, we follow \citep{ghosal1995basic} but consider the
general case including
terms up to $\Delta_\xi^N$ for some $N$. By inverting the definition of $\zeta$,
we can express $y$ as
\begin{equation}
    y = \sum_{i=0}^\infty \Delta_\xi^i \zeta^i y_i
    \label{eq:yseries}
    ,
\end{equation}
where
$y_0 = x$, $y_1 = 1/f'(x)$ and $y_i$ is given by
\begin{equation}
    y_i = - \sum_{n=2}^i 
	\frac{\beta_{n,{i-n}}}{n! f'(x)} \frac{d^nf}{dx^n}
    \label{eq:seriesy}
    ,
\end{equation}
where
\begin{equation}
    \beta_{n,0}= y_1^n, \qquad
    \beta_{n,m} = \frac{1}{m y_1} \sum_{k=1}^m (kn-m+k)y_{k+1}\beta_{n,m-k}
    . 
    \label{eq:betaformula}
\end{equation}
Then $(y-x)$ can be expressed as 
\begin{equation}
    (y-x)^n = 
    \left(\sum_{m=1}^\infty \Delta_\xi^m \zeta^m y_m \right)^n
    = \sum_{j=0}^{N-n} \Delta_\xi^{j+n} \zeta^{j+n}\beta_{n,j} + \dots
    \label{eq:ymxn}
    \;, 
\end{equation}
for $n>0$, which includes all terms with explicit powers of
$\Delta_\xi$ up to some power $N$.
Substitution of (\ref{eq:ymxn}) into a general Taylor series expansion of
$\psi(y)$ about $x$ gives:
\begin{equation} \psi(y) = \psi(x) + \sum_{n=1}^{N}\left( \frac{\psi^{(n)}(x)}{n!}
    \sum_{j=0}^{N-n} \Delta_\xi^{j+n} \zeta^{j+n} \beta_{n,j} \right) + \dots
    \label{eq:genphiy} \; .  
\end{equation} 
Equation (\ref{eq:genphiy}) can be
used to expand each term in (\ref{eq:commerror}) about $x$ so that all the terms
of the commutator with explicit powers of $\Delta_\xi$ up to some order $N$ is 
given by:
\begin{equation}
    \begin{split}
    \C^I(\psi) &=  
    \sum_{m=1}^N 
        \left[ \vphantom{\sum_{j=0}^{N-n}} \right.
        \frac{1}{m!}
        \left[ \psi' \left(\frac{1}{f'}\right)^{(m)} f'\right](x)
        \underset{k+m \in 2 \mathbb{Z}}{\sum_{k=0}^{N-m}} \Delta_\xi^{k+m}
		\beta_{m,k} \int \zeta^{k+m} G(\zeta)
    d\zeta 
\left. \vphantom{\sum_{j=0}^{N-n}} \right]\\
    &+ \sum_{n=1}^N \sum_{m=1}^N \left[ \vphantom{\sum_{j=0}^{N-n}} \right.
        \frac{1}{n!m!}
        \left[\psi^{(n+1)}\left(\frac{1}{f'}\right)^{(m)} f'\right](x)
        \underset{j+k+n+m \in 2\mathbb{Z}}{\underset{j+k+n+m\le
        N}{\sum_{j=0}^{N-n} \sum_{k=0}^{N-m}}}
        \Delta_\xi^{j+k+n+m} \beta_{n,j} \beta_{m,k} \int \zeta^{j+k+n+m} G(\zeta)
    d\zeta \left. \vphantom{\sum_{j=0}^{N-n}} \right]  
    \end{split}
    \label{eq:seriescommerror}
    .
\end{equation}
For example, for $N =2$ we obtain, 
\begin{equation}
	\C^I(\psi) = \left(\frac{f''}{f'^3}\frac{d^2 \psi}{dx^2} + \left(\frac{f'''}{2f'^3}
	- \frac{3f''^2}{2f'^4}\right)\frac{d \psi}{dx} \right)
	\Delta_\xi^2 \int
	\zeta^2 G(\zeta) d\zeta
    \label{eq:unfilteredcommerror}
	\; ,
\end{equation}
which agrees with equation (3.9) in \cite{ghosal1995basic}. 

To express the commutator in terms of $\overline{\psi}$, we follow the same
procedure as (\ref{eq:1Dmulti_invert}). 
Inverting (\ref{eq:filtmap1}) gives
\begin{equation}
    \psi(x) = \overline{\psi}(x) - \sum_{n=1}^N \left( \frac{\psi^{(n)}(x)}{n!}
    \underset{j+n \in 2\mathbb{Z}}{\sum_{j=0}^{N-n}} \Delta_\xi^{j+n} \beta_{n,j}
    \int \zeta^{j+n} G(\zeta) d\zeta \right) + \dots 
    \label{eq:filtunfilt}
     \; . 
\end{equation}
Equation (\ref{eq:filtunfilt}) can be recursively substituted into
(\ref{eq:seriescommerror}) to obtain the commutator in terms of the
filtered velocity field. Moreover, the commutator 
can be expressed in terms of the local grid spacing $\Delta(x)$ using the
relationship $f' = \Delta_\xi/\Delta$. The terms with explicit powers of
$\Delta_\xi$ up to $N=2$ are
\begin{equation}
    \C^I(\psi) = \left(- \left[ \frac{1}{2} \left( \Delta'^2 + \Delta
    \Delta'' \right) \right] \frac{d \pbar}{d x} - \left[ \Delta\Delta'
\right]\frac{d^2 \pbar}{dx^2} \right) \int \zeta^2 G(\zeta) d \zeta  .
    \label{eq:secondcommerror}
\end{equation}
For $N=4$ we obtain, 
\begin{equation}
    \begin{split}
        \C^I(\psi) &= -\left[ \frac{1}{2} \left( \Delta'^2 + \Delta
		\Delta'' \right) \right] \frac{d\pbar}{dx} \momG{2} - \left[ \Delta\Delta'
		\right]\frac{d^2 \pbar}{dx^2}\momG{2} \\
        &+ \left[ \frac{-\Delta'^4 -
            11\Delta\Delta'^2\Delta'' - 7 \Delta^2 \Delta'\Delta'''
	-4\Delta^2\Delta''^2 - \Delta^3\Delta''''}{24} \right]\frac{d\pbar}{dx}\momG{4} \\
        &+ \left[ \frac{\Delta'^4 + 8 \Delta \Delta'^2 \Delta'' + \Delta^2
		\Delta''^2 + 2 \Delta^2 \Delta' \Delta'''}{4} \right]
		\frac{d \pbar}{dx}
        \left(\momG{2}\right)^2
        \\
        &+ \left[ \frac{-7 \Delta \Delta'^3 - 13 \Delta^2 \Delta' \Delta'' - 2
		\Delta^3 \Delta'''}{12} \right]\frac{d^2 \overline{\psi}}{dx^2} \momG{4}\\ 
        &+\left[ \frac{11
		\Delta \Delta'^3 + 11 \Delta^2 \Delta' \Delta''}{4}
	\right]\frac{d^2 \overline{\psi}}{dx^2}
        \left(\momG{2}\right)^2  \\
        &+ \left[ \frac{-3 \Delta^2 \Delta'^2-\Delta^3 \Delta''}{4}
		\right]\frac{d^3 \pbar}{dx^3} \momG{4}
        + \left[ \frac{11 \Delta^2 \Delta'^2 +
		\Delta^3\Delta''}{4} \right]\frac{d^3 \pbar}{dx^3} \left(\momG{2}\right)^2\\
        &+ \left[
		\frac{-\Delta^3 \Delta'}{6} \right]\frac{d^4 \pbar}{dx^4} \momG{4} 
        + \left[
	\frac{\Delta^3 \Delta'}{2} \right]\frac{d^4 \pbar}{dx^4} \left(\momG{2}\right)^2 \end{split}
    .
    \label{eq:fourthcommerror}
\end{equation}

Unlike the analysis in Appendix~\ref{sec:multiscale_deconvolution}, no
ordering has been given to the commutation terms. They are simply expressed
in explicit powers of $\Delta_\xi$ to show the structure of the higher order
terms neglected in (\ref{eq:1Dmulti_ordering}). To get back to this result, 
take the limit $\Delta_\xi\to 0$ and recall that in high Reynolds number
turbulence it makes sense to consider the scaling $d^n \overline{\psi}/dx^n
\sim \Delta_\xi^{1/3-n}$. In this case one obtains
\begin{equation}
	\begin{split}
	\C^I(\psi) &= \sum_{n=1}^{\infty} \left(
		C_{2n}\frac{f''}{f'^{2n+1}}\Delta_\xi^{2n}\frac{\partial^{2n}
		\overline{\psi}}{\partial
x^{2n}} \right) + \mathcal{O}(\Delta_\xi^{4/3})
 =  \sum_{n=1}^{\infty} \left(
		C_{2n}\Delta^{2n-1} \frac{d\Delta}{dx}\frac{\partial^{2n} \overline{\psi}}{\partial
x^{2n}} \right) + \mathcal{O}(\Delta_\xi^{4/3})
\end{split}
\label{eq:ordering1}
,
\end{equation}
which is the same as (\ref{eq:1Dmulti_ordering}). Each term in the sum in (\ref{eq:ordering1})
is of order $\Delta_\xi^{1/3}$, and is proportional to $d\Delta/dx$ and an
even derivative of $\overline{\psi}$. However, the asymptotically
higher order terms (order $\Delta_\xi^{4/3}$ and higher), such as those in (\ref{eq:secondcommerror}) and
(\ref{eq:fourthcommerror}), include higher order derivatives of
$\Delta$, higher powers of $d\Delta/dx$ and odd-order derivatives of
$\overline \psi$. 

To arrive at (5.8) and (5.9) in \citet{ghosal1995basic}, which are the analog of
(\ref{eq:ordering1}), the authors consider
$\Delta_\xi\ll1$, which we interpret in the sense of an asymptotic
analysis for $\Delta_\xi\rightarrow0$. They also introduce the ansatz
$\psi=\exp(i\kappa x)$, along with the assumption that
$\kappa\Delta_\xi\gg\Delta_\xi$, which while dimensionally
inconsistent, arose from the assertion that $\kappa\Delta_\xi$
could be as large as order one. In the context of the current
analysis, this implies a scaling for the derivatives of
$\overline\psi$. Equations (5.8) and (5.9) in \citet{ghosal1995basic}
include only the first term in (\ref{eq:ordering1}) because the
remaining terms would be higher order in $\kappa\Delta_\xi$. The authors do,
however, point out that the series can be extended to higher order in
$\kappa\Delta_\xi$, which would then include more of the terms in
(\ref{eq:ordering1}). We interpret these arguments from
\citep{ghosal1995basic} to be asymptotic for
$\kappa\Delta_\xi\rightarrow 0$, while $\kappa\rightarrow\infty$,
which would be consistent with $\kappa\sim\Delta_\xi^{-p}$ for
$0<p<1$. However, the introduction of the $\psi=\exp(i\kappa x)$
ansatz is essentially ad hoc, and is inconsistent with the scaling of
the derivatives of the filtered velocity for high Reynolds number
turbulence, as described in Appendix~\ref{sec:multiscale_deconvolution}.

\section{Commutation Model Coefficient}
\label{sec:A2}

Here the coefficient for the commutation model developed in
Sec.~\ref{sec:modelform} is evaluated. In LES, the coefficient in a
model of the commutator can in general be
calibrated to match known statistical characteristics \emph{a priori}, based on
the analysis in Sec~\ref{sec:statistics}.
However, for the case of linear convection, 
it is useful to examine changes in the behavior of the model as a function of
the coefficient.
Let $\varepsilon \in (0,1)$
be the maximum allowed fraction of energy at any wavenumber to be
reflected due to resolution variation. 
Now, consider the action of the commutation model defined in
(\ref{eq:largekcomm1}) on the Fourier coefficient $\widehat u(\kappa,t)$, which
is given by
\begin{equation}
    \frac{\partial \hat{u}(\kappa,t)}{\partial t} = (-1)^{\frac{N-2}{2}}C U \frac{\partial \Delta}{\partial x}
    \Delta^{\N-1} {\hat{F}}_\N(\kappa) \hat{u}(\kappa,t)
    \label{eq:ode}
     , 
\end{equation}
where ${\hat{F}}_\N(\kappa)$ is the spectrum of $F_\N$ evaluated at
wavenumber $\kappa$.
After a time $t$, the amplification of $\hat u(\kappa)$ is:
\begin{equation}
    \frac{\hat{u}(\kappa,t)}{\hat{u}(\kappa,0)}= \exp\left({(-1)^{\frac{N-2}{2}}C U \frac{\partial \Delta}{\partial
    x} \Delta^{\N-1} \hat{F}_{\N}(\kappa)t} \right)
     . 
    \label{eq:ampfactork}
\end{equation}
As the resolved turbulence convects through a coarsening grid, we
insist that
$\hat{u}^2(\kappa,t)/\hat{u}^2(\kappa,0) \le \varepsilon$ for
all reflected wavenumbers. This requires that $C$ satisfy
\begin{equation}
    C \ge \frac{ (-1)^{\frac{N-2}{2}}\log(\varepsilon)}{2U \frac{\partial \Delta}{\partial x}
    \Delta^{\N-1} \hat{F}_\N(\kappa) t}
     , 
    \label{eq:C1}
\end{equation}
for all reflected wavenumbers.
If we assume for simplicity that $d\Delta/dx \approx (\max(\Delta) - \min(\Delta))/L$
and that $t = L/U$, for some length of gradual coarsening $L$,
equation~(\ref{eq:C1}) simplifies to
\begin{equation}
    C \ge \frac{(-1)^{\frac{N-2}{2}}\log(\varepsilon)}{2\left( \frac{\max(\Delta) -
    \min(\Delta)}{\Delta} \right)
\left(\Delta^\N \hat{F}_\N(\kappa) \right) }
    \label{eq:C2}
    . 
\end{equation}
Notice that the lower the wavenumber
with positive group velocity, the higher the wavenumber of the reflection with
negative group velocity. Accordingly, the smallest wavenumber with nonpositive
group velocity is dissipated the least by the model.
Therefore, evaluating $\hat{F}_\N$ at $\kappa_a$ associated with the
numerical first derivative operator $B_1$, as defined in
Sec.~\ref{sec:numerics} will ensure (\ref{eq:C2}) is satisfied for all
reflected wavenumbers. Furthermore, because for any numerical
approximations, $\kappa_a\sim 1/\Delta$
and $\hat F_N(\kappa\Delta)\sim 1/\Delta^N$, $\Delta^N\hat
F_N(\kappa_a)$ depends only on the numerical schemes, and is
independent of $\Delta$. Finally, by replacing the remaining $\Delta$
with $\max(\Delta)$ in (\ref{eq:C2}) when evaluating $C$, we ensure
that the inequality is satisfied, and  obtain an expression that
depend only on the numerical schemes involved and the extreme values of
$\Delta$:
\begin{equation}
    C = \frac{(-1)^{\frac{N-2}{2}}\log(\varepsilon)}{ 2\left( \frac{ (\max(\Delta) -
    \min(\Delta)}{\max(\Delta)} \right) \left( \Delta^\N \hat{F}_\N(\kappa_{a}) \right)} 
    .
    \label{eq:C3}
\end{equation}
Note that when using the $B_2-B_1B_1$ model, one can simply substitute
$(-1)^{\frac{N-2}{2}}\Delta^2(\hat B_2-\hat B_1\hat B_1)$ for $\Delta^NF_N$ in
(\ref{eq:C3}) to obtain the coefficient in (\ref{eq:largekcomm2}).
\begin{acknowledgments}
The authors acknowledge the generous financial support from the National
Aeronautics and Space Administration (cooperative agreement number NNX15AU40A),
the National Science Foundation (project number 1904826), and the U.S.
Department of Energy, Exascale Computing Project (subcontract number
XFC-7-70022-01 from contract number DE-AC36-08GO28308 with the National
Renewable Energy Laboratory). Thanks are also due to the Texas Advanced
Computing Center at The University of Texas at Austin for providing HPC
resources that have contributed to the research results reported here.
\end{acknowledgments}

\bibliography{turbulence,grid,commutation}

\begin{thebibliography}{47}%
\makeatletter
\providecommand \@ifxundefined [1]{%
 \@ifx{#1\undefined}
}%
\providecommand \@ifnum [1]{%
 \ifnum #1\expandafter \@firstoftwo
 \else \expandafter \@secondoftwo
 \fi
}%
\providecommand \@ifx [1]{%
 \ifx #1\expandafter \@firstoftwo
 \else \expandafter \@secondoftwo
 \fi
}%
\providecommand \natexlab [1]{#1}%
\providecommand \enquote  [1]{``#1''}%
\providecommand \bibnamefont  [1]{#1}%
\providecommand \bibfnamefont [1]{#1}%
\providecommand \citenamefont [1]{#1}%
\providecommand \href@noop [0]{\@secondoftwo}%
\providecommand \href [0]{\begingroup \@sanitize@url \@href}%
\providecommand \@href[1]{\@@startlink{#1}\@@href}%
\providecommand \@@href[1]{\endgroup#1\@@endlink}%
\providecommand \@sanitize@url [0]{\catcode `\\12\catcode `\$12\catcode
  `\&12\catcode `\#12\catcode `\^12\catcode `\_12\catcode `\%12\relax}%
\providecommand \@@startlink[1]{}%
\providecommand \@@endlink[0]{}%
\providecommand \url  [0]{\begingroup\@sanitize@url \@url }%
\providecommand \@url [1]{\endgroup\@href {#1}{\urlprefix }}%
\providecommand \urlprefix  [0]{URL }%
\providecommand \Eprint [0]{\href }%
\providecommand \doibase [0]{https://doi.org/}%
\providecommand \selectlanguage [0]{\@gobble}%
\providecommand \bibinfo  [0]{\@secondoftwo}%
\providecommand \bibfield  [0]{\@secondoftwo}%
\providecommand \translation [1]{[#1]}%
\providecommand \BibitemOpen [0]{}%
\providecommand \bibitemStop [0]{}%
\providecommand \bibitemNoStop [0]{.\EOS\space}%
\providecommand \EOS [0]{\spacefactor3000\relax}%
\providecommand \BibitemShut  [1]{\csname bibitem#1\endcsname}%
\let\auto@bib@innerbib\@empty
%</preamble>
\bibitem [{\citenamefont {Ghosal}\ and\ \citenamefont
  {Moin}(1995)}]{ghosal1995basic}%
  \BibitemOpen
  \bibfield  {author} {\bibinfo {author} {\bibfnamefont {S.}~\bibnamefont
  {Ghosal}}\ and\ \bibinfo {author} {\bibfnamefont {P.}~\bibnamefont {Moin}},\
  }\bibfield  {title} {\bibinfo {title} {The basic equations for the large eddy
  simulation of turbulent flows in complex geometry},\ }\href@noop {}
  {\bibfield  {journal} {\bibinfo  {journal} {Journal of Computational
  physics}\ }\textbf {\bibinfo {volume} {118}},\ \bibinfo {pages} {24}
  (\bibinfo {year} {1995})}\BibitemShut {NoStop}%
\bibitem [{\citenamefont {van~der Ven}(1995)}]{van1995family}%
  \BibitemOpen
  \bibfield  {author} {\bibinfo {author} {\bibfnamefont {H.}~\bibnamefont
  {van~der Ven}},\ }\bibfield  {title} {\bibinfo {title} {A family of large
  eddy simulation (les) filters with nonuniform filter widths},\ }\href@noop {}
  {\bibfield  {journal} {\bibinfo  {journal} {Physics of Fluids}\ }\textbf
  {\bibinfo {volume} {7}},\ \bibinfo {pages} {1171} (\bibinfo {year}
  {1995})}\BibitemShut {NoStop}%
\bibitem [{\citenamefont {Vasilyev}\ \emph {et~al.}(1998)\citenamefont
  {Vasilyev}, \citenamefont {Lund},\ and\ \citenamefont
  {Moin}}]{vasilyev1998general}%
  \BibitemOpen
  \bibfield  {author} {\bibinfo {author} {\bibfnamefont {O.~V.}\ \bibnamefont
  {Vasilyev}}, \bibinfo {author} {\bibfnamefont {T.~S.}\ \bibnamefont {Lund}},\
  and\ \bibinfo {author} {\bibfnamefont {P.}~\bibnamefont {Moin}},\ }\bibfield
  {title} {\bibinfo {title} {A general class of commutative filters for les in
  complex geometries},\ }\href@noop {} {\bibfield  {journal} {\bibinfo
  {journal} {Journal of Computational Physics}\ }\textbf {\bibinfo {volume}
  {146}},\ \bibinfo {pages} {82} (\bibinfo {year} {1998})}\BibitemShut
  {NoStop}%
\bibitem [{\citenamefont {Marsden}\ \emph {et~al.}(2002)\citenamefont
  {Marsden}, \citenamefont {Vasilyev},\ and\ \citenamefont
  {Moin}}]{marsden2002construction}%
  \BibitemOpen
  \bibfield  {author} {\bibinfo {author} {\bibfnamefont {A.~L.}\ \bibnamefont
  {Marsden}}, \bibinfo {author} {\bibfnamefont {O.~V.}\ \bibnamefont
  {Vasilyev}},\ and\ \bibinfo {author} {\bibfnamefont {P.}~\bibnamefont
  {Moin}},\ }\bibfield  {title} {\bibinfo {title} {Construction of commutative
  filters for les on unstructured meshes},\ }\href@noop {} {\bibfield
  {journal} {\bibinfo  {journal} {Journal of Computational Physics}\ }\textbf
  {\bibinfo {volume} {175}},\ \bibinfo {pages} {584} (\bibinfo {year}
  {2002})}\BibitemShut {NoStop}%
\bibitem [{\citenamefont {Haselbacher}\ and\ \citenamefont
  {Vasilyev}(2003)}]{haselbacher2003commutative}%
  \BibitemOpen
  \bibfield  {author} {\bibinfo {author} {\bibfnamefont {A.}~\bibnamefont
  {Haselbacher}}\ and\ \bibinfo {author} {\bibfnamefont {O.~V.}\ \bibnamefont
  {Vasilyev}},\ }\bibfield  {title} {\bibinfo {title} {Commutative discrete
  filtering on unstructured grids based on least-squares techniques},\
  }\href@noop {} {\bibfield  {journal} {\bibinfo  {journal} {Journal of
  Computational Physics}\ }\textbf {\bibinfo {volume} {187}},\ \bibinfo {pages}
  {197} (\bibinfo {year} {2003})}\BibitemShut {NoStop}%
\bibitem [{\citenamefont {Iovieno}\ and\ \citenamefont
  {Tordella}(2003)}]{iovieno2003variable}%
  \BibitemOpen
  \bibfield  {author} {\bibinfo {author} {\bibfnamefont {M.}~\bibnamefont
  {Iovieno}}\ and\ \bibinfo {author} {\bibfnamefont {D.}~\bibnamefont
  {Tordella}},\ }\bibfield  {title} {\bibinfo {title} {Variable scale filtered
  navier--stokes equations: a new procedure to deal with the associated
  commutation error},\ }\href@noop {} {\bibfield  {journal} {\bibinfo
  {journal} {Physics of Fluids}\ }\textbf {\bibinfo {volume} {15}},\ \bibinfo
  {pages} {1926} (\bibinfo {year} {2003})}\BibitemShut {NoStop}%
\bibitem [{\citenamefont {Sagaut}(2006)}]{sagaut2006large}%
  \BibitemOpen
  \bibfield  {author} {\bibinfo {author} {\bibfnamefont {P.}~\bibnamefont
  {Sagaut}},\ }\href@noop {} {\emph {\bibinfo {title} {Large eddy simulation
  for incompressible flows: an introduction}}}\ (\bibinfo  {publisher}
  {Springer Science \& Business Media},\ \bibinfo {year} {2006})\BibitemShut
  {NoStop}%
\bibitem [{\citenamefont {Meneveau}\ and\ \citenamefont
  {Katz}(2000)}]{meneveau2000scale}%
  \BibitemOpen
  \bibfield  {author} {\bibinfo {author} {\bibfnamefont {C.}~\bibnamefont
  {Meneveau}}\ and\ \bibinfo {author} {\bibfnamefont {J.}~\bibnamefont
  {Katz}},\ }\bibfield  {title} {\bibinfo {title} {Scale-invariance and
  turbulence models for large-eddy simulation},\ }\href@noop {} {\bibfield
  {journal} {\bibinfo  {journal} {Annual Review of Fluid Mechanics}\ }\textbf
  {\bibinfo {volume} {32}},\ \bibinfo {pages} {1} (\bibinfo {year}
  {2000})}\BibitemShut {NoStop}%
\bibitem [{\citenamefont {Girimaji}\ and\ \citenamefont
  {Wallin}(2013)}]{girimaji2013closure}%
  \BibitemOpen
  \bibfield  {author} {\bibinfo {author} {\bibfnamefont {S.~S.}\ \bibnamefont
  {Girimaji}}\ and\ \bibinfo {author} {\bibfnamefont {S.}~\bibnamefont
  {Wallin}},\ }\bibfield  {title} {\bibinfo {title} {Closure modeling in
  bridging regions of variable-resolution (vr) turbulence computations},\
  }\href@noop {} {\bibfield  {journal} {\bibinfo  {journal} {Journal of
  Turbulence}\ }\textbf {\bibinfo {volume} {14}},\ \bibinfo {pages} {72}
  (\bibinfo {year} {2013})}\BibitemShut {NoStop}%
\bibitem [{\citenamefont {Haering}(2015)}]{haering2015anisotropic}%
  \BibitemOpen
  \bibfield  {author} {\bibinfo {author} {\bibfnamefont {S.~W.}\ \bibnamefont
  {Haering}},\ }\emph {\bibinfo {title} {Anisotropic Hybrid Turbulence Modeling
  with Specific Application to the Simulation of Pulse-Actuated Dynamic Stall
  Control}},\ \href@noop {} {Ph.D. thesis},\ \bibinfo  {school} {The University
  of Texas at Austin} (\bibinfo {year} {2015})\BibitemShut {NoStop}%
\bibitem [{\citenamefont {Fureby}\ and\ \citenamefont
  {Tabor}(1997)}]{fureby1997mathematical}%
  \BibitemOpen
  \bibfield  {author} {\bibinfo {author} {\bibfnamefont {C.}~\bibnamefont
  {Fureby}}\ and\ \bibinfo {author} {\bibfnamefont {G.}~\bibnamefont {Tabor}},\
  }\bibfield  {title} {\bibinfo {title} {Mathematical and physical constraints
  on large-eddy simulations},\ }\href@noop {} {\bibfield  {journal} {\bibinfo
  {journal} {Theoretical and Computational Fluid Dynamics}\ }\textbf {\bibinfo
  {volume} {9}},\ \bibinfo {pages} {85} (\bibinfo {year} {1997})}\BibitemShut
  {NoStop}%
\bibitem [{\citenamefont {Hamba}(2011)}]{hamba2011analysis}%
  \BibitemOpen
  \bibfield  {author} {\bibinfo {author} {\bibfnamefont {F.}~\bibnamefont
  {Hamba}},\ }\bibfield  {title} {\bibinfo {title} {Analysis of filtered
  navier--stokes equation for hybrid rans/les simulation},\ }\href@noop {}
  {\bibfield  {journal} {\bibinfo  {journal} {Physics of Fluids}\ }\textbf
  {\bibinfo {volume} {23}},\ \bibinfo {pages} {015108} (\bibinfo {year}
  {2011})}\BibitemShut {NoStop}%
\bibitem [{\citenamefont {Germano}(1986)}]{germano1986differential}%
  \BibitemOpen
  \bibfield  {author} {\bibinfo {author} {\bibfnamefont {M.}~\bibnamefont
  {Germano}},\ }\bibfield  {title} {\bibinfo {title} {Differential filters for
  the large eddy numerical simulation of turbulent flows},\ }\href@noop {}
  {\bibfield  {journal} {\bibinfo  {journal} {The Physics of Fluids}\ }\textbf
  {\bibinfo {volume} {29}},\ \bibinfo {pages} {1755} (\bibinfo {year}
  {1986})}\BibitemShut {NoStop}%
\bibitem [{\citenamefont {Lund}(2003)}]{lund2003use}%
  \BibitemOpen
  \bibfield  {author} {\bibinfo {author} {\bibfnamefont {T.}~\bibnamefont
  {Lund}},\ }\bibfield  {title} {\bibinfo {title} {The use of explicit filters
  in large eddy simulation},\ }\href@noop {} {\bibfield  {journal} {\bibinfo
  {journal} {Computers \& Mathematics with Applications}\ }\textbf {\bibinfo
  {volume} {46}},\ \bibinfo {pages} {603} (\bibinfo {year} {2003})}\BibitemShut
  {NoStop}%
\bibitem [{\citenamefont {Ghosal}(1996)}]{ghosal1996analysis}%
  \BibitemOpen
  \bibfield  {author} {\bibinfo {author} {\bibfnamefont {S.}~\bibnamefont
  {Ghosal}},\ }\bibfield  {title} {\bibinfo {title} {An analysis of numerical
  errors in large-eddy simulations of turbulence},\ }\href@noop {} {\bibfield
  {journal} {\bibinfo  {journal} {Journal of Computational Physics}\ }\textbf
  {\bibinfo {volume} {125}},\ \bibinfo {pages} {187} (\bibinfo {year}
  {1996})}\BibitemShut {NoStop}%
\bibitem [{\citenamefont {Chow}\ and\ \citenamefont
  {Moin}(2003)}]{chow2003further}%
  \BibitemOpen
  \bibfield  {author} {\bibinfo {author} {\bibfnamefont {F.~K.}\ \bibnamefont
  {Chow}}\ and\ \bibinfo {author} {\bibfnamefont {P.}~\bibnamefont {Moin}},\
  }\bibfield  {title} {\bibinfo {title} {A further study of numerical errors in
  large-eddy simulations},\ }\href@noop {} {\bibfield  {journal} {\bibinfo
  {journal} {Journal of Computational Physics}\ }\textbf {\bibinfo {volume}
  {184}},\ \bibinfo {pages} {366} (\bibinfo {year} {2003})}\BibitemShut
  {NoStop}%
\bibitem [{\citenamefont {Kravchenko}\ and\ \citenamefont
  {Moin}(1997)}]{kravchenko1997effect}%
  \BibitemOpen
  \bibfield  {author} {\bibinfo {author} {\bibfnamefont {A.}~\bibnamefont
  {Kravchenko}}\ and\ \bibinfo {author} {\bibfnamefont {P.}~\bibnamefont
  {Moin}},\ }\bibfield  {title} {\bibinfo {title} {On the effect of numerical
  errors in large eddy simulations of turbulent flows},\ }\href@noop {}
  {\bibfield  {journal} {\bibinfo  {journal} {Journal of Computational
  Physics}\ }\textbf {\bibinfo {volume} {131}},\ \bibinfo {pages} {310}
  (\bibinfo {year} {1997})}\BibitemShut {NoStop}%
\bibitem [{\citenamefont {Carati}\ \emph {et~al.}(2001)\citenamefont {Carati},
  \citenamefont {Winckelmans},\ and\ \citenamefont
  {Jeanmart}}]{carati2001modelling}%
  \BibitemOpen
  \bibfield  {author} {\bibinfo {author} {\bibfnamefont {D.}~\bibnamefont
  {Carati}}, \bibinfo {author} {\bibfnamefont {G.~S.}\ \bibnamefont
  {Winckelmans}},\ and\ \bibinfo {author} {\bibfnamefont {H.}~\bibnamefont
  {Jeanmart}},\ }\bibfield  {title} {\bibinfo {title} {On the modelling of the
  subgrid-scale and filtered-scale stress tensors in large-eddy simulation},\
  }\href@noop {} {\bibfield  {journal} {\bibinfo  {journal} {Journal of Fluid
  Mechanics}\ }\textbf {\bibinfo {volume} {441}},\ \bibinfo {pages} {119}
  (\bibinfo {year} {2001})}\BibitemShut {NoStop}%
\bibitem [{\citenamefont {Gullbrand}\ and\ \citenamefont
  {Chow}(2003)}]{gullbrand2003effect}%
  \BibitemOpen
  \bibfield  {author} {\bibinfo {author} {\bibfnamefont {J.}~\bibnamefont
  {Gullbrand}}\ and\ \bibinfo {author} {\bibfnamefont {F.~K.}\ \bibnamefont
  {Chow}},\ }\bibfield  {title} {\bibinfo {title} {The effect of numerical
  errors and turbulence models in large-eddy simulations of channel flow, with
  and without explicit filtering},\ }\href@noop {} {\bibfield  {journal}
  {\bibinfo  {journal} {Journal of Fluid Mechanics}\ }\textbf {\bibinfo
  {volume} {495}},\ \bibinfo {pages} {323} (\bibinfo {year}
  {2003})}\BibitemShut {NoStop}%
\bibitem [{\citenamefont {Bose}\ \emph {et~al.}(2011)\citenamefont {Bose},
  \citenamefont {Moin},\ and\ \citenamefont {Ham}}]{bose2011explicitly}%
  \BibitemOpen
  \bibfield  {author} {\bibinfo {author} {\bibfnamefont {S.}~\bibnamefont
  {Bose}}, \bibinfo {author} {\bibfnamefont {P.}~\bibnamefont {Moin}},\ and\
  \bibinfo {author} {\bibfnamefont {F.}~\bibnamefont {Ham}},\ }\bibfield
  {title} {\bibinfo {title} {Explicitly filtered large eddy simulation on
  unstructured grids},\ }\href@noop {} {\bibfield  {journal} {\bibinfo
  {journal} {Annual Research Briefs, Center for Turbulence Research}\ ,\
  \bibinfo {pages} {87}} (\bibinfo {year} {2011})}\BibitemShut {NoStop}%
\bibitem [{\citenamefont {Bose}\ \emph {et~al.}(2010)\citenamefont {Bose},
  \citenamefont {Moin},\ and\ \citenamefont {You}}]{bose2010grid}%
  \BibitemOpen
  \bibfield  {author} {\bibinfo {author} {\bibfnamefont {S.~T.}\ \bibnamefont
  {Bose}}, \bibinfo {author} {\bibfnamefont {P.}~\bibnamefont {Moin}},\ and\
  \bibinfo {author} {\bibfnamefont {D.}~\bibnamefont {You}},\ }\bibfield
  {title} {\bibinfo {title} {Grid-independent large-eddy simulation using
  explicit filtering},\ }\href@noop {} {\bibfield  {journal} {\bibinfo
  {journal} {Physics of Fluids}\ }\textbf {\bibinfo {volume} {22}},\ \bibinfo
  {pages} {105103} (\bibinfo {year} {2010})}\BibitemShut {NoStop}%
\bibitem [{\citenamefont {Langford}\ and\ \citenamefont
  {Moser}(1999{\natexlab{a}})}]{Langford1999}%
  \BibitemOpen
  \bibfield  {author} {\bibinfo {author} {\bibfnamefont {J.~A.}\ \bibnamefont
  {Langford}}\ and\ \bibinfo {author} {\bibfnamefont {R.~D.}\ \bibnamefont
  {Moser}},\ }\bibfield  {title} {\bibinfo {title} {Optimal {LES} formulations
  for isotropic turbulence},\ }\href@noop {} {\bibfield  {journal} {\bibinfo
  {journal} {Journal of Fluid Mechanics}\ }\textbf {\bibinfo {volume} {398}},\
  \bibinfo {pages} {321} (\bibinfo {year} {1999}{\natexlab{a}})}\BibitemShut
  {NoStop}%
\bibitem [{\citenamefont {Hughes}\ \emph {et~al.}(2000)\citenamefont {Hughes},
  \citenamefont {Mazzei},\ and\ \citenamefont {Jansen}}]{hughes2000large}%
  \BibitemOpen
  \bibfield  {author} {\bibinfo {author} {\bibfnamefont {T.~J.}\ \bibnamefont
  {Hughes}}, \bibinfo {author} {\bibfnamefont {L.}~\bibnamefont {Mazzei}},\
  and\ \bibinfo {author} {\bibfnamefont {K.~E.}\ \bibnamefont {Jansen}},\
  }\bibfield  {title} {\bibinfo {title} {Large eddy simulation and the
  variational multiscale method},\ }\href@noop {} {\bibfield  {journal}
  {\bibinfo  {journal} {Computing and Visualization in Science}\ }\textbf
  {\bibinfo {volume} {3}},\ \bibinfo {pages} {47} (\bibinfo {year}
  {2000})}\BibitemShut {NoStop}%
\bibitem [{\citenamefont {Moser}\ \emph {et~al.}(2021)\citenamefont {Moser},
  \citenamefont {Haering},\ and\ \citenamefont {Yalla}}]{moser2020statistical}%
  \BibitemOpen
  \bibfield  {author} {\bibinfo {author} {\bibfnamefont {R.~D.}\ \bibnamefont
  {Moser}}, \bibinfo {author} {\bibfnamefont {S.~W.}\ \bibnamefont {Haering}},\
  and\ \bibinfo {author} {\bibfnamefont {G.~R.}\ \bibnamefont {Yalla}},\
  }\bibfield  {title} {\bibinfo {title} {Statistical properties of
  subgrid-scale turbulence models},\ }\href@noop {} {\bibfield  {journal}
  {\bibinfo  {journal} {Annual Review of Fluid Mechanics}\ }\textbf {\bibinfo
  {volume} {53}} (\bibinfo {year} {2021})}\BibitemShut {NoStop}%
\bibitem [{\citenamefont {Langford}\ and\ \citenamefont
  {Moser}(1999{\natexlab{b}})}]{langford1999optimal}%
  \BibitemOpen
  \bibfield  {author} {\bibinfo {author} {\bibfnamefont {J.~A.}\ \bibnamefont
  {Langford}}\ and\ \bibinfo {author} {\bibfnamefont {R.~D.}\ \bibnamefont
  {Moser}},\ }\bibfield  {title} {\bibinfo {title} {Optimal les formulations
  for isotropic turbulence},\ }\href@noop {} {\bibfield  {journal} {\bibinfo
  {journal} {Journal of Fluid Mechanics}\ }\textbf {\bibinfo {volume} {398}},\
  \bibinfo {pages} {321} (\bibinfo {year} {1999}{\natexlab{b}})}\BibitemShut
  {NoStop}%
\bibitem [{\citenamefont {Vasilyev}\ and\ \citenamefont
  {Goldstein}(2004)}]{vasilyev2004local}%
  \BibitemOpen
  \bibfield  {author} {\bibinfo {author} {\bibfnamefont {O.~V.}\ \bibnamefont
  {Vasilyev}}\ and\ \bibinfo {author} {\bibfnamefont {D.~E.}\ \bibnamefont
  {Goldstein}},\ }\bibfield  {title} {\bibinfo {title} {Local spectrum of
  commutation error in large eddy simulations},\ }\href@noop {} {\bibfield
  {journal} {\bibinfo  {journal} {Physics of Fluids}\ }\textbf {\bibinfo
  {volume} {16}},\ \bibinfo {pages} {470} (\bibinfo {year} {2004})}\BibitemShut
  {NoStop}%
\bibitem [{\citenamefont {Meneveau}(1994)}]{meneveau1994statistics}%
  \BibitemOpen
  \bibfield  {author} {\bibinfo {author} {\bibfnamefont {C.}~\bibnamefont
  {Meneveau}},\ }\bibfield  {title} {\bibinfo {title} {Statistics of turbulence
  subgrid-scale stresses: Necessary conditions and experimental tests},\
  }\href@noop {} {\bibfield  {journal} {\bibinfo  {journal} {Physics of
  Fluids}\ }\textbf {\bibinfo {volume} {6}},\ \bibinfo {pages} {815} (\bibinfo
  {year} {1994})}\BibitemShut {NoStop}%
\bibitem [{\citenamefont {Trefethen}(1982)}]{trefethen1982group}%
  \BibitemOpen
  \bibfield  {author} {\bibinfo {author} {\bibfnamefont {L.~N.}\ \bibnamefont
  {Trefethen}},\ }\bibfield  {title} {\bibinfo {title} {Group velocity in
  finite difference schemes},\ }\href@noop {} {\bibfield  {journal} {\bibinfo
  {journal} {SIAM review}\ }\textbf {\bibinfo {volume} {24}},\ \bibinfo {pages}
  {113} (\bibinfo {year} {1982})}\BibitemShut {NoStop}%
\bibitem [{\citenamefont
  {Vichnevetsky}(1981{\natexlab{a}})}]{vichnevetsky1981energy}%
  \BibitemOpen
  \bibfield  {author} {\bibinfo {author} {\bibfnamefont {R.}~\bibnamefont
  {Vichnevetsky}},\ }\bibfield  {title} {\bibinfo {title} {Energy and group
  velocity in semi discretizations of hyperbolic equations},\ }\href@noop {}
  {\bibfield  {journal} {\bibinfo  {journal} {Mathematics and Computers in
  Simulation}\ }\textbf {\bibinfo {volume} {23}},\ \bibinfo {pages} {333}
  (\bibinfo {year} {1981}{\natexlab{a}})}\BibitemShut {NoStop}%
\bibitem [{\citenamefont {Vichnevetsky}(1983)}]{vichnevetsky1983group}%
  \BibitemOpen
  \bibfield  {author} {\bibinfo {author} {\bibfnamefont {R.}~\bibnamefont
  {Vichnevetsky}},\ }\bibfield  {title} {\bibinfo {title} {Group velocity and
  reflection phenomena in numerical approximations of hyperbolic equations},\
  }\href@noop {} {\bibfield  {journal} {\bibinfo  {journal} {Journal of the
  Franklin Institute}\ }\textbf {\bibinfo {volume} {315}},\ \bibinfo {pages}
  {307} (\bibinfo {year} {1983})}\BibitemShut {NoStop}%
\bibitem [{\citenamefont
  {Vichnevetsky}(1981{\natexlab{b}})}]{vichnevetsky1981propagation}%
  \BibitemOpen
  \bibfield  {author} {\bibinfo {author} {\bibfnamefont {R.}~\bibnamefont
  {Vichnevetsky}},\ }\bibfield  {title} {\bibinfo {title} {Propagation through
  numerical mesh refinement for hyperbolic equations},\ }\href@noop {}
  {\bibfield  {journal} {\bibinfo  {journal} {Mathematics and Computers in
  Simulation}\ }\textbf {\bibinfo {volume} {23}},\ \bibinfo {pages} {344}
  (\bibinfo {year} {1981}{\natexlab{b}})}\BibitemShut {NoStop}%
\bibitem [{\citenamefont {Long}\ and\ \citenamefont
  {Thuburn}(2011)}]{long2011numerical}%
  \BibitemOpen
  \bibfield  {author} {\bibinfo {author} {\bibfnamefont {D.}~\bibnamefont
  {Long}}\ and\ \bibinfo {author} {\bibfnamefont {J.}~\bibnamefont {Thuburn}},\
  }\bibfield  {title} {\bibinfo {title} {Numerical wave propagation on
  non-uniform one-dimensional staggered grids},\ }\href@noop {} {\bibfield
  {journal} {\bibinfo  {journal} {Journal of Computational Physics}\ }\textbf
  {\bibinfo {volume} {230}},\ \bibinfo {pages} {2643} (\bibinfo {year}
  {2011})}\BibitemShut {NoStop}%
\bibitem [{\citenamefont
  {Vichnevetsky}(1987{\natexlab{a}})}]{vichnevetsky1987wave2}%
  \BibitemOpen
  \bibfield  {author} {\bibinfo {author} {\bibfnamefont {R.}~\bibnamefont
  {Vichnevetsky}},\ }\bibfield  {title} {\bibinfo {title} {Wave propagation and
  reflection in irregular grids for hyperbolic equations},\ }\href@noop {}
  {\bibfield  {journal} {\bibinfo  {journal} {Applied Numerical Mathematics}\
  }\textbf {\bibinfo {volume} {3}},\ \bibinfo {pages} {133} (\bibinfo {year}
  {1987}{\natexlab{a}})}\BibitemShut {NoStop}%
\bibitem [{\citenamefont {Frank}\ and\ \citenamefont
  {Reich}(2004)}]{frank2004spurious}%
  \BibitemOpen
  \bibfield  {author} {\bibinfo {author} {\bibfnamefont {J.}~\bibnamefont
  {Frank}}\ and\ \bibinfo {author} {\bibfnamefont {S.}~\bibnamefont {Reich}},\
  }\bibfield  {title} {\bibinfo {title} {On spurious reflections, nonuniform
  grids and finite difference discretizations of wave equations. cwi report
  mas-e0406},\ }\href@noop {} {\bibfield  {journal} {\bibinfo  {journal}
  {Center for Mathematics and Computer Science}\ } (\bibinfo {year}
  {2004})}\BibitemShut {NoStop}%
\bibitem [{\citenamefont {Ascher}\ and\ \citenamefont
  {McLachlan}(2004)}]{ascher2004multisymplectic}%
  \BibitemOpen
  \bibfield  {author} {\bibinfo {author} {\bibfnamefont {U.~M.}\ \bibnamefont
  {Ascher}}\ and\ \bibinfo {author} {\bibfnamefont {R.~I.}\ \bibnamefont
  {McLachlan}},\ }\bibfield  {title} {\bibinfo {title} {Multisymplectic box
  schemes and the korteweg--de vries equation},\ }\href@noop {} {\bibfield
  {journal} {\bibinfo  {journal} {Applied Numerical Mathematics}\ }\textbf
  {\bibinfo {volume} {48}},\ \bibinfo {pages} {255} (\bibinfo {year}
  {2004})}\BibitemShut {NoStop}%
\bibitem [{\citenamefont {Kravchenko}\ \emph {et~al.}(1999)\citenamefont
  {Kravchenko}, \citenamefont {Moin}, \citenamefont {Shariff}, \citenamefont
  {Kravchenko}, \citenamefont {Moin},\ and\ \citenamefont
  {Shariff}}]{kravchenko1999b}%
  \BibitemOpen
  \bibfield  {author} {\bibinfo {author} {\bibfnamefont {A.}~\bibnamefont
  {Kravchenko}}, \bibinfo {author} {\bibfnamefont {P.}~\bibnamefont {Moin}},
  \bibinfo {author} {\bibfnamefont {K.}~\bibnamefont {Shariff}}, \bibinfo
  {author} {\bibfnamefont {A.}~\bibnamefont {Kravchenko}}, \bibinfo {author}
  {\bibfnamefont {P.}~\bibnamefont {Moin}},\ and\ \bibinfo {author}
  {\bibfnamefont {K.}~\bibnamefont {Shariff}},\ }\bibfield  {title} {\bibinfo
  {title} {B-spline method and zonal grids for simulations of complex turbulent
  flows},\ }in\ \href@noop {} {\emph {\bibinfo {booktitle} {35th Aerospace
  Sciences Meeting and Exhibit}}}\ (\bibinfo {year} {1999})\ p.\ \bibinfo
  {pages} {433}\BibitemShut {NoStop}%
\bibitem [{\citenamefont {Shariff}\ and\ \citenamefont
  {Moser}(1998)}]{shariff1998two}%
  \BibitemOpen
  \bibfield  {author} {\bibinfo {author} {\bibfnamefont {K.}~\bibnamefont
  {Shariff}}\ and\ \bibinfo {author} {\bibfnamefont {R.~D.}\ \bibnamefont
  {Moser}},\ }\bibfield  {title} {\bibinfo {title} {Two-dimensional mesh
  embedding for b-spline methods},\ }\href@noop {} {\bibfield  {journal}
  {\bibinfo  {journal} {Journal of Computational Physics}\ }\textbf {\bibinfo
  {volume} {145}},\ \bibinfo {pages} {471} (\bibinfo {year}
  {1998})}\BibitemShut {NoStop}%
\bibitem [{\citenamefont {Kwok}\ \emph {et~al.}(2001)\citenamefont {Kwok},
  \citenamefont {Moser},\ and\ \citenamefont {Jim{\'e}nez}}]{kwok2001critical}%
  \BibitemOpen
  \bibfield  {author} {\bibinfo {author} {\bibfnamefont {W.~Y.}\ \bibnamefont
  {Kwok}}, \bibinfo {author} {\bibfnamefont {R.~D.}\ \bibnamefont {Moser}},\
  and\ \bibinfo {author} {\bibfnamefont {J.}~\bibnamefont {Jim{\'e}nez}},\
  }\bibfield  {title} {\bibinfo {title} {A critical evaluation of the
  resolution properties of b-spline and compact finite difference methods},\
  }\href@noop {} {\bibfield  {journal} {\bibinfo  {journal} {Journal of
  Computational Physics}\ }\textbf {\bibinfo {volume} {174}},\ \bibinfo {pages}
  {510} (\bibinfo {year} {2001})}\BibitemShut {NoStop}%
\bibitem [{\citenamefont {Bazilevs}\ \emph {et~al.}(2007)\citenamefont
  {Bazilevs}, \citenamefont {Calo}, \citenamefont {Cottrell}, \citenamefont
  {Hughes}, \citenamefont {Reali},\ and\ \citenamefont
  {Scovazzi}}]{bazilevs2007variational}%
  \BibitemOpen
  \bibfield  {author} {\bibinfo {author} {\bibfnamefont {Y.}~\bibnamefont
  {Bazilevs}}, \bibinfo {author} {\bibfnamefont {V.}~\bibnamefont {Calo}},
  \bibinfo {author} {\bibfnamefont {J.}~\bibnamefont {Cottrell}}, \bibinfo
  {author} {\bibfnamefont {T.}~\bibnamefont {Hughes}}, \bibinfo {author}
  {\bibfnamefont {A.}~\bibnamefont {Reali}},\ and\ \bibinfo {author}
  {\bibfnamefont {G.}~\bibnamefont {Scovazzi}},\ }\bibfield  {title} {\bibinfo
  {title} {Variational multiscale residual-based turbulence modeling for large
  eddy simulation of incompressible flows},\ }\href@noop {} {\bibfield
  {journal} {\bibinfo  {journal} {Computer Methods in Applied Mechanics and
  Engineering}\ }\textbf {\bibinfo {volume} {197}},\ \bibinfo {pages} {173}
  (\bibinfo {year} {2007})}\BibitemShut {NoStop}%
\bibitem [{\citenamefont {Lee}\ and\ \citenamefont
  {Moser}(2015)}]{lee2015direct}%
  \BibitemOpen
  \bibfield  {author} {\bibinfo {author} {\bibfnamefont {M.}~\bibnamefont
  {Lee}}\ and\ \bibinfo {author} {\bibfnamefont {R.~D.}\ \bibnamefont
  {Moser}},\ }\bibfield  {title} {\bibinfo {title} {Direct numerical simulation
  of turbulent channel flow up to $re_\tau \approx 5200$},\ }\href@noop {}
  {\bibfield  {journal} {\bibinfo  {journal} {Journal of Fluid Mechanics}\
  }\textbf {\bibinfo {volume} {774}},\ \bibinfo {pages} {395} (\bibinfo {year}
  {2015})}\BibitemShut {NoStop}%
\bibitem [{\citenamefont {Spalart}\ \emph {et~al.}(1991)\citenamefont
  {Spalart}, \citenamefont {Moser},\ and\ \citenamefont
  {Rogers}}]{spalart1991spectral}%
  \BibitemOpen
  \bibfield  {author} {\bibinfo {author} {\bibfnamefont {P.~R.}\ \bibnamefont
  {Spalart}}, \bibinfo {author} {\bibfnamefont {R.~D.}\ \bibnamefont {Moser}},\
  and\ \bibinfo {author} {\bibfnamefont {M.~M.}\ \bibnamefont {Rogers}},\
  }\bibfield  {title} {\bibinfo {title} {Spectral methods for the navier-stokes
  equations with one infinite and two periodic directions},\ }\href@noop {}
  {\bibfield  {journal} {\bibinfo  {journal} {Journal of Computational
  Physics}\ }\textbf {\bibinfo {volume} {96}},\ \bibinfo {pages} {297}
  (\bibinfo {year} {1991})}\BibitemShut {NoStop}%
\bibitem [{\citenamefont
  {Vichnevetsky}(1987{\natexlab{b}})}]{vichnevetsky1987wave}%
  \BibitemOpen
  \bibfield  {author} {\bibinfo {author} {\bibfnamefont {R.}~\bibnamefont
  {Vichnevetsky}},\ }\bibfield  {title} {\bibinfo {title} {Wave propagation
  analysis of difference schemes for hyperbolic equations: a review},\
  }\href@noop {} {\bibfield  {journal} {\bibinfo  {journal} {International
  Journal for Numerical Methods in Fluids}\ }\textbf {\bibinfo {volume} {7}},\
  \bibinfo {pages} {409} (\bibinfo {year} {1987}{\natexlab{b}})}\BibitemShut
  {NoStop}%
\bibitem [{\citenamefont {Vichnevetsky}\ and\ \citenamefont
  {Scheidegger}(1991)}]{vichnevetsky1991nonlocal}%
  \BibitemOpen
  \bibfield  {author} {\bibinfo {author} {\bibfnamefont {R.}~\bibnamefont
  {Vichnevetsky}}\ and\ \bibinfo {author} {\bibfnamefont {T.}~\bibnamefont
  {Scheidegger}},\ }\bibfield  {title} {\bibinfo {title} {The nonlocal nature
  of internal reflection in computational fluid dynamics with spectral
  methods},\ }\href@noop {} {\bibfield  {journal} {\bibinfo  {journal} {Applied
  Numerical Mathematics}\ }\textbf {\bibinfo {volume} {8}},\ \bibinfo {pages}
  {533} (\bibinfo {year} {1991})}\BibitemShut {NoStop}%
\bibitem [{\citenamefont {Waleffe}(1992)}]{waleffe1992nature}%
  \BibitemOpen
  \bibfield  {author} {\bibinfo {author} {\bibfnamefont {F.}~\bibnamefont
  {Waleffe}},\ }\bibfield  {title} {\bibinfo {title} {The nature of triad
  interactions in homogeneous turbulence},\ }\href@noop {} {\bibfield
  {journal} {\bibinfo  {journal} {Physics of Fluids A: Fluid Dynamics}\
  }\textbf {\bibinfo {volume} {4}},\ \bibinfo {pages} {350} (\bibinfo {year}
  {1992})}\BibitemShut {NoStop}%
\bibitem [{\citenamefont {Cook}\ and\ \citenamefont
  {Cabot}(2005)}]{cook2005hyperviscosity}%
  \BibitemOpen
  \bibfield  {author} {\bibinfo {author} {\bibfnamefont {A.~W.}\ \bibnamefont
  {Cook}}\ and\ \bibinfo {author} {\bibfnamefont {W.~H.}\ \bibnamefont
  {Cabot}},\ }\bibfield  {title} {\bibinfo {title} {Hyperviscosity for
  shock-turbulence interactions},\ }\href@noop {} {\bibfield  {journal}
  {\bibinfo  {journal} {Journal of Computational Physics}\ }\textbf {\bibinfo
  {volume} {203}},\ \bibinfo {pages} {379} (\bibinfo {year}
  {2005})}\BibitemShut {NoStop}%
\bibitem [{\citenamefont {Grinstein}\ \emph {et~al.}(2007)\citenamefont
  {Grinstein}, \citenamefont {Margolin},\ and\ \citenamefont
  {Rider}}]{grinstein2007implicit}%
  \BibitemOpen
  \bibfield  {author} {\bibinfo {author} {\bibfnamefont {F.~F.}\ \bibnamefont
  {Grinstein}}, \bibinfo {author} {\bibfnamefont {L.~G.}\ \bibnamefont
  {Margolin}},\ and\ \bibinfo {author} {\bibfnamefont {W.~J.}\ \bibnamefont
  {Rider}},\ }\href@noop {} {\emph {\bibinfo {title} {Implicit large eddy
  simulation: computing turbulent fluid dynamics}}}\ (\bibinfo  {publisher}
  {Cambridge university press},\ \bibinfo {year} {2007})\BibitemShut {NoStop}%
\bibitem [{\citenamefont {Haering}\ \emph {et~al.}(2020)\citenamefont
  {Haering}, \citenamefont {Oliver},\ and\ \citenamefont
  {Moser}}]{haering2020active}%
  \BibitemOpen
  \bibfield  {author} {\bibinfo {author} {\bibfnamefont {S.~W.}\ \bibnamefont
  {Haering}}, \bibinfo {author} {\bibfnamefont {T.~A.}\ \bibnamefont
  {Oliver}},\ and\ \bibinfo {author} {\bibfnamefont {R.~D.}\ \bibnamefont
  {Moser}},\ }\bibfield  {title} {\bibinfo {title} {Active model split hybrid
  rans/les},\ }\href@noop {} {\bibfield  {journal} {\bibinfo  {journal} {arXiv
  preprint arXiv:2006.13118}\ } (\bibinfo {year} {2020})}\BibitemShut {NoStop}%
\end{thebibliography}%

\end{document}